\documentclass[english,12pt]{article}
\usepackage[T1]{fontenc}
\usepackage[latin9]{inputenc}
\usepackage{geometry}
\usepackage{color}
\usepackage{babel}
\usepackage{verbatim}
\usepackage{amsmath}
\usepackage{amssymb}
\usepackage{cancel}
\usepackage{subcaption}
\usepackage{esint}
\usepackage{csquotes}
\usepackage{xargs}[2008/03/08]

\usepackage{hyperref}
\hypersetup{unicode=true,bookmarksnumbered=true,bookmarksopen=true,bookmarksopenlevel=5,bookmarksdepth=4,
 breaklinks=false,pdfborder={0 0 1},
 colorlinks,
	citecolor=red,
	filecolor=blue,
	linkcolor=blue,
	urlcolor=blue}

\makeatletter
\usepackage{braket}
\usepackage{faktor}
\usepackage{simpler-wick}
\usepackage[T1]{fontenc}
\usepackage{graphicx}
\usepackage{authblk}

\numberwithin{equation}{section}

\usepackage[backend=biber,sorting=none,style=numeric-comp,url=false,maxbibnames=4]{biblatex}
\renewbibmacro{in:}{}
\usepackage{babel}
\bibliography{references}{}

\newcommandx\kappah{{\hat \kappa}}
\newcommandx\lambdah{{\hat \lambda}}
\newcommandx{\todo}[1]{\textcolor{red}{#1}}
\newcommandx\ab{{\alpha \beta}}
\newcommandx\al{{\alpha}}
\newcommandx\be{{\ensuremath{\beta}}}
\newcommandx\phisq{{\bar{\Phi}\Phi}}
\newcommandx\phisqbr{{\left(\phisq\right)}}
\newcommandx\phisqexp{{\left\langle\phisq\right\rangle}}
\newcommandx\lambdasigph{{\lambda_{\sigma \Phi}}}
\newcommandx\lambdasig{{\lambda_{\sigma}}}

\newcommandx\incgraph[2][usedefault, addprefix=\global, 1=text, 2=.3]{\includegraphics[width=#2\linewidth]{images/#1.pdf}}%

\renewcommandx\incgraph[2][usedefault, addprefix=\global, 1=text, 2=.3]{\ensuremath{\begin{gathered}\includegraphics[width=#2\linewidth]{images/#1.pdf}\end{gathered}}}

\makeatother

\title{Full Phase Diagram of a UV Completed \texorpdfstring{$\mathcal{N}=1$}{N=1} Yang-Mills-Chern-Simons Matter Theory}
\author{Adar Sharon}
\author{Tal Sheaffer}
\affil{Department of Particle Physics and Astrophysics, \\
Weizmann Institute of Science, Rehovot 7610001, Israel \\
\small \tt \href{mailto:adar.sharon@weizmann.ac.il}{adar.sharon@weizmann.ac.il}, \small \tt \href{mailto:tal.sheaffer@weizmann.ac.il}{tal.sheaffer@weizmann.ac.il}
}

\begin{document}

\maketitle

\abstract{We study the large $N$ phase diagram of an asymptotically free UV completion of $\mathcal{N}=1$ $SU(N)$ super-Yang-Mills-Chern-Simons theory coupled to a single massive fundamental scalar multiplet with a quartic superpotential coupling. We compute the effective superpotential at small gauge coupling $\lambda\equiv N/k$, and combine this with previous results in the literature to obtain the full phase diagram in this regime. We find that tuning the UV parameters allows us to reach various phases and fixed points of Chern-Simons theory that were recently discovered using large $N$ techniques, as well as new phases that characterize the Yang-Mills theory. We also conjecture the form of the phase diagram for general values of $\lambda$ and for finite $N$.}

\newpage

\tableofcontents{}

\newpage

\section{Introduction}

Supersymmetry (SUSY) has proven to be an invaluable tool for studying quantum field theories (QFTs). 
In particular, studying SUSY theories has lead to a better understanding of non-SUSY theories as well. It is then useful to study theories with minimal SUSY, since these will hopefully be as similar as possible to non-SUSY theories.

To this end, in this paper we continue the study of $3d$ $\mathcal{N}=1$ supersymmetric Chern-Simons-matter theories. These theories have been discussed recently in the literature \cite{Choi:2018ohn,Gaiotto:2018yjh,Dey:2019ihe,Gomis:2017ixy,Eckhard:2018raj,Benini:2018umh,Benini:2018bhk,Bashmakov:2018wts,Inbasekar:2015tsa,Aharony:2019mbc}, mostly in relation to the $3d$ Bosonization dualities. In fact, we will study a more general \textbf{Yang-Mills}-Chern-Simons theory, as in \cite{Choi:2018ohn}, which flows in the IR to such Chern-Simons theories. Indeed, there exist $3d$ $\mathcal{N}=1$ versions of the Bosonization dualities which are under slightly more control than the non-SUSY versions.

$3d$ $\mathcal{N}=1$ SUSY is more similar to non-SUSY theories than to theories with higher SUSY. This is mainly due to the fact that the superpotential for $\mathcal{N}=1$ theories is not holomorphic, which takes away most of the power of SUSY. Despite this shortcoming, some exact results can still be extracted using this minimal amount of SUSY. A recent example is a ``non-renormalization theorem'', which requires only a discrete $\mathbb{Z}_2$ R-symmetry which is related to $\mathcal{N}=1$ SUSY \cite{Gaiotto:2018yjh}.

There are a couple of important aspects of SUSY which do not depend on the holomorphicity of the superpotential. First, the Witten index \cite{Witten:1982df} still exists, and is a useful non-perturbative tool. Second, since SUSY-preserving vacua must have zero energy, phase transitions between SUSY-preserving vacua must be second-order (or higher), and so they correspond to conformal field theories (CFTs). This fact is very important when discussing IR dualities, since most dualities require (at least) second-order transitions. These facts will simplify some of the calculations in this paper.

The theory we will be discussing in this paper is a $3d$ $\mathcal{N}=1$ $SU(N)$ Yang-Mills-Chern-Simons (YMCS) theory coupled to a single fundamental Bosonic superfield $\Phi$. We will add a quartic superpotential for $\Phi$:
\begin{equation}\label{eq:quadratic_W}
    W=m\bar \Phi\Phi +\frac{\omega}{2}(\bar\Phi\Phi)^2\;.
\end{equation}
This theory has five parameters - the rank $N$, the CS level $k$, the gauge coupling $g^2$, the mass $m$ and the quartic coupling $\omega$. We will focus on the large-$N$ limit, where we also define the 't Hooft coupling $\lambda=\frac{N}{k+\frac{N}{2}}$ (see appendix \ref{app:CS_conventions} for our CS conventions). We will discuss the phase diagram and the various phase transitions as a function of these parameters. 

More precisely, we will be studying a small variation of this theory. The theory described above is not UV-complete, and must be regulated carefully. To fix this, we will add an additional singlet field $\sigma$, and consider the most general super-renormalizable - and most general asymptotically free - superpotential:\footnote{\label{footnote:super-renormalizable} The quartic superpotential couplings omitted from this Lagrangian correspond to the classically marginal operators. In 3d these are sextic Bosonic self-interactions and certain Bose-Fermi interactions. Quantum mechanically, these are marginally irrelevant, and so \eqref{eq:UV_superpotenial} is in fact the most general asymptotically free superpotential.}
\begin{equation}\label{eq:UV_superpotenial}
    W=m\bar \Phi\Phi + \sigma \bar\Phi\Phi-\frac{\sigma^2}{2\omega}+\lambda_3\sigma^3\;.
\end{equation}
We will mostly be interested in the theory with $\lambda_3=0$. In this case, naively integrating out $\sigma$ gives us the quadratic superpotential \eqref{eq:quadratic_W}. In this sense this theory is a UV completion of our original CS-matter theory.

Some limits of this theory have already been discussed in the past. In particular, the phase diagram was found for two specific cases:
\begin{enumerate}
    \item In the ``conformal limit'' at infinite $N$ and $kg^2$ \cite{Dey:2019ihe}. By this we mean that at infinite $N$ the theory \eqref{eq:quadratic_W} has a line of fixed points (schematically parametrized by $\omega$) which can be reached by tuning the mass. The phase diagram for all $\omega$ and $m$ was found around this conformal manifold at infinite $N$ by computing the free energy in the large-$N$ limit in both the ``Higgs branch'' (i.e. assuming $\langle\Phi\rangle)\neq 0$) and in the ``unHiggsed branch'' (i.e. assuming $\langle\Phi\rangle)= 0$). The IR CFT at this line of fixed points is the CS-matter theory of \cite{Dey:2019ihe,Aharony:2019mbc}. We note that at one point in the parameter space of this IR theory ($\omega_{\rm IR}=2\pi\lambda,m_{\rm IR} =0$), SUSY enhances to $\mathcal{N}=2$. Hence there is a corresponding point in the theory \eqref{eq:UV_action_defs} which has emergent IR $\mathcal{N}=2$ SUSY. At large but finite $N$, where a beta function for $\omega$ emerges this point becomes a stable fixed point of the flow. This is of no consequence to our calculations. The full YM-CS theory cannot have enhanced SUSY with the field content of \eqref{eq:UV_action_defs}, as an additional gluino is needed.
    \item In the limit $\omega=0$ \cite{Choi:2018ohn}. In this case the phase diagram was computed by calculating the 1-loop effective superpotential in the ``Higgsed'' phase for the theory \eqref{eq:quadratic_W}. The full 1-loop superpotential takes the form:
    \begin{equation}\label{eq:res}
        m\phisq-\frac{1}{2N}\kappa{\rm Tr}_{{\rm adj}}\sqrt{\kappa^2+16\pi\lambda\kappa\bar\Phi T^{(a}T^{b)}\Phi}\;,
    \end{equation}
    where $a,b$ are color indices, and we have ignored unimportant constant factors. This superpotential can be used to find the phase diagram of the theory. To leading order in small $\omega$, the only correction that is required is adding a term $\frac{\omega}{2}|\Phi|^4$ to \eqref{eq:res}.
\end{enumerate}
We will review both of these results in the next section.

In this paper we will generalize these results. We will extend the phase diagram to all values of the parameters, including large but finite $N$. In order to achieve this, we will calculate the 2-loop effective superpotential for $\sigma$ in the theory \eqref{eq:UV_superpotenial} at large $N$ in the unHiggsed phase (this is equivalent to an expansion in small 
$\lambda$). We will then combine this with previous results to find the complete phase diagram. At leading order in small $\lambda$ we can compute the full phase diagram, which appears in figure \ref{fig: phase diagram infinite N leading lambda}. This is the main result of this paper. We then discuss possible extensions of this phase diagram to the strongly-coupled regions where $\lambda$ is not necessarily small, and also large but finite values of $N$. 

This analysis solves some puzzles that appear when considering the previous results more carefully. For example, there is an apparent mismatch in the number of vacua in certain regions in parameter space. On one hand, the analysis near the conformal limit in \cite{Dey:2019ihe} finds up to two vacua for any value of the parameters. On the other hand, for small $\omega$ away from the conformal limit we can find the vacua by adding the term $\frac{\omega}{2}|\Phi|^4$ to the superpotential \eqref{eq:res}. It is then easy to show that there exists a region with three vacua for small negative $\omega$ and small positive $m$ (assuming $k>0$). We will show that the resolution to this puzzle is that the additional vacuum is at a distance of order $kg^2$ in parameter space near the conformal limit, so that in the conformal limit (where $g^2\to\infty$) this vacuum is at infinity and so is not visible. 

Another interesting phenomenon discussed in the conformal limit was the appearance of ``walls'' in parameter space across which the Witten index jumps. As discussed in \cite{Witten:1982df}, this is due to vacua escaping to infinity in field space. We will show that once the UV theory is regulated carefully, some of these vacua actually escape to a distance $kg^2$ in field space, where they meet an additional vacuum that is invisible in the conformal limit.

The rest of this paper is organized as follows. We begin in section \ref{sec:bg} by reviewing known results about this theory, and use them to obtain the phase diagram in some limits. Next, in section \ref{sec: computing the superpotential} we compute the 2-loop effective potential for the $\sigma$ field in the CS-matter theory with superpotential \eqref{eq:UV_superpotenial}. Finally, in section \ref{sec:Phase Diagram} we put together these results to plot the full phase diagram. The result for infinite $N$ and small $\lambda$ can be computed exactly, and appears in figure \ref{fig: phase diagram infinite N leading lambda}. We also discuss possible forms for the phase diagram away from finite $\lambda$ and infinite $N$. Various computations are described in detail in the Appendices.

\section{Background}\label{sec:bg}

In this section we review known results about $3d$ $\mathcal{N}=1$ CS-matter theories. The theory we are interested in discussing is a $3d$ $\mathcal{N}=1$ YMCS matter theory with gauge group $SU(N)$ with one fundamental matter field $\Phi$. The action of the theory is given by (our superspace and CS conventions are summarized in appendix \ref{app:conventions}):
\begin{equation}\label{eq:IR_action}
S_{IR}=N\int d^5z\left(\mathcal{L}_{{\rm YM}}+\mathcal{L}_{{\rm CS}}+\mathcal{L}_{{\rm matter,IR}}\right)\;,
\end{equation}
where
\begin{equation}\label{eq:IR_action_defs}
\begin{split}
\mathcal{L}_{{\rm YM}}&=\frac{1}{g^{2}N}{\rm tr}W^{2}\;,\\
\mathcal{L}_{{\rm CS}}&=\frac{k}{4\pi N}{\rm tr}\frac{1}{2}\Gamma^{\alpha}\left(W_{\alpha}-\frac{1}{6}\left\{ \Gamma^{\beta},\Gamma_{\alpha\beta}\right\} \right)\;,\\
\mathcal{L}_{{\rm matter,IR}}&=\bar{\Phi}\nabla^{2}\Phi+m\phisq+\frac{\omega}{2}\left(\phisq\right)^{2}\;.
\end{split}
\end{equation}

As discussed in the introduction, this theory requires a UV completion at finite $N$ due to the marginally irrelevant $\Phi^4$ coupling. The UV completion we choose requires adding an additional singlet field $\sigma$. The action of the theory is now
\begin{equation}\label{eq:UV_action}
S_{UV}=N\int d^5z\left(\mathcal{L}_{{\rm YM}}+\mathcal{L}_{{\rm CS}}+\mathcal{L}_{{\rm matter,UV}}\right)\;,
\end{equation}
where
\begin{equation}\label{eq:UV_action_defs}
\mathcal{L}_{{\rm matter,UV}}=\bar{\Phi}\nabla^{2}\Phi+\frac{1}{2\lambda_{\sigma\Phi}^{2}}\sigma D^{2}\sigma+\left(m+\sigma\right)\phisq-\frac{\sigma^{2}}{2\omega}+\lambda_3\sigma^3\;.
\end{equation}
We will refer to the two theories above schematically as the ``IR'' and ``UV'' theories. We will also make frequent use of the $F$-term equation for $\sigma$ at $\lambda_3=0$:
\begin{align}
    \left<\sigma\right> = \omega \phisq.\label{eq:sigma EOM relation to Phi}
\end{align}
Note that this equation can be corrected quantum-mechanically, unlike the case with more SUSY. 

Some comments are in order. First, we have explicitly normalized our theory such that there is a factor of $N$ outside of the action, since we will be focusing on a large-$N$ expansion of the theory where $N$ will be the loop-counting parameter of our theory. Next, we will mostly be interested in the theory with $\lambda_3=0$, and will discuss $\lambda_3$ towards the end of this paper (note that integrating out the field $\sigma$ at $\lambda_3=0$ reproduces the $\Phi^4$ term). We have also ignored terms that are linear in $\sigma$ in \eqref{eq:UV_action_defs} since these can be removed by a shift of $\sigma$ and a redefinition of some couplings. 
Finally, the non-canonical normalization of $\sigma$ in \eqref{eq:UV_action_defs} has been chosen to make contact with the case $\lambdasigph\to \infty,\lambda_3=0$ where it is genuinely auxiliary.

We will mainly consider the theory at large $N$ and $k$, in which case the theory is a function of the 't Hooft coupling $\lambda=\frac{N}{k+\frac{N}{2}}$ and the YM scale $\kappa=\frac{kg^2}{4\pi}$. At infinite $N$, the ``IR'' theory \eqref{eq:IR_action} has a conformal limit where $\kappa\to\infty,m\to 0$ while $\lambda,\omega$ can be varied. The theory thus lives on a conformal manifold, and does not require a UV completion. At order $1/N$ the theory is no longer conformal, and we need to discuss the UV theory \eqref{eq:UV_action}.

The main point of this paper is to find the phase diagram of this theory as a function of the parameters $\lambda,m/\kappa,\omega$ at large $N$. In general, this is done by computing the Coleman-Weinberg effective superpotential. We will focus on some specific limts:
\begin{enumerate}
    \item The conformal limit, where we are close to a CS-matter fixed point (near $m=0$).
    \item The small $\lambda$ limit, where the theory is weakly coupled.
    \item The small $\omega$ limit, where we can perform perturbation theory in $\omega$.
\end{enumerate}
Some of these calculations have already been done. In particular, the effective potential in the conformal limit was found in \cite{Dey:2019ihe}, and the effective action for $\omega=0$ in the Higgsed phase was found in \cite{Choi:2018ohn}. We will review these results in this section. In the rest of this paper we will generalize these computations and also perform the computation for general $\omega$ in the unHiggsed phase. We will find agreement with previous results in the relevant limits.

\subsection{\label{subsec: witten index and vacua}Vacua and the Witten Index}

We start by discussing the possible low-energy theories in our model and the behavior of the Witten index. The theories we will consider will have four possible low-energy theories, characterized by two signs $(\delta k,\delta N)$ which are summarized in Table \ref{table:phases} (following most of the notation in \cite{Dey:2019ihe}, but working with a different CS level regularization).
\begin{table}[ht!]
\centering
\begin{tabular}{ c c c c }
 Phase & CS level shift ($\delta k$) & Higgsed/UnHiggsed ($\delta N$) & Low-energy $\mathcal{N}=1$ theory \\
 \hline
 $(+,+)$ & +ve & unHiggsed & $SU(N)_{k+\frac12}$ \\ 
 $(-,+)$ & -ve & unHiggsed & $SU(N)_{k-\frac12}$ \\  
 $(+,-)$ & +ve & Higgsed & $SU(N-1)_{k}$ \\   
 $(-,-)$ & -ve & Higgsed & $SU(N-1)_{k-1}$ \end{tabular}
\caption{Possible phases in the IR.}
\label{table:phases}
\end{table}

Semiclassically\footnote{More generally, since $\Phi$ is not gauge invariant, Higgsing can only be defined by the rank of the IR topological theory.}, the vacua are characterized by whether the vev $\langle\bar\Phi\Phi\rangle$  is zero (which leads to either a Higgsed or an unHiggsed vacuum, meaning that the rank of the gauge group $N$ changes by $0$ (unHiggsed) or $-1$ (Higgsed)), and whether the effective mass of the Fermion is positive or negative (which corresponds to a negative or a positive shift for the CS level $k$, $\pm 1/2$). Note that due to our CS conventions, when the gauge group is Higgsed the CS level is shifted as well (this is easy to see since the combination $k+\frac{N}{2}$ must remain an integer). 

Up to a possible sign, the Witten index of a low-energy $\mathcal{N}=1$ $SU(N)_k$ CS-theory is given by \cite{Witten:1999ds}
\begin{equation}\label{eq:Witten_Index}
I=\begin{pmatrix}k+\frac{N}{2}-1\\N-1\end{pmatrix}\;.
\end{equation}
Note that this vanishes for $k<\frac{N}{2}$, which is consistent with SUSY breaking.
We will thus modify this notation by adding a sign $s$, so that a full low-energy theory is completely characterized by the notation:
\begin{align}
\left(\delta k,\delta N\right)_s.\label{eq:vacuum notation}
\end{align}
Here $s=\pm$ denotes whether this vacuum contributes to the Witten index with a positive or negative sign. We will omit $s$ if it is positive and only include it if it is negative. For example, by definition the Witten index of these theories obeys the identity
\begin{align}
I\left(\left(\delta k,\delta N\right)_-\right)=-I\left(\left(\delta k,\delta N\right)\right).\label{eq:Witten index statistics flip}
\end{align}
The sign $s$ depends nontrivially  on
gapped singlet excitations over the vacua. For instance, if the gauge group is Higgsed due to some matter field $\Phi$, the sign of the mass of the \textbf{longitudinal} mode of $\Phi$ (fluctuations in the VEV of $\Phi$) has no bearing on the IR level or rank, but does affect the statistics. The ground states will have the form $ \left| 0 \right>_{\rm singlet} \otimes \left| 0 \right>_{\rm CS} $, so the contribution to the Witten index will factorize. If the mass of the singlet mode is negative, then $ \left| 0 \right>_{\rm singlet}$ is a Fermionic vacuum, so we'll get:
\begin{align}
\left( -1 \right)^{\rm F} \left| 0 \right>_{\rm singlet} \otimes \left| 0 \right>_{\rm CS}=-  \left| 0 \right>_{\rm singlet} \otimes \left(\left( -1 \right)^{\rm F} \left| 0 \right>_{\rm CS}\right).
\end{align}
Therefore, one could say that the statistics of the vacuum are flipped. 

In this paper we will follow the Witten index very carefully, and so the sign $s$ will be very important. In particular, we will find many codimension-1 ``wall-crossings'' as we vary the parameters, across which the Witten index jumps. Such wall crossings occur only as a function of real parameters, and so they are more common in $3d$ $\mathcal{N}=1$ theories than in theories with higher SUSY. For simple cases like a single matter multiplet, such a wall crossing is attributed to a change in the asymptotics of the superpotential \cite{Witten:1982df}. For example, the Witten index of a single $3d$ $\mathcal{N}=1$ real multiplet $\Phi$ with superpotential $m\Phi^2$ is $\text{sign}(m)$ (up to an overall sign ambiguity). We thus find a wall-crossing at $m=0$ where the Witten index jumps, which is related to a change of the large-vev behavior of the superpotential at $m=0$. See \cite{Ghim:2019rol} for a discussion about more complicated superpotentials. 

The relation between the position of wall-crossings and the asymptotics of the superpotential has a very intuitive explanation. A change in the asymptotics of the superpotential allows for a flat direction to appear, and as a result new vacua can appear (or disappear) ``from infinity'' (for a specific example see \cite{Bashmakov:2018wts}). We will find many examples where this occurs in this paper.

The situation is very different for phase transitions, which we will also encounter. When two (or more) SUSY vacua merge, we find a second-order phase transition, and the Witten index does not jump. This will give us a nice consistency check on the corresponding phase diagram. 
The basic Witten index identity that will ensure continuity when crossing a phase transition in our phase diagram is
\begin{align}
I\left(\left( +,+\right)\right) = I\left(\left( -,+\right)\right)+I\left(\left( +,-\right)\right),\label{eq:Witten index continuity identitiy}
\end{align}
and variants thereof obtained with \eqref{eq:Witten index statistics flip}. This can be proven easily by plugging in the result \eqref{eq:Witten_Index}. Equation \eqref{eq:Witten index continuity identitiy} is simply the $N_{\rm f}=1$ version of equation (D.5) of \cite{Choi:2018ohn}.

To summarize, the interesting parts of our phase diagrams will be codimension-1 wall crossings and phase transitions. In the former vacua escape to infinity (or appear from infinity) and the Witten index jumps, while in the latter vacua merge together and the Witten index does not change.

\subsection{\label{subsec: classical phase diagram}Tree Level Phase Diagram}

To familiarize ourselves with the notation, we begin by analyzing the phases of the classical theory. We will only be able to trust the result when quantum corrections are negligible, but it is a good starting point nevertheless.
The superpotential of our UV theory \eqref{eq:UV_action} at $\lambda_3=0$ is given by: 
\begin{align}
W=\left(m+\sigma\right)\phisq-\frac{\sigma^{2}}{2\omega},\label{eq:tree level superpotential}
\end{align}
which leads to the F-term equations
\begin{align}
\sigma & =\omega\phisq\\
0 & =\Phi^{i}\left(m+\sigma\right),\,i=1,\dots,N.
\end{align}
We find an unHiggsed vacuum always exists at $\Phi=\sigma=0$, and a Higgsed
vacuum exists for ${\rm sign}\left(m\omega\right)=-1$ at:
\begin{align}
\phisq=\frac{\sigma}{\omega}=-\frac{m}{\omega}.
\end{align}

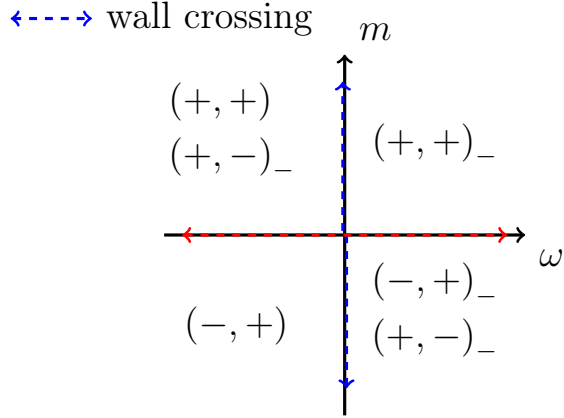
\begin{figure}
\centering
\begin{tikzpicture}[scale=1.2, every node/.style={transform shape}]


\draw[very thick,->] (-2,0) -- (2,0) node[anchor=north west] {$\omega$};
\draw[ very thick,->] (0,-2) -- (0,2) node[black, anchor=south west] {$m $};

\draw[blue,dashed, very thick,<->] (0.02,-1.7) -- (0.02,0) -- (-0.02,0) -- (-0.02,1.7);

\draw[red,dashed, very thick,<->] (-1.8,0) -- (1.8,0);

\draw (1,1) node[black]{$\left( +,+\right)_-$};
\draw (1,-1) node[black,text width=1.5cm,align=center]{$\left( -,+\right)_- $ \vspace{.3cm} $\left( +,-\right)_-$};
\draw (-1.2,-1) node[black]{$\left( -,+\right)$};
\draw (-1.2,1) node[text width=1.5cm,align=left,black]{$\left( +,+\right)$ \vspace{.3cm} $ \left( +,-\right)_-$};

\draw[ very thick,<->,blue,dashed] (-3.7,2.4) -- (-2.8,2.4) node[black, anchor=west ] {wall crossing};

\end{tikzpicture}
\caption{Phase diagram of the classical theory. The blue line denotes a wall crossing and the red line a phase transition.}
\label{fig: classical phase diagram}  
\end{figure}

The phase diagram is presented in figure \ref{fig: classical phase diagram}. $\lambda$ is taken to be positive in the figure. As expected a wall crossing appears when the leading interaction term $\frac{\omega}{2} \left(\phisq\right)^2\sim  \frac{\sigma^2}{2\omega}$ vanishes, and upon restricting to the $\omega=0$ line, there is a ``wall crossing within the wall crossing'' at $m=0$ - the one in the analysis of \cite{Choi:2018ohn}. The Higgsed vacuum (dis-)appears (to) from infinity at the wall crossing, so that the Witten index can be viewed as continuous from the right (left) for $m>0$ ($m<0$). Classically we also find a second-order phase transition along the line $m=0$, denoted by a red line. Note that due to the identity \eqref{eq:Witten index continuity identitiy}, the Witten index does not jump across the phase transition at $m=0$. Precisely at $m=\omega=0$ there is classically a moduli space of vacua and the Witten index is ill-defined. 

The statistics of all vacua in the $\omega>0$ half-plane are flipped, because of the mass of $\sigma$ - $-\frac{1}{\omega}$. The Higgsed vacuum $\left(+,-\right)_-$ in the second quadrant is also flipped. The graph of \eqref{eq:tree level superpotential} in this quadrant makes clear that the  longitudinal mode of $\Phi$ around the Higgsed vacuum (fluctuations in the VEV) has a negative mass, thereby flipping the statistics. This assignment of statistics makes the Witten index continuous wherever it should be.

Note that if we take $\sigma$ to be purely auxiliary (by taking $\lambdasigph\to \infty$ in \eqref{eq:UV_action_defs}) so as to arrive at the theory \eqref{eq:IR_action_defs}, then we should undo this statistics-flipping in the $\omega>0$ half-plane.

\subsection{\label{subsec:The Conformal Limit}The Conformal Limit}

Next we review the conformal limit of $SU(N)$ CS-matter theories with a single fundamental matter field (with no YM term). This corresponds to the ``IR'' theory \eqref{eq:IR_action} close to $\kappa=\infty,m=0$. 
In this section we follow the notation of \cite{Dey:2019ihe}, and we will explicitly give the mapping to our ``IR'' parameters \eqref{eq:IR_action}. The theory is defined by the following action:
\begin{equation}
S=\int d^5z\left(\mathcal{L}_{{\rm CS}}+\mathcal{L}_{{\rm matter}}\right)\;,
\end{equation}
where
\begin{equation}\label{eq:Minwalla_Defs}
\begin{split}
\mathcal{L}_{{\rm matter}}&=\bar \Phi\nabla^2\Phi+\mu |\Phi|^2 +\frac{\pi w}{\kappah}|\Phi|^4\;,\\
\mathcal{L}_{{\rm CS}}&=\frac{\hat\kappa}{4\pi }{\rm tr}\frac{1}{2}\Gamma^{\alpha}\left(W_{\alpha}-\frac{1}{6}\left\{ \Gamma^{\beta},\Gamma_{\alpha\beta}\right\} \right)\;.
\end{split}
\end{equation}
and is regularized using dimensional regularization (see appendix \ref{app:CS_conventions} for a discussion about the precise definition of the CS level for different regularizations).
The parameters of the theory are an 't Hooft parameter $\lambdah=\frac{N}{\kappah}$, along with a mass term $\mu$ and a $\Phi^4$ coupling $w$. These are related to the parameters of the ``IR'' theory \eqref{eq:IR_action_defs} by $\mu=m$ and $\omega=\frac{w}{2\pi\hat\lambda}$. In our discussion we will focus on positive $\lambdah$; the results for negative $\lambdah$ are obtained by performing a parity transformation, which flips the sign of $\mu,w$. At infinite $N$, both $\lambdah$ and $w$ are exactly marginal couplings, and the theory has a conformal manifold. Turning on a mass $\mu$ takes us away from this conformal manifold.

\begin{figure}
  \centering
  \begin{tikzpicture}


\draw[very thick,->] (-3,0) -- (3,0) node[anchor=north west] {$w$};
\draw[ very thick,->] (0,-3) -- (0,3) node[black, anchor=south west] {$\mu $};

\draw[very thick,dashed,<->,red] (-2.6,0) -- (2.6,0);

\draw[blue,dashed, very thick,<->] (1.5,-2.5) -- (1.5,2.5);
\draw[blue,dashed, very thick,<->] (-0.5,-2.5) -- (-0.5,2.5);
\draw[blue,dashed, very thick,<->] (-2,-2.5) -- (-2,2.5);

\draw (2.5,1.5) node[black]{\begin{tabular}{c}$( +,+)$\\$(-,+)$ \end{tabular}};
\draw (2.5,-1.5) node[black]{$( +,-)$};

\draw (0.7,1.5) node[black]{$(+,+)$};
\draw (0.7,-1.5) node[black]{\begin{tabular}{c}$( +,-)$\\$(-,+)$ \end{tabular}};

\draw (-1.2,1.5) node[black]{\begin{tabular}{c}$( +,+)$\\$(+,-)$ \end{tabular}};
\draw (-1.2,-1.5) node[black]{$( -,+)$};	

\draw (-2.8,1.5) node[black]{$(+,-)$};
\draw (-2.8,-1.5) node[black]{\begin{tabular}{c}$( +,+)$\\$(-,+)$ \end{tabular}};

\draw[blue,thick,fill=blue] (1.5,0) circle (0.07cm);
\draw (1.5,0) node[blue,anchor=north east]{$a_3$};
\draw[blue,thick,fill=blue] (-0.5,0) circle (0.07cm);
\draw (-0.5,0) node[blue,anchor=north east]{$a_2$};
\draw[blue,thick,fill=blue] (-2,0) circle (0.07cm);
\draw (-2,0) node[blue,anchor=north east]{$a_1$};

	\draw[ very thick,<->,blue,dashed] (-3.5,3) -- (-2.7,3) node[black, anchor=west ] {wall crossing};

\end{tikzpicture}
  \label{fig:Minwalla_phase}
    \caption{Phase diagram in the conformal limit for $\lambdah>0$. The vacua are defined in Table \ref{table:phases}. The values of the wall crossings $a_i$ are $a_1=a_2^{-1}=-\frac{2-|\lambdah|}{|\lambdah|}$ and $a_3=\frac{2+|\lambdah|}{|\lambdah|}$. There is a second order phase transition at $\mu=0$ with a CS-matter fixed point.}
\label{fig:Minwalla_phase_diagram}
\end{figure}
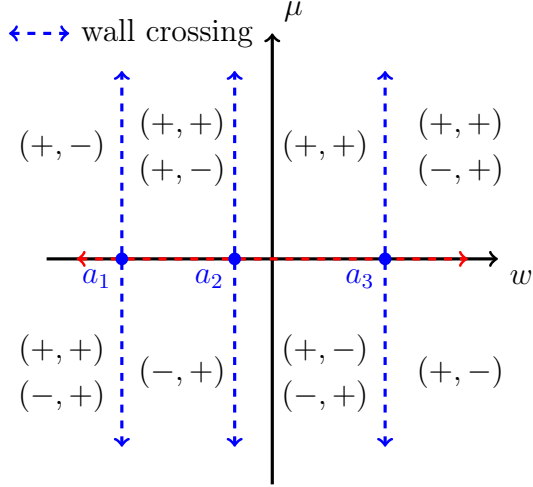

The authors of \cite{Dey:2019ihe} calculated the effective potential for this theory to all orders in the parameter $\lambdah$ at infinite $N$, which allows them to find the vacuum structure of the theory as a function of $\mu,w$. There are four possible vacua which can appear, which are summarized in Table \ref{table:phases}.
We present the result for positive $\lambdah$ in figure \ref{fig:Minwalla_phase_diagram}.

Note that in going to negative $\lambdah$ using a parity transformation, some of the signs of the CS level shift are also flipped, so that some vacua change. It turns out that the result for negative $\lambdah$ is obtained by flipping the phase diagram around the $\mu=0$ axis.

There are a few important things to note in the phase diagram figure \ref{fig:Minwalla_phase_diagram}. First, the phase diagram includes ``wall crossings'' at $w=a_1,a_2,a_3$, denoted in blue in the diagram, where the Witten index jumps as a result of vacua escaping to infinity in field space. This is in contrast to the line $\mu=0$, where there is no jump in the Witten index; there, vacua merge at a finite distance in parameter space and we find a second-order phase transition. In the following, we will discuss a UV completion of the theory at finite $\mu$, which will change the behavior of the vacua at infinity and of the wall crossings. 

Second, the vacuum $(-,-)$ does not appear. The $\left(-,-\right)$ vacua are not only absent from the phase diagram figure \ref{fig:Minwalla_phase_diagram} of the $\mathcal{N}=1$ locus of the theory of \cite{Dey:2019ihe}, but also from all of the phase diagrams we draw for the YM-CS theory  \eqref{eq:UV_action}, in section \ref{sec:Phase Diagram}. We do not expect these vacua to appear anywhere in the SUSY theory. Since \cite{Dey:2019ihe} worked in a more general non-SUSY theory with fundamental Fermions and Bosons, it was possible to vary the Fermion mass independently from the Boson mass, making the appearance of $\left(-,-\right)$ vacua outside the $\mathcal{N}=1$ locus natural. But in the \textbf{manifestly} SUSY theory, as can be seen from the SUSY unitary gauge discussed in subsection \ref{subsubsec: N counting Higgsed}, the fundamental Fermion in the Higgsed phase joins a super-multiplet with the fundamental $W$-Bosons, tying its mass to theirs. Hence the appearance of a SUSY $\left(-,-\right)$ vacuum seems to entail that the $W$-Boson mass changes sign. This could not happen smoothly, since the small mass regime is governed by the theory of \cite{Dey:2018ykx}, nor could such a vacuum appear from infinity - since the deep UV is governed by the ungauged theory, where again no such vacua are present. We cannot, however, rule out that at large coupling $\lambda$, at intermediate VEVs $\phisq\approx |\kappa|$, $\left(-,-\right)$ vacua appear in statistics-reversed pairs. In fact, it is not clear that the manifestly SUSY procedure we employ to compute the phase diagram - the effective superpotential of section \ref{sec: computing the superpotential} - is able to detect such vacua at large $\lambda$, but at least at small $\lambda$ we can trust it.

\begin{figure}
\caption{Phase diagram for $\omega=0$. At $m=0$ there is a wall across which the Witten index jumps, while at $m=m_*$ there is a second-order phase transition.}
\centering
\includegraphics[width=0.5\textwidth]{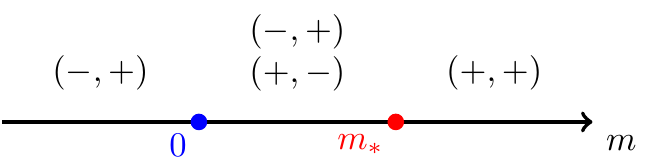}
\label{fig:Choi_phase}
\end{figure}

\subsubsection{\label{subsubsec: The Bosonization Dualities}The Bosonization Dualities}

CS-matter theories with fundamental matter are conjectured to enjoy an IR Bosonization duality (the literature on this subject is vast by now and so we cannot refer to every paper; for the precise conjecture see \cite{Aharony:2015mjs}). Schematically, this duality interchanges Bosons with Fermions, and interchanges gauge groups with their level-rank duals \cite{Hsin:2016blu}. There is an abundance of evidence for these dualities at large $N$, including the matching of correlation functions (see e.g. \cite{Aharony:2012nh}) and of the free energies of the theories (e.g. \cite{Jain:2013gza}). There is also some evidence at finite $N$, which includes matching the phase diagrams, symmetries and anomalies of the theories (see e.g. \cite{Seiberg:2016gmd}).

These dualities also have an $\mathcal{N}=1$ SUSY version. For example, there exists a duality relating an $SU$ gauge group with $N_{\rm f}$ fundamental matter fields to a $U$ gauge group  with $N_{\rm f}$ fundamental matter fields:
\begin{equation}
    U(N)_{k+\frac{N+N_{\rm f}}{2},k+\frac{N_{\rm f}}{2}} + N_{\rm f} \Phi \longleftrightarrow SU(k+N_{\rm f})_{-N-\frac{k}{2}}+N_{\rm f}\tilde \Phi\;.
\end{equation}
There also exist other dualities relating $U$ gauge groups to $U$ gauge groups, see \cite{Choi:2018ohn} for details. At infinite $N$, where the theory is conformal for all values of the couplings $\lambda,w$ (and where we do not distinguish $SU$ and $U$ gauge groups), the precise mapping between the parameters of the theories for the theory \eqref{eq:Minwalla_Defs} was found in \cite{Jain:2013gza}:
\begin{equation}\label{eq:bosonization_transformation}
    \hat\lambda\to \hat\lambda-\text{sgn}\lambda,\quad \mu \to -\frac{2\mu}{1+\omega},\quad w\to \frac{3-w}{1+w}\;.
\end{equation}
Note that in these conventions, $|\lambdah|\leq 1$ and so strong coupling corresponds to $\lambdah$ close to one.

Our theory should also display this duality in the conformal limit. Indeed, it was discussed in \cite{Dey:2019ihe} how the phase diagram figure \ref{fig:Minwalla_phase_diagram} is consistent with this duality. However, since this is an IR duality, it applies only to the conformal limit. In particular, it only relates vacua which appear at the CFT. We thus do not expect to be able to say anything new about the duality. However, the duality can help us constrain the phase diagram; assuming we know the phase diagram in the conformal limit for small $\lambda$, the duality can map it to the conformal limit for $\lambda$ close to 1. This was also used in \cite{Aharony:2019mbc} to find the non-trivial CS-matter fixed points that will inform our phase diagram at ``large but finite $N$'' - figure \ref{fig:finite_N}.

\subsection{No Bare \texorpdfstring{$\Phi^4$}{Phi to the fourth} Term}

Next we discuss the phase diagram for the theory \eqref{eq:IR_action} when we set the UV value of $\omega$ to zero. The authors of \cite{Choi:2018ohn} calculated the 1-loop effective superpotential for $\Phi$. Combining this with the tree-level mass term, the full superpotential takes the form appearing in equation \eqref{eq:res}. The corresponding phase diagram appears in figure \ref{fig:Choi_phase}. Although this diagram includes only a 1-loop computation, its results can be trusted whenever the vev of $\Phi$ is large enough \cite{Bashmakov:2018wts}. In particular, this calculation does not require $N$ to be large. 

There are two interesting points in this phase diagram. The first is at $m=0$, where the Witten index jumps. As discussed in \cite{Witten:1982df}, this is due to the fact that the asymptotics of the superpotential change at $m=0$ (provided that $\omega=0$). The second is at $m=m_*$, where there is a second-order phase transition or a CFT. This point thus corresponds to the conformal limit, and indeed the phase diagram around this point agrees with the phases in figure \ref{fig:Minwalla_phase_diagram} around $w=0$. 

For nonzero $\omega$, at leading order in $\omega$ one can just add a term $\frac{\omega}{2}(\bar\Phi\Phi)^2$ to this action, and compute the corresponding vacuum structure. It turns out that at leading order in $1/N$, this term is enough to find the full $\omega$-dependence. We will describe the result in this paper.

\section{Computing the Superpotential\label{sec: computing the superpotential}}

In this section we compute the Higgsed and unHiggsed branches of the effective superpotential $W$ of the
theory \eqref{eq:UV_action} at leading order in $\lambda$. The main results of the section are equations \eqref{eq:final unHiggsed superpotential} and \eqref{eq: Higgs branch superpotential small lambda}, which represent the two ``branches'' of the superpotential (Higgsed and unHiggsed - see below). We discuss the regime of validity of the approximation, and argue that it encompasses the entire parameter space. This superpotential is then studied in section \ref{sec:Phase Diagram} to give the full phase diagram of the theory appearing in figure \ref{fig: phase diagram infinite N leading lambda}.

We compute the unHiggsed branch \eqref{eq:final unHiggsed superpotential} directly. The leading contribution of the gauge sector to the Higgsed branch was computed in \cite{Choi:2018ohn} in the case where the quartic coupling vanishes $\omega=0$, and we argue that (at leading order in $N$ and all orders in $\lambda$) the modification needed to include it is simply to add it in at tree level, giving \eqref{eq: Higgs branch superpotential small lambda}.
The derivation in \cite{Choi:2018ohn} had some minor numerical and sign errors that serendipitously cancelled. We rederive this result in a slightly different method in the appendix \ref{sec:Detailed Higgsed Phase Computations}, and obtain the same result.

\subsection{General Considerations}

The effective superpotential is a supersymmetric analog
of the effective potential of Coleman and Weinberg \cite{coleman:1973}. Extrema of
the effective superpotential occur at VEVs of operators in a supersymmetric
vacuum. The F-term equations, $W'=0$ therefore play the role of a
consistency condition for the VEV. If no solution exists, one has
to conclude that no supersymmetric vacua exists, and if multiple solutions
exist, then multiple SUSY vacua exist. 

Just as the effective potential can be obtained by evaluating the
effective action at a uniform field configuration $\partial_{\mu}\phi=0$
and dividing by the volume of spacetime, so can the effective superpotential
be found by choosing field values that preserve the entire super-Poincare
group, $D_{\alpha}\Phi=0$, and ``dividing by the volume of superspace''.
In practice, this division simply means that in all calculations one
should fix the superspace coordinate $z$ of precisely one of the
operator insertions.

Rather than a single superpotential, the calculation splits into two
``branches'' -  a
Higgsed branch $W^{{\rm H}}\left(\phisq\right)$ and an unHiggsed branch $W^{{\rm uH}}\left(M\right)$:
\begin{itemize}
    \item Higgsed vacua are characterized by having a nonzero vev for the fundamental
    superfield $\left\langle \Phi_{i}\right\rangle =\Phi_{i,\,{\rm cl}}$.
    Hence one can compute the effective superpotential for the operator
    $\Phi_{i}$ in order to find Higgsed vacua. By symmetry, the operator
    will only depend on the invariant combination $\phisq$. In
    a Higgsed vacuum the leading contribution to $\left\langle \phisq\right\rangle $
    is the disconnected one:
    \begin{equation}
    \left\langle \bar{\Phi}\right\rangle \left\langle \Phi\right\rangle =\bar{\Phi}_{{\rm cl}}\Phi_{{\rm cl}},
    \end{equation}
    so in a way we can think of the superpotential as a function of $\left\langle \phisq\right\rangle $
    as well.
    \item In unHiggsed vacua $\left\langle \Phi_{i}\right\rangle =0$ and so
    to find F-term equations one must use a gauge-singlet operator. The
    simplest such operator is the ``mass deformation'' $\phisq$.
    However, due to equation \eqref{eq:sigma EOM relation to Phi}, we can instead use $\sigma$. In practice,
    as elaborated on in subsubsection \ref{subsubsec: set up unhiggsed}, it's more convenient to use an operator $M\equiv m+\left\langle \sigma\right\rangle +\dots$
    which is shifted relative to $\sigma$ by conveniently chosen constants.
\end{itemize}

Two mass scales are present in the calculation - the VEV and the Yang-Mills
mass-scale $\kappa\equiv \frac{kg^2}{4\pi}$. The parameter $\lambda_{\sigma\Phi}$ will not appear at leading order
in $N$, while the mass $m$ of $\Phi$ only appears in loop calculations
through the combination $M$. Hence the calculation has two regimes\footnote{We emphasize that although we ``define'' the regimes in terms of the scale of the VEVs $M$ and $\phisq$, in practice they are determined by $m$. The off-shell VEVs have no meaning and the on-shell VEVs are only large or small when $m$ is tuned to make them such. Thus, for instance, the IR regime will correspond to when $|m_{\rm IR}|\ll|\kappa|$, where $m_{\rm IR}$ is the distance of $m$ to the CFT. However, since only the scales of the VEVs appear explicitly in the detailed large $N$ computations, we treat them as undetermined by $m$.}:
\begin{enumerate}
\item A UV regime $\left|M\right|,\phisq\gg\kappa$: The masses
of excitations above the vacuum are much heavier then $\kappa$, and
so loop corrections to the superpotential are dominated by modes of
energy $\gg\kappa$. Since the YM term is a classically relevant deformation,
the effects of the gauge sector are weak and the theory should resemble
the ``ungauged'' vector model. In particular, although generically
the loop contributions to the superpotential due to the gauge fields
should generate every possible order of interaction $\kappa^{2-n}M^{n},\,\kappa^{2-n}\left(\phisq\right)^{n}$ with $n\in \mathbb{Z}$,
these should re-sum to give a slow growth. In particular, we find in
the Higgsed phase a sub-sub-leading growth $\sim\kappa^{3/2}\left(\phisq\right)^{1/2}\ll m\phisq$
and in the unHiggsed a sub-leading growth $\sim M\log M\ll\frac{M^{2}}{\omega}$.
Wall crossings - inherently UV aspects of the theory - are therefore insensitive
to gauge dynamics.
\item An IR regime $\left|M\right|,\phisq\ll\kappa$: The dominant
contribution will be from energies $\ll\kappa$, where the YM mode
is unexcited. Hence one is operating effectively in a Chern-Simons-matter
theory, with the YM term acting as a regulator. Results in this regime
are sensitive to gauge dynamics, and should match those reported in \cite{Dey:2019ihe,Inbasekar:2015tsa}.
If one takes the limit $\kappa\to\infty$, the superpotential obtained
is simply the superpotential of the Chern-Simons theory in Yang-Mills
regularization. If the limit is taken prior to loop integration then
new UV divergences appear, and the propagators become identical to
those in Chern-Simons theory. Then if those new divergences are dimensionally
regulated, then one is working in Chern-Simons theory with dimensional regularization.
Hence there are essentially two distinct ways of taking the IR limit corresponding to two ways of regulating the IR theory, which lead to the superpotentials $W_{{\rm IR}}^{{\rm YM-reg}}$
and $W_{{\rm IR}}^{{\rm dim-reg}}$. The results obtained  in dim-reg,
should match directly with those found by \cite{Dey:2019ihe,Inbasekar:2015tsa}, and we'll use this fact to check our results at various stages.
\end{enumerate}
The section is organized as follows: We begin with the calculation
of the unHiggsed branch in subsection \ref{subsec:The UnHiggsed Branch of the Superpotential} and then of the Higgsed branch in subsection \ref{subsec: Higgesd branch of the superpotential}. In each, we perform
checks that compare our results in the UV and (dim-reg) IR limits
to known results. We then discuss the IR limit in subsection \ref{subsec: Dimensional vs Yang-Mills Regularization} in more detail,
and demonstrate the equivalence of the two regularization schemes. Finally, in subsection \ref{subsec: Validity of the Calculation} we argue that the perturbative (in $\lambda$) superpotential we obtain can be trusted over the entire parameter space, so that it gives the full phase diagram at small $\lambda$. 

\subsection{\label{subsec:The UnHiggsed Branch of the Superpotential}The UnHiggsed Branch of the Superpotential}

In this subsection we'll compute the effective superpotential governing
unHiggsed vacua up to 2 loops - or first order in the gauge coupling
$\lambda\equiv\frac{N}{k}$ - and at leading order in $N$, for the theory \eqref{eq:UV_superpotenial}. For simplicity,
we will work in the special case $\lambda_3=0$. As we will argue in
subsection \ref{subsec: adding a sigma cubed coupling}, the generalization to the case $\lambda_3\neq 0$ is
straightforward.

Over the unHiggsed vacuum, the fundamental superfield has vanishing
VEV $\left\langle \Phi\right\rangle =0$\footnote{The operator $\Phi$ is not gauge-invariant, and hence ill defined. This definition is ``good enough'' at weak coupling, and more generally we define unHiggsed and Higgsed by the rank of the gauge group in the IR topological theory.}, so the superpotential will
be a function only of the VEV of $\sigma$\footnote{In principle, we could compute the superpotential for both $\sigma$ and $\phisq$, giving two F-term equations, however, one of those will just amount to \eqref{eq:sigma EOM relation to Phi}, so that $\phisq$ can be eliminated. \eqref{eq:sigma EOM relation to Phi} could receive quantum corrections, but since $\sigma$ is effectively auxilliary at large $N$, we assume those aren't significant, or at least don't modify the relation \eqref{eq:sigma EOM relation to Phi} into one that isn't one-to-one.}. In fact we will find it
useful to work with a shifted variable $M\equiv m+\left\langle \sigma\right\rangle +\dots$,
the precise definition of which will be chosen so that it corresponds
roughly to the (signed) pole mass of $\Phi$ over the vacuum. More
precisely, $M$ will be defined so that it coincides with the pole
mass when $\left|M\right|\ll\left|\kappa\right|$, but in general
may differ from the pole mass while having the same sign. This choice
will be motivated further in subsubsection \ref{subsubsec:Feynman-Rules}.

This superpotential $W\left(M\right)$ is a supersymmetric analog
of the effective potentials \\ $U_{{\rm eff}}^{\left(+,+\right)},U_{{\rm eff}}^{\left(-,+\right)}$
of \cite{Dey:2019ihe}, although we find it simpler in this case to incoporate
both branches into one function $W\left(M\right)$ where each branch
corresponds to a half-line ${\rm sign}\left(M\lambda\right)=\pm1$
in the domain. We will find that although from purely classical analysis one expects
a single unHiggsed vacuum to always exist, quantum corrections create
phases with 0, 2 and even 3 distinct unHiggsed vacua, as well as introduce
wall crossings where an unHiggsed vacuum escapes to, or appears from,
infinity, and the Witten index jumps. This is because, while the classical
superpotential is necessarily ``confining'' $\sim M^{2}$, quantum
corrections can introduce terms with the asymptotic form $\sim M\left|M\right|$.

We find \eqref{eq:final unHiggsed superpotential}:
\begin{align}
W & =-\frac{\left(M-\left(m-\frac{1}{2}\lambda\left|\kappa\right|\right)\right)^{2}}{2\omega}-\frac{1}{8\pi}M\left|M\right|\label{eq:final unHiggsed superpotential-1}\\
 & +\frac{1}{16\pi}\left(2M+\kappa\right)\lambda\kappa\log\left(1+\frac{2\left|M\right|}{\left|\kappa\right|}\right)-\frac{1}{8\pi}\lambda\left|\kappa\right|\left|M\right|.\nonumber 
\end{align}
The superpotential at leading order in $\lambda$ depends
on the YM mass $\kappa$, and has an IR limit $\kappa\to\infty$,
corresponding to when the gap of $\Phi$, and the VEV of $\sigma$,
are well below the YM scale. In this limit, one is effectively computing
the superpotential in a \textbf{pure Chern-Simons-matter} theory,
albeit in Yang-Mills regularization. This is the theory studied in
section 5 of \cite{Dey:2019ihe} and in \cite{Inbasekar:2015tsa}, modulo
the regularization scheme. As a sanity check, we can compare our computations
with those found in these papers. We will find that the answers match
trivially when one takes $\kappa\to\infty$ \textbf{prior} to loop-integration,
in which case one is effectively working without YM regularization.
If $\kappa$ is taken to infinity \textbf{after} integration, then
the answers match provided certain renormalizations of parameters
are included, as expected when different regularization schemes are
employed.

We'll begin by setting up the computation. Next, we'll perform the
1-loop and 2-loop computations. Finally, we'll discuss how to slightly
modify the computation to subtract the divergences that arise at 2-loops
and obtain a physically reasonable result.

\subsubsection{\label{subsubsec: set up unhiggsed}Set-Up}

The theory is defined by the action \eqref{eq:UV_action},\eqref{eq:UV_action_defs}. 
Note that we find it useful to normalize $\sigma$ non-canonically, so
it's defined by the equation of motion \eqref{eq:sigma EOM relation to Phi}:
\begin{equation}
\left\langle \sigma\right\rangle =\omega\left\langle \phisq\right\rangle .\label{eq:sigma equation of motion}
\end{equation}
Although this means that the propagator of $\sigma$ has the unconventional
form:
\begin{equation}
\left\langle \sigma\sigma\right\rangle _{{\rm free}}=-\frac{1}{N}\frac{1}{\frac{1}{\lambda_{\sigma\Phi}^{2}}D^{2}-\frac{1}{\omega}}=-\frac{1}{N}\lambda_{\sigma\Phi}^{2}\frac{1}{D^{2}+m_{\sigma}},
\end{equation}
such propagators don't appear anyway at leading order in $N$, as will be explained below. 

The free energy is defined by:
\begin{equation}
E\left[J\right]=\frac{1}{N}\log Z\left[J\right]
\end{equation}
\begin{equation}
Z\left[J\right]=\intop D\sigma D\bar{\Phi}^{i}D\Phi_{i}D\Gamma\exp\left(N\left(S+\intop d^{5}zJ\sigma\right)\right).
\end{equation}
The quantum corrections to $E\left[J\right]$ consist of the sum of
all \textbf{connected} supergraphs, and satisfies:
\begin{equation}
\frac{\delta E}{\delta J\left(z\right)}=\left\langle \sigma\left(z\right)\right\rangle _{J}.
\end{equation}
The effective action:
\begin{equation}
\Gamma\left[\sigma_{{\rm cl}}\right],
\end{equation}
is then simply the Legendre transform, which consists of all \textbf{connected
1PI} supergraphs, and satisfies:
\begin{equation}
\frac{\delta\Gamma}{\delta\sigma_{{\rm cl}}\left(z\right)}=-J\left[\sigma_{{\rm cl}}\right]\left(z\right),
\end{equation}
where $J\left[\sigma_{{\rm cl}}\right]$ is defined by:
\begin{equation}
\sigma_{{\rm cl}}\left(w\right)=\left\langle \sigma\left(w\right)\right\rangle _{J\left[\sigma_{{\rm cl}}\right]},\,\forall w\;.
\end{equation}
In analogy with the Coleman-Weinberg effective potential, the effective
superpotential is the part of the effective action that involves no
superderivatives. The standard Coleman-Weinberg potential can be obtained
by evaluating the effective action on a uniform background, and dividing
by the volume of spacetime. The superpotential is quite similar -
we evaluate the effective action with $\sigma_{{\rm cl}}$ that is
independent of both $x^{\mu},\theta^{\alpha}$. This naive procedure
actually yields 0. This is because all Grasmann integrals evaluate
to 0, and reflects the fact that the chosen vacuum is supersymmetric,
and thus has vanishing energy. The fix is to ``divide by the volume
of superspace'' (which we can denote ${\rm Vol}^{3\mid2}$) - that
is, fix precisely one ``supercoordinate'' $z$ in every computation.
This procedure then yields the superpotential. 

Thus, we write:
\begin{equation}
\sigma\left(z\right)=\sigma_{{\rm cl}}+\sigma_{{\rm qu}}\left(z\right)
\end{equation}
\begin{equation}
J\left(\sigma_{{\rm cl}}\right):\,0=\left\langle \sigma_{{\rm qu}}\right\rangle _{J\left(\sigma_{{\rm cl}}\right)}
\end{equation}
\begin{align}
NW\left(\sigma_{{\rm cl}}\right) & =\frac{1}{{\rm Vol}^{3\mid2}}\log\left(\intop D\sigma D\bar{\Phi}^{i}D\Phi_{i}D\Gamma\exp\left(NS\left[\sigma,\dots\right]+\intop d^{5}zJ\left(\sigma_{{\rm cl}}\right)\sigma\right)\right)\label{eq:unHiggsed superpotential path integral definition}\\
 & -J\left(\sigma_{{\rm cl}}\right)\sigma_{{\rm cl}}\nonumber \\
 & =\frac{1}{{\rm Vol}^{3\mid2}}\log\left(\intop D\sigma D\bar{\Phi}^{i}D\Phi_{i}D\Gamma\exp\left(NS\left[\sigma_{{\rm qu}},\dots\right]+\intop d^{5}zJ\left(\sigma_{{\rm cl}}\right)\sigma_{{\rm qu}}\right)\right).\nonumber 
\end{align}
The definition of $J\left(\sigma_{{\rm cl}}\right)$ simply ensures
that the non 1PI diagrams vanish. 

\subsubsection{\texorpdfstring{$N$}{N} - Counting\label{subsubsec:N Counting}}

The leading contribution to $NW$ is technically $O\left(N^{2}\right)$,
from $\Gamma$ loops, but is independent of $\sigma_{{\rm cl}}$
and so can be ignored. Hence the true leading order is $O\left(N\right)$,
since at least one $\Phi$ loop must be included. In the absence of
$\sigma_{{\rm qu}}$, the $N$ counting is the usual one in the 't
Hooft limit of gauge theory - a diagram with $g$ handles and $h$
holes is $O\left(N^{2-2g-h}\right)$. The leading contribution is
then simply the sum of all ``disc graphs'' - a single matter loop
with planar pure-gauge interactions in its ``interior''. The interactions
with the gauge fields can always be thought of as propagator corrections
to the matter loop. For a review of large $N$ diagrammatics, see \cite{Coleman:1985rnk}.

What happens when we include $\sigma_{{\rm qu}}$ propagators in the
diagram? A single $\sigma_{{\rm qu}}$ loop is $O\left(N^{0}\right)$
and so can be neglected. The interaction vertices $N\sigma_{{\rm qu}}\phisq$
must come in pairs, and are the only way in which $\sigma_{{\rm qu}}$
could have an effect at leading order. If there are $n$ pairs, then
that changes the $N$-power of the diagram by:
\begin{equation}
\underbrace{N^{2n}}_{\text{vertices}}\times\underbrace{\frac{1}{N^{n}}}_{\sigma_{{\rm qu}}\,\text{propagators}}\times\underbrace{\frac{1}{N^{2n}}}_{\Phi\,\text{propagators}}\times N^{\text{\# of color loops}}.
\end{equation}
This means that each new $\sigma_{{\rm qu}}$ must come with a new
$\Phi$ - loop. Furthermore, interactions with $\Gamma$ are only
leading when they do not couple \textbf{different} $\Phi$ loops,
as that would produce a diagram with 2 holes. Hence the \textbf{general}
$O\left(N\right)$ diagram involving an interaction with $\sigma_{{\rm qu}}$
has the form of a tree graph, whose edges are $\sigma_{{\rm qu}}$-propagators
and whose vertices are ``disc graphs'' - a single $\Phi$ loop with
arbitrary planar $\Gamma$ interactions. Hence, cutting \textbf{any}
$\sigma_{{\rm qu}}$ propagator would give 2 disconnected graphs.
Hence the 1PI contributions involve no $\sigma_{{\rm qu}}$ propagators. 

The reason for this can also be understood in a different way. Suppose
one keeps $\sigma_{{\rm qu}}$ as a background, and integrates out
only $\Phi,\Gamma$. The resulting action for $\sigma_{{\rm qu}}$,
although highly nonlocal, has $N$ as an overall constant outside
the action, and appearing nowhere else. Hence the leading order in
$N$ is simply the semiclassical approximation. This means this action
is just evaluated at $\sigma_{{\rm cl}}$.

Hence going forward we can ``forget'' about $\sigma_{{\rm qu}}$
and just write:
\begin{align}
W & =-\frac{\sigma_{{\rm cl}}^{2}}{2\omega}+\frac{1}{N}\times\left(\text{all disc diagrams}\right).
\end{align}
When expanding around $\sigma_{{\rm cl}}$ the action for $\Phi$
becomes:
\begin{equation}
\intop d^{5}z\left(\bar{\Phi}\nabla^{2}\Phi+\left(m+\sigma_{{\rm cl}}\right)\phisq\right),
\end{equation}
so if we define $M\equiv m+\sigma_{{\rm cl}}$ then:
\begin{align}
W\left(M\right) & =-\frac{\left(M-m\right)^{2}}{2\omega}+\frac{1}{N}\log\Big[\intop D\bar{\Phi}^{i}D\Phi_{i}D\Gamma\\
 & \exp\left(N\intop d^{5}z\left(\bar{\Phi}\nabla^{2}\Phi+M\phisq\right)+N\left(S_{{\rm YM}}+S_{{\rm CS}}\right)\right)\Big].
\end{align}

\subsubsection{Feynman Rules\label{subsubsec:Feynman-Rules}}
\begin{enumerate}
\item The $\Phi$ propagator is given by:
\begin{equation}
\left\langle \bar{\Phi}^{i}\Phi_{j}\right\rangle _{{\rm free}}\incgraph[freescalarprop][.25]=-\frac{1}{N}\frac{1}{D^{2}+M}\delta_{j}^{i}=\frac{1}{N}\frac{D^{2}-M}{p^{2}+M^{2}}\delta_{j}^{i}.\label{eq:free_scalar_prop}
\end{equation}
\item We will work in Landau gauge (see appendix \ref{app:Gauge_propagator} for details) where the gauge propagator becomes:
\begin{equation}
\left\langle \Gamma^{\alpha a}\Gamma^{\beta b}\right\rangle _{{\rm free}}=\incgraph[freegaugeprop][.25]=\frac{4\pi\lambda}{N}\kappa\frac{D^{\alpha}D^{\beta}\left(\kappa-D^{2}\right)}{p^{2}\left(\kappa^{2}+p^{2}\right)}\delta^{ab},\label{eq:Landau_gauge_propagator}
\end{equation}
where $\kappa\equiv\frac{kg^{2}}{4\pi}$ is the Yang Mills mass. This
choice satisfies ``transversality of the propagator'':
\begin{equation}
\Delta^{\alpha\beta}D_{\alpha}=D_{\beta}\Delta^{\alpha\beta}=0.
\end{equation}
\item The gauge coupling $\lambda$ appears only behind the gauge propagator
\eqref{eq:Landau_gauge_propagator} and so will act as a loop-counting
parameter. As previously mentioned, we will compute the 1-loop ($\lambda^{0}$)
and the 2-loop ($\lambda$) contributions to the effective superpotential.
\item The interactions with $\Phi$ come from the terms:
\begin{equation}
-\frac{1}{2}\left(D^{\alpha}+i\Gamma^{\alpha}\right)\bar{\Phi}\left(D_{\alpha}-i\Gamma_{\alpha}\right)\Phi.
\end{equation}
So we get a quartic vertex:
\begin{equation}
-\bar{\Phi}\Gamma^{2}\Phi,
\end{equation}
and a cubic vertex which (using propagator transversality and integration by parts) can be
written in a few different ways:
\begin{equation}
\frac{i}{2}D^{\alpha}\bar{\Phi}\Gamma_{\alpha}\Phi-\frac{i}{2}\bar{\Phi}\Gamma^{\alpha}D_{\alpha}\Phi\sim-i\bar{\Phi}\Gamma^{\alpha}D_{\alpha}\Phi\sim-iD_{\alpha}\bar{\Phi}\Gamma^{\alpha}\Phi.\label{eq:cubic vertex choices}
\end{equation}
\item Gauge field self-interactions and interactions with ghosts will not
enter in our calculations, which are at leading order in the gauge
coupling $\lambda$.
\item We will find it useful to include terms in the Lagrangian that will
act as counterterms.
\begin{enumerate}
\item A linear term in $M$ (equivalently $\sigma$) can be included:
\begin{equation}
|\kappa|\delta J M,\label{eq: linear M counter term}
\end{equation}
which will allow us to subtract divergences, and will also come in
handy in subsection \ref{subsec: Dimensional vs Yang-Mills Regularization}, when we compare the IR superpotential in different
regularization schemes. This term will simply be added to the tree-level
superpotential $W_{{\rm tree}}=-\frac{\sigma_{{\rm cl}}^{2}}{2\omega}$
and will not affect the diagramatics. 
\item The definition of $M$ should coincide with the pole mass of $\Phi$.
When $\lambda=0$ it is precisely $m+\sigma_{{\rm cl}}+O\left(1/N\right)$.
However, at leading order in $\lambda$, the mass might receive further
corrections $m+\sigma_{{\rm cl}}-q\lambda\left|\kappa\right|$ for
some numerical $q$ (at higher orders, $q$ will just become a function $q(|\lambda|)$). In fact, we'll find that $q=\frac{1}{2}$. For
perturbation theory to be self-consistent, we cannot use the propagator
\eqref{eq:free_scalar_prop} when $M\ll\lambda\kappa$, since the
perturbative corrections to the mass are larger than the mass. To
fix this we will have to redefine:
\begin{equation}
M=\left(m-q\lambda\left|\kappa\right|\right)+\sigma_{{\rm cl}},\label{eq:Redfinition of M}
\end{equation}
which will have the effect of introducing a ``mass'' counter-term
to the Lagrangian:
\begin{equation}
q\lambda\left|\kappa\right|\phisq,
\end{equation}
which would cancel corrections to $M$ at $M\ll\lambda\kappa$, giving
a consistent perturbation theory.
\item Such counterterms invalidate $\lambda$ as a loop-counting parameter,
but we will still refer to $\sim\lambda^{l-1}$ diagrams as $l$-loop
diagrams in the following. 
\end{enumerate}
\item Thus we can write the full (all orders in $\lambda$) superpotential at infinite $N$ as:
\begin{align}
W\left(M\right) & =-\frac{\left(M-\left(m-q(|\lambda|)\lambda |\kappa|\right)\right)^{2}}{2\omega}+|\kappa| \delta J(|\lambda|)M+\frac{1}{N}\log\Big[\intop D\bar{\Phi}^{i}D\Phi_{i}D\Gamma\label{eq: formal all orders unHiggsed superpotentail}\\
 & \exp\left(N\intop d^{5}z\left(\bar{\Phi}\nabla^{2}\Phi+M\phisq+q(|\lambda|)\lambda\left|\kappa\right|\phisq\right)+N\left(S_{{\rm YM}}+S_{{\rm CS}}\right)\right)\Big].\nonumber
\end{align}

\end{enumerate}

\subsubsection{1-Loop Analysis}

The 1-loop contribution is simply the 1-loop determinant:
\begin{align}
W_{1-{\rm loop}}= & \frac{1}{N}\log\left(\intop D\bar{\Phi}^{i}D\Phi_{i}\exp\left(N\intop d^{5}z\left(\bar{\Phi}D^{2}\Phi+M\phisq\right)\right)\right)\\
= & \frac{1}{N}\log\left(1/\det\left(\left(D^{2}+M\right)\delta_{j}^{i}\right)\right)\\
= & -{\rm Tr}\log\left(D^{2}+M\right)\\
= & -\frac{1}{8\pi}M\left|M\right|,
\end{align}
where in the second line we used the standard Gaussian integration
forumla for a complex scalar and in the last line we used \eqref{eq:trace-log-scalar-formula}.
The superpotential:
\begin{align}
W_{\lambda=0} & =W_{{\rm tree}}+W_{1-{\rm loop}} =-\frac{\left(M-m\right)^{2}}{2\omega}-\frac{1}{8\pi}M\left|M\right|,\label{eq: 1-loop unHiggsed superpotential}
\end{align}
is exact at large $N$ and $\lambda=0$ (that is - in the ungauged
theory). This gives an F-term equation $0=\partial W_{\lambda=0}/\partial M$ which can be solved for $M$:
\begin{equation}
M=\frac{m}{1+\frac{\omega}{4\pi}{\rm sign}\left(M\right)}.\label{eq:ungauged_pole_mass}
\end{equation}
We can immediately see that wall crossings exist at $\omega=\pm4\pi$,
where the $\left(\mp,+\right)$ vacua (dis-)appear (to)from infinity.
Since the YM interactions are relevant, and therefore unimportant
in the deep UV, we expect this position of the wall crossings to be
unaffected by the 2-loop analysis. Put differently we expect higher-loop
contributions to be subleading at large $M$.\footnote{There is no symmetry prohibiting terms of arbitrary order $\sim\kappa^{2-n}M^{n}$, with $n\in \mathbb{Z}$,
from appearing, so we clarify that the \textbf{re-summed} expression
will be subleading.} As we will see below, we do in fact find $W_{2-{\rm loop}}\sim M\log\left(M\right)$
at large $M$, which is subleading.

A simple sanity check is to compare the F-term equation \eqref{eq:ungauged_pole_mass} to the value of the pole mass in equation (2.13) of \cite{Inbasekar:2015tsa}.
The authors of \cite{Inbasekar:2015tsa} use a parameter which they call
$w$ in place of our $\omega$. From the Lagrangian, equation (2.1)
of \cite{Inbasekar:2015tsa}, we can infer:
\begin{equation}
w=\frac{\omega}{2\pi\lambda}.
\end{equation}
Plugging this into the aforementioned equation (2.13) and taking $\lambda\to0$,
we arrive at \eqref{eq:ungauged_pole_mass}. Although the theory studied
in \cite{Inbasekar:2015tsa} is a Chern-Simons matter theory without a
YM interaction, its $\lambda\to0$ limit is indistinguishable from
that of the CS-YM theory, being simply an ``ungauged'' vector model.

\subsubsection{\label{subsubsec:2-Loop-Analysis}2-Loop Analysis}

The 2-loop contribution is given by:
\begin{equation}
W_{2-{\rm loop}}=\frac{1}{N}\begin{gathered}\incgraph[vacuum1]\end{gathered}
+\frac{1}{N}\begin{gathered}\incgraph[vacuum2]\end{gathered}.\label{eq: 2-loop diagrams unHiggsed}
\end{equation}
These diagrams are computed in appendix \ref{app:supergraphs}.
Using equation \eqref{eq:2-loop_log_divergent_vacuum_bubble}
we find:
\begin{align}
W_{2-{\rm loop}} & =-\left(2M+\kappa\right)\pi\lambda\kappa\intop\frac{d^{3}p}{\left(2\pi\right)^{3}}\frac{d^{3}k}{\left(2\pi\right)^{3}}\frac{1}{\left(\kappa^{2}+k^{2}\right)\left(\left(p-k\right)^{2}+M^{2}\right)\left(p^{2}+M^{2}\right)}\label{eq:2-loop superpotential unevaluated}\\
 & =\frac{1}{16\pi}\left(2M+\kappa\right)\lambda\kappa\left(\frac{1}{\epsilon}+\log\left(\frac{\left|\kappa\right|}{\Lambda}\right)+\log\left(1+\frac{2\left|M\right|}{\left|\kappa\right|}\right)-\frac{\log(4\pi)+\psi\left(\frac{3}{2}\right)}{2}\right),\nonumber 
\end{align}
where $\psi$ is the digamma function and we evaluate the convergent
$k$ integral in 3 dimensions and then the resulting $p$ integral
is evaluated in dim-reg. 

To subtract the divergence, as discussed in subsubsection \ref{subsubsec:Feynman-Rules},
we need to include a ``counterterm'' $\lambda\delta JM$. Although
this is technically a tree-level contribution, it is proportional
to $\lambda$ and so we'll count it as 2-loop. With the appropriate
choice of $\delta J$ we arrive at:
\begin{equation}
W_{2-{\rm loop}}=\frac{1}{16\pi}\left(2M+\kappa\right)\lambda\kappa\log\left(1+\frac{2\left|M\right|}{\left|\kappa\right|}\right).
\end{equation}

\subsubsection{A Small Modification}

To probe the IR dynamics we should take $\left|\kappa\right|\gg\left|M\right|$.
This immediately gives:
\begin{equation}
W_{{\rm IR},\,2-{\rm loop}}=\frac{1}{8\pi}\lambda\left(2M{\rm sign}\left(\kappa\right)\left|M\right|-M^{2}+\left|\kappa\right|\left|M\right|\right).\label{eq:W IR 2-loop YM reg}
\end{equation}
The term $\sim\left|M\right|$ is problematic. It causes a discontinuity
in the F-term equation $W'=0$, which could give rise to unexpected
jumps in the Witten index, with vacua suddenly disappearing. The presence
of this term indicates we have made some unwarranted assumption, and
in fact we have. As discussed in subsubsection \ref{subsubsec:Feynman-Rules}, linear terms in $M$ arise from mass correction to
$\Phi$, and here we find that for $\left|M\right|\ll\left|\kappa\right|$
we have a correction of order $\lambda\left|\kappa\right|$, which is larger than $M$, which is the mass of $\Phi$. To fix this, we add the counterterm
\begin{equation}
-\frac{1}{4}q\lambda\left|\kappa\right|\left|M\right|,
\end{equation}
which shifts $M$ to
$M=\left(m-q\lambda\left|\kappa\right|\right)+\sigma_{{\rm cl}}$. We see that for the choice $q=1/2$, the linear term cancels in the
IR. Hence perturbation theory is consistent for this choice. We arrive
at the final expression for the superpotential:
\begin{align}
W & =-\frac{\left(M-\left(m-\frac{1}{2}\lambda\left|\kappa\right|\right)\right)^{2}}{2\omega}-\frac{1}{8\pi}M\left|M\right|\label{eq:final unHiggsed superpotential}\\
 & +\frac{1}{16\pi}\left(2M+\kappa\right)\lambda\kappa\log\left(1+\frac{2\left|M\right|}{\left|\kappa\right|}\right)-\frac{1}{8\pi}\lambda\left|\kappa\right|\left|M\right|.\nonumber 
\end{align}

\subsubsection{The IR Limit\label{subsubsec:The-IR-Limit}}

The authors of \cite{Dey:2019ihe} investigated the phase diagram of Chern-Simons-matter
theories with fundamental Bosons and Fermions in the large $N$ limit,
to all orders in $\omega$ and $\lambda$. In chapter 5 of their paper,
they focus on the supersymmetric $\mathcal{N}=1$ locus in parameter
space. They worked with the effective potential rather than a superpotential,
given by equation (5.9) of \cite{Dey:2019ihe}. In the limit $\kappa\to\infty$,
our results should match theirs, to leading order in $\lambda$, at
least when we use the same regularization scheme (dimensional reduction).
The gauge propagator \eqref{eq:xi gauge propagator} reduces to the
pure Chern-Simons gauge propagator (which can be found in the literature, e.g. in \cite{Gomes:2012qv}) in the $\kappa\to\infty$ limit:
\begin{equation}
\kappa\frac{D^{\alpha}D^{\beta}\left(\kappa-D^{2}\right)+\xi\left(\kappa+D^{2}\right)D^{\beta}D^{\alpha}}{p^{2}\left(\kappa^{2}+p^{2}\right)}\to\frac{D^{\alpha}D^{\beta}+\xi D^{\beta}D^{\alpha}}{p^{2}},
\end{equation}
so we can check our results against those of \cite{Dey:2019ihe} simply
by taking $\kappa\to\infty$ \textbf{prior} to any loop integration,
and then using dimensional regularization to deal with any new divergences.
Applying this to \eqref{eq:2-loop superpotential unevaluated}, and
using \eqref{eq:tadpole_1-loop}, we get:
\begin{align}
W_{2-{\rm loop}} & =-\left(2M+\kappa\right)\pi\lambda\kappa\intop\frac{d^{3}p}{\left(2\pi\right)^{3}}\frac{d^{3}k}{\left(2\pi\right)^{3}}\frac{1}{\left(\kappa^{2}+k^{2}\right)\left(\left(p-k\right)^{2}+M^{2}\right)\left(p^{2}+M^{2}\right)}\\
 & \to-\pi\lambda\intop\frac{d^{3}p}{\left(2\pi\right)^{3}}\frac{d^{3}k}{\left(2\pi\right)^{3}}\frac{1}{\left(\left(p-k\right)^{2}+M^{2}\right)\left(p^{2}+M^{2}\right)}\\
 & =-\pi\lambda\intop\frac{d^{3}k}{\left(2\pi\right)^{3}}\frac{1}{k^{2}+M^{2}}\intop\frac{d^{3}p}{\left(2\pi\right)^{3}}\frac{1}{p^{2}+M^{2}}\\
 & =-\pi\lambda\left(-\frac{\left|M\right|}{4\pi}\right)^{2}\\
 & =-\frac{1}{16\pi}\lambda M^{2}.
\end{align}
Since no mass shift to $\Phi$ is generated, we should pick $q=0$
in \eqref{eq:Redfinition of M}, so we get:
\begin{equation}
W^{{\rm dim-reg}}=-\frac{\left(M-m\right)^{2}}{2\omega}-\frac{1}{8\pi}M\left|M\right|-\frac{1}{16\pi}\lambda M^{2},
\end{equation}
which gives F-term equations:
\begin{equation}
M=\frac{m}{1+\frac{\omega}{4\pi}\left({\rm sign}\left(M\right)+\frac{1}{2}\lambda\right)}.
\end{equation}
 To compare with \cite{Dey:2019ihe}, we use the relation $\omega=2\pi\lambda w$
(which applies here just as it did when comparing to \cite{Inbasekar:2015tsa}
as we have before), $\sigma_{{\rm cl}}=M-m$ and the smallness of
$\lambda$:
\begin{align}
\sigma_{{\rm cl}} & =-\frac{m\frac{\omega}{4\pi}\left({\rm sign}\left(M\right)+\frac{1}{2}\lambda\right)}{1+\frac{\omega}{4\pi}\left({\rm sign}\left(M\right)+\frac{1}{2}\lambda\right)}\nonumber \\
 & =-\frac{1}{2}\lambda w\frac{m}{\left({\rm sign}\left(M\right)+\frac{1}{2}\lambda\right)^{-1}+\frac{1}{2}\lambda w}\nonumber \\
 & \approx-\frac{1}{2}\lambda w\frac{m}{{\rm sign}\left(M\right)-\frac{1}{2}\lambda+\frac{1}{2}\lambda w}\nonumber \\
 & =-w\frac{m}{\frac{2{\rm sign}\left(M\right)-\lambda}{\lambda}+w}\nonumber \\
 & =-w\frac{m}{w-\frac{-2{\rm sign}\left(M\lambda\right)+\left|\lambda\right|}{\left|\lambda\right|}}.\label{eq: sigma IR solution dim reg}
\end{align}
Now also recall the the EOM for $\sigma$ \eqref{eq:sigma equation of motion}:
\begin{equation}
\left\langle \sigma\right\rangle =\omega\left\langle \phisq\right\rangle =2\pi\lambda w\left\langle \phisq\right\rangle ,
\end{equation}
which, together with \eqref{eq: sigma IR solution dim reg}, gives:
\begin{equation}
2\pi\left\langle \phisq\right\rangle =-\frac{m}{\lambda}\frac{1}{w-\frac{-2{\rm sign}\left(M\lambda\right)+\left|\lambda\right|}{\left|\lambda\right|}}.\label{eq:Phi^2 IR solution dim reg}
\end{equation}
How does this compare to the results reported in \cite{Dey:2019ihe}?
The authors of \cite{Dey:2019ihe} work with a variable $\sigma$ which
is distinct from ours, and is essentially the left-hand side of \eqref{eq:Phi^2 IR solution dim reg}
(up to a factor of $N$, which comes from different normalization
choices for $\Phi$) - see equation (2.22) of \cite{Dey:2019ihe}, and
the discussion around it.

From equations (5.9) and (5.10) of \cite{Dey:2019ihe}, the effective
potential in the $\mathcal{N}=1$ case, we can see that the supersymmetric
vacua in the $\left(\pm,+\right)$ branches are indeed located at:
\begin{equation}
2\pi\left\langle \phisq\right\rangle =-\frac{m}{\lambda}\frac{1}{w-\frac{\mp2+\left|\lambda\right|}{\left|\lambda\right|}},
\end{equation}
which is simply \eqref{eq:Phi^2 IR solution dim reg} with the indentification
${\rm sign}\left(M\lambda\right)=\pm$, as expected.

Had we instead taken $\kappa\to\infty$ \textbf{after} loop-integration,
the form of the IR superpotential would be different (see \eqref{eq:W IR 2-loop YM reg}).
This is because one is working with the same theory in a different
regularization scheme. A further sanity check would be to find a redefinition
of parameters that matches between the two regularization schemes.
We will do this for both the Higgsed and unHiggsed branches of the
superpotential in subsection \ref{subsec: Dimensional vs Yang-Mills Regularization}.

\subsection{\label{subsec: Higgesd branch of the superpotential}The Higgsed Branch of the Superpotential}

In this subsection we discuss the Higgsed branch of the superpotential,
which we denote $W\left(\phisq\right)$, at leading order in $\lambda$. The authors of \cite{Choi:2018ohn} computed the 1-loop superpotential in the theory with $\omega=0$. In practice, this computation corresponds to a Higgsed phase computation, since it involves giving a VEV to the fundamental field. Our main result is to argue that the all orders in $\omega$ superpotential is simply that obtained by \cite{Choi:2018ohn} with the addition of the \textbf{tree level} quartic term at leading order in $N$. 

The reason for this is essentially $N$-counting. By working in a supersymmetric analog of unitary gauge, most of the fundamental field $\Phi$'s d.o.f. are absorbed by the gauge field, and what remains is a singlet ``longitudinal'' mode. By the arguments of subsubsection \ref{subsubsec:N Counting}, at leading order in $N$ it is simply ``classical''. The singlet field $\sigma$ of \eqref{eq:UV_action} is also present and similarly enters at tree level, but we can set it equal to $\omega \phisq$ by means of its EOM \eqref{eq:sigma EOM relation to Phi}, thereby generating the quartic term $\frac{\omega}{2}\phisqbr^2$. Effectively, we integrate it out.

$W\left(\phisq\right)$ is a function
of the VEV of the fundamental superfield $\left\langle \Phi_{{\rm dyn}}\right\rangle =\Phi$
(we will denote the dynamical superfield that appears in the path
integral $\Phi_{{\rm dyn}}\left(z\right)$ in this section and let
$\Phi$ represent the VEV). Naively the F-term equations are $N$
equations given by:
\begin{equation}
0=\frac{\partial W}{\partial\Phi_{i}},\,i=1,\dots,N,
\end{equation}
however, since:
\begin{equation}
\frac{\partial W}{\partial\Phi_{i}}=\bar{\Phi}^{i}W',
\end{equation}
then if any of the components of $\Phi$ are non-zero, the F-term
equations reduce to:
\begin{equation}
0=W',\label{eq:Higgsed F term equation general form}
\end{equation}
and we'll ignore the solution at $\Phi=0$ which corresponds to an
unHiggsed vacuum and is not self-consistent with the assumptions of
the computation. Hence we can regard the superpotential as a function
of a single variable - the VEV, which is restricted to be positive.

Solutions to \eqref{eq:Higgsed F term equation general form} are
vacua of the type $\left(+,-\right)$, so $W$ is the supersymmetric
analog of the effective potential $U_{{\rm eff}}^{\left(+,-\right)}$
of \cite{Dey:2019ihe}, described in subsection \ref{subsec:The Conformal Limit}. In \cite{Dey:2019ihe} a further Higgsed branch of the potential
exists, $U_{{\rm eff}}^{\left(-,-\right)}$, but no vacua are ever
found in it in the $\mathcal{N}=1$ locus. In general it is likely irrelevant in the SUSY theory \eqref{eq:UV_action}. We discuss this in the background subsection \ref{subsec:The Conformal Limit}. 

Unlike in the unHiggsed branch, the leading order in $\lambda$ due
to gauge-dynamics depends on the order at which $\Phi$ appears, namely
it has the form:
\begin{equation}
\left(\lambda\phisq\right)^{n},\,n=1,2,\dots
\end{equation}
Furthermore, the quantum corrections grow at large $\Phi$ no faster
then $\sim\sqrt{\lambda\phisq}$ \cite{Bashmakov:2018wts}, and so never affect the
position of wall crossings. In fact, they are sub-sub-leading, unlike
in the unHiggsed branch where there was a sub-leading correction $\sim M\log M$
which overtakes the classical ``mass term'' and which appeared due
to log divergences. We will argue in subsection \ref{subsec: Validity of the Calculation} that this should persist
at higher orders in $\lambda$.

We will begin (subsubsection \ref{subsubsec: N counting Higgsed}) by justifying that the all orders in $\omega$ result is given by formally by \eqref{eq: formal all orders higgsed superpotential} and explicitly at leading order in $\lambda$ by \eqref{eq: Higgs branch superpotential small lambda}. Then (subsubsection \ref{subsubsec: The IR Limit higgsed}) we'll analyze the IR limit $\kappa\to\infty$.

\subsubsection{\label{subsubsec: N counting Higgsed}Large \texorpdfstring{$N$}{N} Analysis}

The theory is defined by \eqref{eq:UV_action}, except that we think of $\Phi$ in \eqref{eq:UV_action} as $\Phi_{\rm dyn}$.

The superpotential can be computed in a ``supersymmetric unitary gauge''. Write:
\begin{equation}
\Phi_{{\rm dyn},i}\left(z\right)=H\left(z\right)\delta_{j}^{1}\left(e^{iK\left(z\right)}\right)_{i}^{j},\,H\left(z\right)\equiv\sqrt{\phisq}+\frac{1}{\sqrt{2}}h\left(z\right),\label{eq: decomposition of phi to h and K}
\end{equation}
where $H,h$ are real superfields and $K$ are real ``Goldstone''
superfields valued in 
\begin{equation}
SU\left(N\right)\Big/SU\left(N-1\right),
\end{equation}
which are absorbed by the gauge superfields $\Gamma$ in this choice
of gauge. Hence the gauge amounts to the criterion $K=0$. The total number of real degrees of freedom is:
\begin{equation}
\underbrace{1}_{H}+\underbrace{\left(N^{2}-1\right)-\left(\left(N-1\right)^{2}-1\right)}_{K\in SU\left(N\right)\Big/SU\left(N-1\right)}=2N,
\end{equation}
which is the same as the number of real degrees of freedom in $\Phi$,
as expected. Of these $2N$ d.o.f., $2N-1$ are absorbed by the gauge
superfields. They become new polarizations of 1 singlet ``Z-Boson''
superfield and $N-1$ complex fundamental ``W-Boson'' superfields.
In component-field language, the Bosonic components are absorbed,
while the Fermionic components undergo ``mass mixing'' with the
Fermions in the $\Gamma$ superfield.

The path integral becomes:
\begin{equation}
\intop DhD\Gamma\exp\left(NS\left[H\delta_{i}^{1},\Gamma,\dots\right]\right),
\end{equation}
with:
\begin{equation}
\mathcal{L}_{{\rm matter}}=HD^{2}H-H^{2}\left(\Gamma^{2}\right)_{1}^{1}+mH^{2}+\frac{\omega}{2}H^{4}.\label{eq:matter Lagrangian for Higgs phase}
\end{equation}
The cubic coupling to the gauge field vanishes:
\begin{equation}
-i\frac{1}{2}HD^{\alpha}H\left(\Gamma_{\alpha}\right)_{1}^{1}+i\frac{1}{2}\left(\Gamma^{\alpha}\right)_{1}^{1}HD_{\alpha}H=0,
\end{equation}
because the product of 2 Fermionic spinors is symmetric $\psi^{\alpha}\chi_{\alpha}=-\chi_{\alpha}\psi^{\alpha}=\chi^{\alpha}\psi_{\alpha}$. 

In analogy with the unHiggsed phase \eqref{eq:unHiggsed superpotential path integral definition},
the superpotential can be defined as:
\begin{equation}
NW=\frac{1}{{\rm Vol}^{3\mid2}}\log\left(\intop DhD\Gamma\exp\left(NS\left[H\delta_{i}^{1},\Gamma,\dots\right]+\intop d^{5}zJ\left(\phisq\right)\frac{1}{\sqrt{2}}h\right)\right),
\end{equation}
where $J\left(\phisq\right)$ is defined so that:
\begin{equation}
\left\langle h\right\rangle _{J\left(\phisq\right)}=0,
\end{equation}
or, equivalently:
\begin{equation}
\left\langle H\right\rangle _{J\left(\phisq\right)}=\sqrt{\phisq}.
\end{equation}

The $N$-counting of diagrams is the same as in the unHiggsed phase - as dicussed in subsubsection
\ref{subsubsec:N Counting} - except now the singlet fields are $h$,
the ``longitudinal'' superfield; and $Z$ - the ``Z-Boson'' superfield.
Similarly the fundamental fields are now the $\left(N-1\right)$ $W$-Boson
superfields, and the adjoint fields are the $\left(\left(N-1\right)^{2}-1\right)$
- $\Gamma$ unbroken gauge fields. The $W$ superfields have two polarizations with different masses which can be found in the literature \cite{Dunne:1998qy} or read off from the propagator \eqref{eq:broken gauge propagator}:
\begin{equation}
\frac{\kappa}{2}\left(1\pm\sqrt{1+8\pi\lambda\frac{\phisq}{\kappa}}\right).\label{eq:Higgsed gauge masses}
\end{equation}
Hence the leading contribution
to the superpotential will involve a single loop of the gauge superfields.
The singlet fields only enter the leading order in $N$ through non
1PI diagrams which do not affect the superpotential. Hence all matter fields will enter at tree level.
Thus we can write the full (all orders in $\lambda,\omega$) superpotential at infinite $N$ as:
\begin{align}
    W\phisqbr&= m \phisq+ \frac{\omega}{2}\phisqbr^2  +\frac{1}{N{\rm Vol}^{3\mid2}}\log\left(\intop D\Gamma\exp\left(NS_{\rm YM+CS}- N\intop d^{5}z \phisq\left(\Gamma^{2}\right)_{1}^{1}\right)\right).\label{eq: formal all orders higgsed superpotential}
\end{align}
Further using the above discussion we can write:
\begin{align}
    & \frac{1}{N{\rm Vol}^{3\mid2}}\log\left(\intop D\Gamma\exp\left(NS_{\rm YM+CS}- N\intop d^{5}z \phisq\left(\Gamma^{2}\right)_{1}^{1}\right)\right)\nonumber\\
    = & \frac{1}{N{\rm Vol}^{3\mid2}}\log \Big( \intop D\Gamma_{\rm unbroken} D\bar{W}DW\exp \Big(NS_{\rm YM+CS}\left[ W,\Gamma_{\rm unbroken}  \right] \nonumber\\ &- N\intop d^{5}z \frac{1}{2}\phisq \bar{W}W{\Big)}{\Big)}+O(1/N) \\
    = & \left(\rm \, sum\, of\, all\, disc\, diagrams \right)+O(1/N),
\end{align}
Furthermore, we will not need to go further than 1-loop to obtain the leading small $\lambda$ approximation, as the coupling
$\lambda$ is already present there, unlike in the unHiggsed phase.

The 1-loop contribution from the gauge fields was computed in \cite{Choi:2018ohn}:
\begin{equation}
W_{\text{1-loop}}=-\kappa\left|\kappa\right|\sqrt{1+8\pi\lambda\frac{\phisq}{\kappa}},\label{eq: W 1 loop Higgsed adar chang-ha}
\end{equation}
which means that the all-orders in $\omega$ superpotential is simply:
\begin{align}
W=m\phisq+\frac{\omega}{2}\left(\phisq\right)^{2}-\kappa\left|\kappa\right|\sqrt{1+8\pi\lambda\frac{\phisq}{\kappa}}.\label{eq: Higgs branch superpotential small lambda}
\end{align}

The authors of \cite{Choi:2018ohn} treated the Higgs-mass term as a perturbation and worked in Landau gauge. Their computation which yielded \eqref{eq: W 1 loop Higgsed adar chang-ha} had a few numerical and sign errors which serendipitously cancelled. We repeat the computation in unitary gauge in a slightly different manner in the appendix \ref{sec:Detailed Higgsed Phase Computations} and obtain the same result. Note that the 1-loop result \eqref{eq: W 1 loop Higgsed adar chang-ha} can also be written in the form:
\begin{align}
W_{\text{1-loop}}  = & -{\rm Tr}\left(D^{2}+\frac{\kappa}{2}\left(1+\sqrt{1+8\pi\lambda\frac{\phisq}{\kappa}}\right)\right)\label{eq:W 1 loop Higgsed unevaluated}\\
   & -{\rm Tr}\left(D^{2}+\frac{\kappa}{2}\left(1-\sqrt{1+8\pi\lambda\frac{\phisq}{\kappa}}\right)\right),\nonumber 
\end{align}
where the trace is over superfield indices ($\theta,p$). \eqref{eq:W 1 loop Higgsed unevaluated} has the form of a pair of 1-loop determinants, each from a different
polarization state, with the anticipated physical masses \eqref{eq:Higgsed gauge masses}.  We will return to this form when we analyze the IR limit in subsubsection \ref{subsubsec: The IR Limit higgsed}. The evaluated form \eqref{eq: W 1 loop Higgsed adar chang-ha} can be readily obtained from \eqref{eq:W 1 loop Higgsed unevaluated} using \eqref{eq:trace-log-scalar-formula}.

\subsubsection{\label{subsubsec: The IR Limit higgsed}The IR Limit}

As in subsubsection \ref{subsubsec:The-IR-Limit}, we can match expression \eqref{eq: Higgs branch superpotential small lambda} to those
of \cite{Dey:2019ihe} by taking $\kappa\to\infty$ \textbf{prior} to
loop integration. Starting from \eqref{eq:W 1 loop Higgsed unevaluated}
and using \eqref{eq:trace-log-scalar-formula}:
\begin{align}
W_{\text{1-loop}}= & -{\rm Tr}\left(D^{2}+\frac{\kappa}{2}\left(1+\sqrt{1+8\pi\lambda\frac{\phisq}{\kappa}}\right)\right)\nonumber \\
 & -{\rm Tr}\left(D^{2}+\frac{\kappa}{2}\left(1-\sqrt{1+8\pi\lambda\frac{\phisq}{\kappa}}\right)\right)\nonumber \\
\to & -{\rm Tr}\left(D^{2}+\kappa\right)-{\rm Tr}\left(D^{2}-2\pi\lambda\phisq\right)\label{eq:decoupling of very massive polarization}\\
= & \frac{1}{2}\pi\lambda\left|\lambda\right|\left(\phisq\right)^{2}+\text{infinite constant},\nonumber 
\end{align}
which gives:
\begin{equation}
W^{{\rm dim-reg}}=m\phisq+\frac{1}{2}\left(\omega+\pi\lambda\left|\lambda\right|\right)\left(\phisq\right)^{2}.\label{eq:Higgsed IR superpotential dim reg}
\end{equation}
This is the same as the classical tree-level potential with a shift
to $\omega$. Starting from equations (5.9) and (5.10) of \cite{Dey:2019ihe},
and expanding the effective potential in the Higgsed branch at small
$\lambda$, we obtain:
\begin{equation}
U_{{\rm eff}}^{\left(\pm,-\right)}=N\left(\left(\omega\pm\pi\lambda^{2}\right)\phisq+\mu\right)^{2}\left(\phisq\pm\frac{1}{6\pi}\mu\right),
\end{equation}
which has almost exactly the form of a potential derived from the
superpotential \eqref{eq:Higgsed IR superpotential dim reg} in the
$\left(+,-\right)$ branch, and has the same F-term equation for supersymmetric
vacua. As discussed in subsection \ref{subsec:The Conformal Limit}, the $\left(-,-\right)$ branch is irrelevant.
Note that we used the realation $w=\frac{\omega}{2\pi\lambda}$ that
was discussed in subsubsection \ref{subsubsec:The-IR-Limit}.

Had we taken $\kappa\to\infty$ \textbf{after} integration, we would
obtain:
\begin{equation}
W^{\text{YM-reg}}=\left(m-\frac{1}{2}\lambda\left|\kappa\right|\right)\phisq+\frac{1}{2}\left(\omega+2\pi\lambda\left|\lambda\right|\right)\left(\phisq\right)^{2}.
\end{equation}
We see that the IR mass is shifted relative to its UV value by $\frac{1}{2}\lambda\left|\kappa\right|$,
the same as in the unHiggsed result \eqref{eq:final unHiggsed superpotential}.

Interestingly, there is a simple physical interpretation to the difference
between the two regularizations in this case. When taking $\kappa\to\infty$
\textbf{prior} to integration, we saw (in \eqref{eq:decoupling of very massive polarization})
that one of the two polarizations (the more massive one) decouples.
This is because the Higgsed pure Chern-Simons matter theory has only a single
propagating mode, the less massive of the two polarization states
in the CS-YM case. In fact, one can also obtain \eqref{eq:Higgsed IR superpotential dim reg}
by taking $\kappa\to\infty$ \textbf{after} integration, provided
one simply \textbf{ignores} the more massive polarization:
\begin{align}
 & \cancel{-{\rm Tr}\left(D^{2}+\frac{\kappa}{2}\left(1+\sqrt{1+8\pi\lambda\frac{\phisq}{\kappa}}\right)\right)}\\
 & -{\rm Tr}\left(D^{2}+\frac{\kappa}{2}\left(1-\sqrt{1+8\pi\lambda\frac{\phisq}{\kappa}}\right)\right)\\
= & \frac{1}{8\pi}\frac{\kappa\left|\kappa\right|}{4}\left(1-\sqrt{1+8\pi\lambda\frac{\phisq}{\kappa}}\right)^{2}\\
\kappa\to\infty\to & \frac{1}{2}\pi\lambda\left|\lambda\right|\left(\phisq\right)^{2}.
\end{align}

\subsection{\label{subsec: Dimensional vs Yang-Mills Regularization}Dimensional vs Yang-Mills Regularization}

The results of this section so far are:
\begin{align}
W^{{\rm uH}}= & -\frac{\left(M-\left(m-\frac{1}{2}\lambda\left|\kappa\right|\right)\right)^{2}}{2\omega}-\frac{1}{8\pi}M\left|M\right|\\
 & +\frac{1}{16\pi}\left(2M+\kappa\right)\lambda\kappa\log\left(1+\frac{2\left|M\right|}{\left|\kappa\right|}\right)-\frac{1}{8\pi}\lambda\left|\kappa\right|\left|M\right|\\
W^{{\rm H}}= & m\phisq+\frac{\omega}{2}\left(\phisq\right)^{2}-\kappa\left|\kappa\right|\sqrt{1+8\pi\lambda\frac{\phisq}{\kappa}}.
\end{align}
By taking $\kappa\to\infty$ we obtain:
\begin{align}
W_{{\rm IR,\,YM-Reg}}^{{\rm uH}}= & -\frac{\left(M-\left(m-\frac{1}{2}\lambda\left|\kappa\right|\right)\right)^{2}}{2\omega}-\frac{1-2\left|\lambda\right|}{8\pi}M\left|M\right|-\frac{1}{8\pi}\lambda M^{2}\\
W_{{\rm IR,\,YM-Reg}}^{{\rm H}}= & \left(m-\frac{1}{2}\lambda\left|\kappa\right|\right)\phisq+\frac{1}{2}\left(\omega+2\pi\lambda\left|\lambda\right|\right)\left(\phisq\right)^{2},
\end{align}
which is essentially the superpotential in Chern-Simons theory in
Yang Mills regularization. Note that the limit only makes sense if
one absorbs the mass renormalization factor of $\frac{1}{2}\lambda\left|\kappa\right|$
into the definition of $m$ - the usual procedure of subtracting divergences.
As we saw in subsubsection \ref{subsubsec:The-IR-Limit} and subsubsection \ref{subsubsec: The IR Limit higgsed}, we reproduce known results directly by taking
$\kappa\to\infty$ \textbf{prior} to loop integration giving:
\begin{align}
W_{{\rm IR,\,dim-Reg}}^{{\rm uH}}= & -\frac{\left(M-m\right)^{2}}{2\omega}-\frac{1}{8\pi}M\left|M\right|-\frac{1}{16\pi}\lambda M^{2}\\
W_{{\rm IR,\,dim-Reg}}^{{\rm H}}= & m\phisq+\frac{1}{2}\left(\omega+\pi\lambda\left|\lambda\right|\right)\left(\phisq\right)^{2}.
\end{align}
The IR physics, however, should not depend on the choice of regularization. We find that the following redefinition of parameters matches the two schemes to leading order in $\lambda$:
\begin{align}
M_{{\rm IR}}= & \sqrt{1-2\left|\lambda\right|}M\label{eq: M_IR definition}\\
m_{{\rm IR}}= & \sqrt{1-2\left|\lambda\right|}\left(m-\frac{1}{2}\lambda\left|\kappa\right|\right)\label{eq:m_IR definition}\\
\omega_{{\rm IR}}= & \left(\omega+\pi\lambda\left|\lambda\right|\right)\frac{1-2\left|\lambda\right|}{1+\frac{\omega\lambda}{8\pi}}\label{eq:omega_IR definition}\\
\Phi_{{\rm IR}}= & \left(1-2\left|\lambda\right|\right)^{-1/4}\Phi,\label{eq:Phi_IR definition}
\end{align}
provided we also modify the ``counterterm'' $\delta J$ from \eqref{eq: linear M counter term} so as to
add the term \\ $\frac{1}{8\pi}\lambda\left(m-\frac{1}{2}\lambda\left|\kappa\right|\right)M$
to the unHiggsed superpotential.

This redefinition suggests that the field amplitude renormalization differs between the two schemes by a factor of $\sqrt{1-2|\lambda|}$. As a sanity check, we compute it directly and find agreement to leading order in $\lambda$ in the appendix \ref{sec:Field-Amplitude-Renormalization}.

\subsection{\label{subsec: Validity of the Calculation}Validity of the Calculation}

In this section we argue that at small $\lambda$ the calculations of this section give the dominant contribution to the superpotential for every value of the parameters. We worked non-perturbatively in the quartic coupling $\omega$, since it appears only in the mass of $\sigma$, which, as discussed in subsubsection \ref{subsubsec:N Counting}, is essentially non-dynamical in the calculation. Nevertheless, the assumption that higher loop corrections are smaller may fail for some values of the VEV.\footnote{Note that while the VEV of the field is not an adjustable parameter - as it is determined by the F-term equations - we treat it as such. In reality, one should think of adjusting the couplings $\omega,m$, and for those values for which the solutions $M(\omega,m),\phisq(\omega,M)$ are large / small, one asks if the approximation used to determine them is consistent with their size.} The canonical example of this is in Coleman and Weinberg's original paper \cite{coleman:1973}. There the effective potential in $\varphi^4$ theory receives a correction $\sim\varphi^4 \log \varphi$ which is dominant at small field values. This reflects the RG flow of the quartic coupling. When the VEV is small relative to the dynamically generated scale of the theory, then so is the mass, making it so that the RG flow is not truncated and the value of the coupling changes. Since higher loop corrections may carry factors of e.g. $\varphi^4 \log^2 (\varphi)$, which is more dominant at small $\varphi$, perturbation theory cannot be trusted.

One might similarly wonder whether terms like $\log|M|,\log\phisqbr$ appear at higher orders in $\lambda$ in the theory \eqref{eq:UV_action}. Those should dominate in the deep IR. For instance $M^2\log|M|$ would dominate over $M^2$ and $M\log|M|$ over $M$, so both would qualitatively change the phase diagram. Similarly in the UV we found in the unHiggsed phase a sub-leading contribution $\sim M\log|M|$, which dominates over $M$, and changes the phase diagram significantly. But could higher powers of the logarithm be generated in perturbative calculation? In the Higgsed phase, we found that the 1-loop correction is sub-sub leading in the UV, but could higher loop contributions become sub-leading as they are in the unHiggsed phase? 

In this subsection we argue that the answer to the above questions is ``no'', and we can trust the leading superpotential in both branches in both the deep IR and deep UV, at leading order in $N$. As we will discuss in subsection \ref{subsec: finite N}, at large but \textbf{finite} $N$, the couplings of the theory do get a small beta function, making much of this analysis inapplicable there.

\subsubsection{UnHiggsed branch}

\paragraph{In the IR} regime $|M|\ll|\kappa|$ we are effectively in the Chern-Simons-matter theory. Since this theory is conformal at leading order in $1/N$ \cite{Aharony:2011jz,Aharony:2019mbc}, the superpotential in the strict $\kappa\to\infty$ limit can only have terms $\sim M^2$ by dimensional analysis.\footnote{Apart from linear terms $\sim \kappa M$ which need to be cancelled using counterterms as discussed in subsubsection \ref{subsubsec:Feynman-Rules}} More generally at finite $\kappa$, since no logarithmic terms are expected, we can expand the superpotential in the IR as:
\begin{align}
    \sum_{n=-\infty}^\infty \left(\lambda a_n (|\lambda|) \left|\kappa^{2-n}M^n\right|+|\lambda| b_n (|\lambda|) \left|\kappa^{2-n}M^{n-1}\right|M \right),
\end{align}
since the superpotential has mass dimension 2 and odd parity (see e.g. \cite{Gaiotto:2018yjh}). The problematic terms with $n<0$ would imply that the IR theory is non-renormalizable, since from the IR point of view $\kappa$ is simply a UV regulator, and these terms are divergences that cannot be subtracted by local counter-terms. Since we know that the IR theory is, in fact, renormalizable - they should vanish.

This can also be understood from a diagrammatic perspective. As discussed in subsubsection \ref{subsubsec:N Counting}, all diagrams contributing to the effective superpotential at leading order in $N$ have the form of a single $\Phi$ loop with various propagator corrections. Each 1PI propagator correction is made of arbitrarily many gauge propagators, and at order $\lambda^n$ has $n$ loops. At $|M|\ll|\kappa|$ the $\Phi$ loop integral will be dominated by the regime where the loop momentum $p$ is much smaller than the Yang-Mills scale $|p|\ll |\kappa|$. Then since the propagator correction loop integrals are convergent (see below), they will all become dominated by the low energy regime where they are insensitive to $\kappa$. Hence terms that depend strongly on $\kappa$ are avoided.

\paragraph{In the UV} $|M|\gg |\kappa|$ we found at 1-loop the subdominant term $\sim M\log M$, which arose due to a logarithmic divergence. Both the linear term $\sim M$ and the YM interaction are classically relevant with mass dimension 1, so in hindsight it is no surprise that the winner of the competition between them in the UV was not a forgone conclusion, but rather is determined by quantum corrections. The question is then whether additional logarithmic divergences can occur that would give rise to terms $\sim M\log^n M$, for $n>1$. Generically this is expected to happen, but in a super-renormalizable theory it can be averted. In fact, since all the propagator corrections are convergent (see below), one can always evaluate them first and leave the integral over the momentum running in the $\Phi$ loop last. A logarithmic divergence can only occur there, so we conclude that at leading order in $N$ there will be at most one.

That the propagator corrections are UV convergent (in dim reg) follows from the super-renormalizability of the theory. We find that they are convergent at 1-loop in subsection \ref{subsec: phi propagator corrections appendix}. At higher loops their convergence can only improve.

We can justify this using a power-counting argument. First, consider an arbitrary 1PI correction that has only a \textbf{single} $\Phi$ propagator. After carrying out all loop integrals that do not involve this $\Phi$ propagator, one arrives (by dimensional analysis) at a single loop that has the following \textbf{large momentum} ($p$) behavior:
\begin{align}
    \intop d^3p \underbrace{d^2\theta D^{-2}}_{\sim p^-2} f\left( \frac{g^2}{|p|},\lambda \right),
\end{align}
for some function $f$. In the argument of $f$ there is also $\frac{g^2}{D^2}$ but we consider $D^2\sim |p|$ for power counting purposes. $f$ should vanish as $g^2\to0$. One can think of this as the UV limit where the gauge field is effectively a pure-YM field and dimensionful couplings only enter multiplicatively (can be used as an expansion parameter). In that case a UV divergence can only appear from the linear term $f\sim \frac{g^2}{|p|}$. This is the 1-loop diagram \eqref{eq:rainbow propagator correction diagram} that we computed, which is actually convergent, as all 1-loop diagrams are in dim-reg.

For diagrams with $n$ $\Phi$-propagators, the argument changes a bit. For $n=0$ the entire 1-PI correction evaluates to some constant, and so is unimportant. For $n>1$ we consider two cases:
\begin{itemize}
    \item The ``special'' case. No loop contains more than one $\Phi$ propagator. The 1PI correction is simply a sequence of $\Phi$ propagators, each with its own loop. In this case the 1-loop argument carries through for each of the $n$ loops (after carrying out whatever loop integrals contain no $\Phi$ propagators).
    \item The ``general'' case. If multiple $\Phi$ propagators appear in a single loop, then at large $N$ they can be viewed as a ``single'' $\Phi$-propagator with various 1PI corrections. This is essentially the case of \textbf{nested} 1PI corrections. By dimensional analysis, the large $p$ behavior of the 1PI corrections is at most $\sim|p|$ (the amplitude renormalizations). Since each comes with an additional factor of the $\Phi$ propagator, which goes like $|p|^{-1}$, this doesn't worsen the UV convergence of the graph. Stated differently, the ``full'' propagator has the same large $p$ behavior $\sim |p|^-1$. Hence the previous arguments also apply to a loop containing multiple $\Phi$ propagators.
\end{itemize}

\subsubsection{Higgsed branch}

In the IR, the same arguments carry through with the replacement $M\to\phisq$.

As discussed in subsubsection \ref{subsubsec: N counting Higgsed} The Feynman rules and leading $N$ diagrams generating this branch of the superpotential are the same as those in the unHiggsed branch, except that the fundamental fields are now the $W$-Boson superfields, with two polarization states with masses:
\begin{align}
    \frac{\kappa}{2}\left(1 \pm \sqrt{1+8\pi\lambda\frac{\phisq}{\kappa}}\right),\label{eq: higgsed polarization masses}
\end{align}
while the gauge superfields are now the \textbf{unbroken} gauge fields only. The gauge superfields will have the same propagator as in the unHiggsed branch in whatever gauge one chooses, e.g. Landau gauge. The $W$ superfields will have the propagator \eqref{eq:broken gauge propagator} with the replacement of $\phi^{ab}$ with its eigenvalue $\frac{1}{2} \phisq$:
\begin{equation}
\left\langle \bar{W}^{\alpha}W^{\beta}\right\rangle =-\frac{1}{N}\frac{2D^{\beta}D^{\alpha}\left(\kappa+D^{2}\right)-8\pi\lambda\kappa\phisq C^{\alpha\beta}}{\left(\left(D^{2}+\frac{1}{2}\kappa\right)^{2}-\frac{1}{4}\left(\kappa^{2}+8\pi\lambda\kappa\phisq\right)\right)\phisq}.\label{eq:w Boson propagator}
\end{equation}

The main difference in the Higgsed branch concerns the UV regime. At leading order in $\lambda$ we found no logarithmic divergences and the UV behavior was $\propto \sqrt{\phisq}$. But could higher corrections give, as was the case in the unHiggsed branch, a term $\propto \phisq \log\phisqbr$, or even a term $\propto\phisq$? Indeed, at leading order a term $\propto\phisq$ almost did appear, but was cancelled between the contributions from the two terms in \eqref{eq:W 1 loop Higgsed unevaluated}, corresponding to the two polarizations. Would a similar thing happen at higher loops? Furthermore, the second diagram in \eqref{eq: 2-loop diagrams unHiggsed}, that gave rise to the $M\log M$ term in the unHiggsed phase, also exists in the Higgsed phase at the next order in $\lambda$. 

Fortunately, the authors of \cite{Bashmakov:2018wts}, showed by a power-counting argument, that the $\sim \phisqbr^{1/2}$ behavior is, in fact, the fastest growth rate. This has to do with the UV behavior of the propagators in the Higgsed phase. For large $\phisq$ the only effective mass in the problem is the $W$-s masses $\approx \sqrt{\lambda \kappa \phisq}$, while the effective couplings are $\kappa$ and $\phisqbr^{-1}$ (as seen from the $W$ propagator \eqref{eq:w Boson propagator}). This arrangement ensures that corrections are suppressed at large $\phisq$.

\section{\label{sec:Phase Diagram}Phase Diagram}

In this section we will compute the phase diagram of our YM-CS-matter theory \ref{eq:UV_action}. The main result of this section is the phase diagram figure \ref{fig: phase diagram infinite N leading lambda}.

Much of this section will rest on the discussion in the background section \ref{sec:bg}. In particular, the Witten index and the definition of the various vacua \eqref{eq:vacuum notation} are reviewed in subsection \ref{subsec: witten index and vacua}, and the classical phase diagram - figure \ref{fig: classical phase diagram} is reviewed in subsection \ref{subsec: classical phase diagram}.

We build the phase diagram in two steps. We start by analyzing the theory at $\lambda=0$, in subsection \ref{subsec: lambda equals 0 phase diagram}. Then, we use the results of section \ref{sec: computing the superpotential} to compute the full phase diagram - figure \ref{fig: phase diagram infinite N leading lambda} for small $\lambda$ in subsection \ref{subsec: small lambda diagram}. We then conjecture the phase diagram of the theory at general $\lambda$ in subsection \ref{subsec:Conjectured Phase diagram for General lambda} and at large but finite $N$ (figure \ref{fig:finite_N}) in subsection \ref{subsec: finite N}. We also analyze the full theory with $\lambda_3 \neq 0$ in subsection \ref{subsec: adding a sigma cubed coupling}.

\subsection{\label{subsec: lambda equals 0 phase diagram}\texorpdfstring{$\lambda=0$}{lambda = 0}}

In the main model studied in this paper, given by the Lagrangian \eqref{eq:UV_action}, the gauge sector decouples from the matter sector in the deep UV. This is because the Yang-Mills term is a relevant deformation, and it is reflected by the fast fall-off behavior of the gauge propagator \eqref{eq:Landau_gauge_propagator} at large momentum. Hence the effective superpotential at $\lambda=0$ should be a good approximation when the VEVs are much larger than the Yang-Mills scale $\kappa$.

\begin{figure}
\centering
\begin{subfigure}{.45\textwidth}
  \centering
  \includegraphics[width=0.95\textwidth]{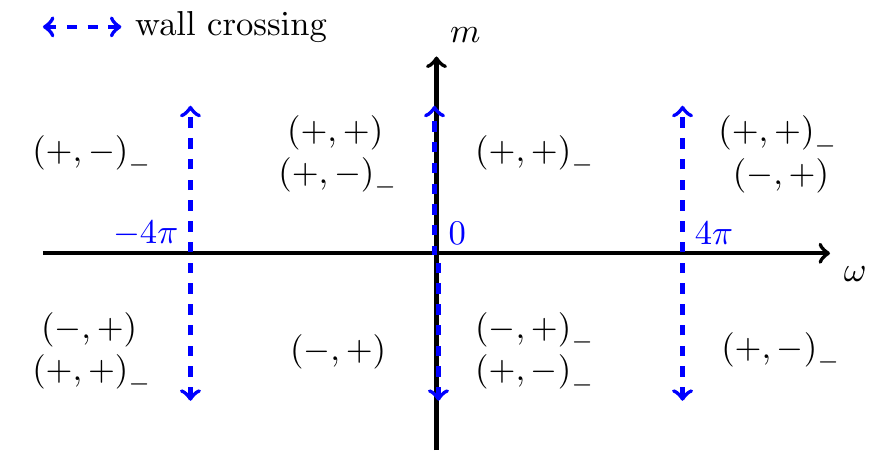}
  \caption{}
  \label{fig: lambda equals 0 phase diagram}
\end{subfigure}
\begin{subfigure}{.45\textwidth}
  \centering
  \includegraphics[width=0.95\textwidth]{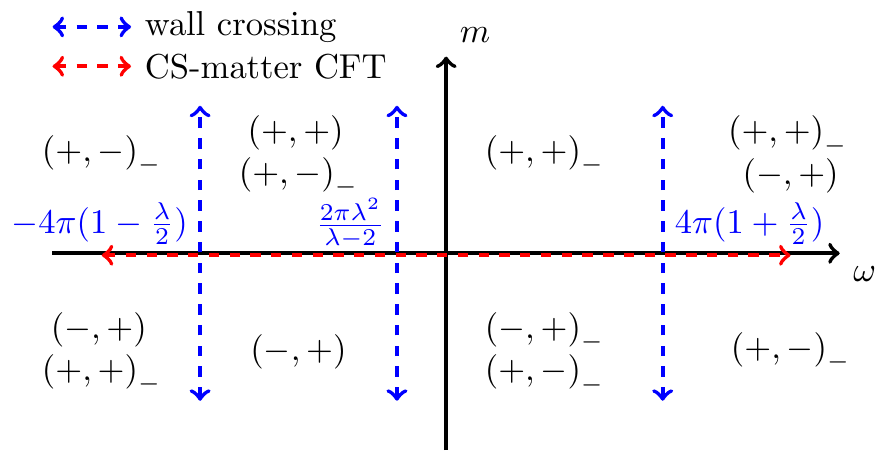} 
  \caption{}
  \label{fig: IR phase diagram minwalla}
\end{subfigure}
\caption{(a) Phase diagram at $\lambda=0$, which governs the ``UV limit'' $\kappa\to0$. (b) Phase diagram of the IR theory studied by \cite{Dey:2019ihe} at arbitrary $\lambda\in \left[0,1\right]$, governing the ``IR limit'' $\kappa\to\infty$.}
\label{fig:lambda=0_phase_diagram}
\end{figure}

When $\lambda=0$ the theory becomes essentially a standard large N vector model. Ironically, we can obtain the same theory also in the IR. If one first takes $\kappa\to\infty$, one arrives at the pure Chern-Simons-matter theory studied in \cite{Dey:2019ihe,Inbasekar:2015tsa} and discussed in subsection \ref{subsec:The Conformal Limit}. Then, the $\lambda\to0$ limit gives us the same vector model. Hence the ``UV phase diagram'' should be almost the same as the ``IR phase diagram'' - figure \ref{fig:Minwalla_phase_diagram}. Indeed, the IR phase diagram, reproduced in figure \ref{fig: IR phase diagram minwalla}, with the coupling $\omega=2\pi\lambda w$ in place of $w$, can readily be seen to give the UV phase diagram - figure \ref{fig: lambda equals 0 phase diagram} - in the $\lambda\to 0$ limit. Note that the IR phase diagram figure \ref{fig: IR phase diagram minwalla} exhibits the Bosonization duality discussed in subsubsection \ref{subsubsec: The Bosonization Dualities}. This is described in detail in \cite{Dey:2019ihe}. When $\lambda\to0$ this reduces to the parity symmetry (or discrete $R$-symmetry) $m\to-m,\omega\to-\omega$.

At leading order in $N$ we computed in section \ref{sec: computing the superpotential} the full quantum effective superpotential when $\lambda=0$. At large $\omega$ the operator $\phisq$, or equivalently (through equation \eqref{eq:sigma EOM relation to Phi}) $\sigma$ takes on a life of its own, distinct from $\Phi$. Hence as discussed in section \ref{sec: computing the superpotential}, the superpotential splits into an unHiggsed branch $W^{\rm uH}(M)$ and a Higgsed branch $W^{\rm H}\phisqbr$. The variable $M$ (see the discussion in subsection \ref{subsec:The UnHiggsed Branch of the Superpotential}) is defined by $M=m+\sigma$ and is equal to the pole mass of $\Phi$ at 0th order in $\lambda$. Hence a solution with ${\rm sign}(M)=\pm 1$ corresponds to a $\left( \pm,+ \right)$ vacuum (We defined the vacua according to ${\rm sign}(M\lambda)$ so one could think of the above statement as applying to ``$\lambda=0^+ $'').

At $\lambda =0$ the Higgsed branch receives no correction, so that the superpotential governing it is simply the classical \eqref{eq:tree level superpotential}. However, in the unHiggsed branch, the superpotential  becomes \eqref{eq: 1-loop unHiggsed superpotential}:
\begin{align}
W^{\rm uH}_{\lambda=0} &  =-\frac{\left(M-m\right)^{2}}{2\omega}-\frac{1}{8\pi}M\left|M\right|,\label{eq: 1-loop unHiggsed superpotential 1}
\end{align}
with F-term equation \eqref{eq:ungauged_pole_mass}:
\begin{equation}
M=\frac{m}{1+\frac{\omega}{4\pi}{\rm sign}\left(M\right)}.
\end{equation}
We see that the diagram splits into:
\begin{enumerate}
\item a ``classical'' regime $\left|\omega\right|<4\pi$ where a self
consistent solution exists only when ${\rm sign}\left(M\right)={\rm sign}\left(m\right)$
(and so there is always one unHiggsed vacuum, as in the classical
analysis), and
\item a ``quantum'' regime $\left|\omega\right|>4\pi$ where a self-consistent
solution has ${\rm sign}\left(M\right)={\rm sign}\left(m\omega M\right)$
(and so there are either no unHiggsed vacua (${\rm sign}\left(m\omega\right)=-1$)
or 2 unHiggsed vacua (${\rm sign}\left(m\omega\right)=1$).
\end{enumerate}
This is summarized in figure \ref{fig: lambda equals 0 phase diagram} which presents the full phase diagram of the theory at $\lambda=0$. As expected, the classical regime is still well described by figure \ref{fig: classical phase diagram}, but in the unHiggsed phase we see the appearance of two new wall crossings, one for each type of unHiggsed vacuum. Naively, one might expect only a single wall crossing to exist, corresponding to the vanishing of the leading term in the superpotential $\frac{\omega}{2} \left(\phisq\right)^2~ \frac{\sigma^2}{2\omega}$, but we see in \eqref{eq: 1-loop unHiggsed superpotential 1} that the the coefficient of this operator vanishes at different points for different branches of the superpotential. We expect these additional wall crossings to disappear when we add the $\lambda_3\sigma^3$ term to the superpotential.

Also, just as in figure \ref{fig: classical phase diagram}, there are ``wall crossings within wall crossings'' at $m=0$, where a moduli space of vacua develops. 

Note that if one follows, for instance, the $\left(+,+\right)$ vacuum as one crosses the wall at negative $\omega$ in figure \ref{fig: IR phase diagram minwalla}, and picks $m$ so that the vacuum continues to exist ($m>0$ to the right of the crossing, and $m<0$ to its left), the sign of the mass of $\sigma$ around this vacuum changes (this is responsible for the statistics flip). This is in accordance with the findings of \cite{Inbasekar:2015tsa}. The authors of \cite{Inbasekar:2015tsa} studied the S-matrix of this theory, and found a particular pole in its singlet channel - corresponding to some propagating singlet state. Indeed, the value of $\omega$ for which they found that the mass of the singlet vanishes is precisely the position of the leftmost wall crossing of figure \ref{fig: IR phase diagram minwalla}. Although a singlet field $\sigma$ wasn't explicitly included in their theory, it's natural to identify it with $\sigma$. One can imagine adding $\sigma$ as a purely auxiliary field, which is imbued with dynamics at low energies when the fundamental field $\Phi$ is integrated out. Probably a similar massless singlet would be found at the rightmost wall, if one analyzed the S-matrix around the $\left(-,+\right)$ vacuum.

\subsection{The Small \texorpdfstring{$\lambda$}{lambda} Regime}\label{subsec: small lambda diagram}

\begin{figure}
\begin{centering}



\definecolor{cadmiumgreen}{rgb}{0.0, 0.42, 0.24}

\begin{tikzpicture}[scale=.8, every node/.style={scale=.8}]


\draw[very thick,->] (-8,0) -- (8,0) node[anchor=north west] {\LARGE $\omega$};
\draw[ very thick,->] (0,-1) -- (0,4.5+1.5) node[black, anchor=south east] {\LARGE  $a \equiv m/ \kappa $};

\draw[blue,dashed, very thick,<->] (0.02,-1.5) -- (0.02,0) -- (-0.02,0) -- (-0.02,4+1.5);
\draw[blue,dashed, very thick,<->] (-6.5,-1.5)  -- (-6.5,4);
\draw[] (-6.5,2.5) node[blue,anchor=west]{$\omega = -4 \pi $};
\draw[blue,dashed, very thick,<->] (6.5,-1.5)  -- (6.5,4+1.5);
\draw[] (6.5,2.5) node[blue,anchor=west]{$\omega = 4 \pi $};

\draw (3,3) node[black] {$ \left(+,+ \right)_- $};
\draw (-3,3) node[black] {$ \left(+,+ \right) \, , \, \left(+,- \right)_{-} $};
\draw (-8,3) node[black] {$  \left(+,- \right)_{-} $};
\draw (3,-1) node[black] {$ \left(-,+ \right)_- \, , \, \left(+,- \right)_-  $};
\draw (-3,-1) node[black] {$ \left(-,+ \right) $};
\draw (-8,1) node[black] {$\left(-,+\right)\,,\,\left(+,+\right)_{-}$};
\draw (8.5,3.5) node[black] {$\left(+,+\right)_-\,,\,\left(-,+\right)$};
\draw (8.5,1) node[black] {$\left(+,-\right)_-$};
\draw (3,2) node[red,anchor=north] {$ a = \frac{\lambda}{2} $};

\draw[very thin, ->] (-.8,.4) node[rectangle,anchor=north east,fill=cadmiumgreen!20, draw=gray,very thin, text=black] {$ \left(-,+ \right) \, , \, \left(+,- \right)_{-}\, , \, \left(+,- \right) $} -- (-.2,1.7);

\draw[very thin, ->] (-6,.8+.8) node[rectangle,anchor=north west,fill=orange!20, draw=gray,very thin, text=black] {$ \left(+,- \right)_{-} \, , \, \left(+,+ \right)_{-}\, , \, \left(+,+ \right) $} -- (-6.7,2.2);

\draw[very thin, ->] (5.5,4.5-1) node[rectangle,anchor=south east,fill=orange!20, draw=gray,very thin, text=black] {$ \left(+,- \right)_- \, , \, \left(-,+ \right)\, , \, \left(-,+ \right)_- $} -- (6.9,1.7);


\draw[red,thick,dashed, <->] (-10,2) -- (10,2);
\draw[cadmiumgreen,thick,dashed] (-1,2) node[blue,circle,fill,inner sep=2pt] {} .. controls (-.5,2) and (0,1) .. (0,0)node[blue,circle,fill,inner sep=2pt] {};
\draw[orange,thick,dashed,->] (8.5,2) node[blue,circle,fill,inner sep=2pt] {} .. controls (6.7,2) and (6.5,.5) .. (6.5,-1.3);
\draw[orange,thick,dashed,->] (-7.5,2) node[blue,circle,fill,inner sep=2pt] {} .. controls (-6.8,2) and (-6.5,2.5) .. (-6.5,3.8);

\draw[thick,<->,blue,dashed] (-9,6) -- (-7.8,6) node[anchor=west] {wall crossing};
\draw[thick,<->,red,dashed] (-9,6-.4) -- (-7.8,6-.4) node[anchor=west] {CS-matter CFT};
\draw[thick,<->,orange,dashed] (-9,6-.8) -- (-7.8,6-.8) node[anchor=west] {pair of unHiggsed vacua (dis-)appear};
\draw[thick,<->,cadmiumgreen,dashed] (-9,6-1.2) -- (-7.8,6-1.2) node[anchor=west] {pair of Higgsed vacua (dis-)appear};

\end{tikzpicture}

\par\end{centering}
\caption{Complete phase diagram of the theory \eqref{eq:UV_action} at infinite $N$, small $\lambda$ and $\lambda_3=0$. }
\label{fig: phase diagram infinite N leading lambda} 
\end{figure}
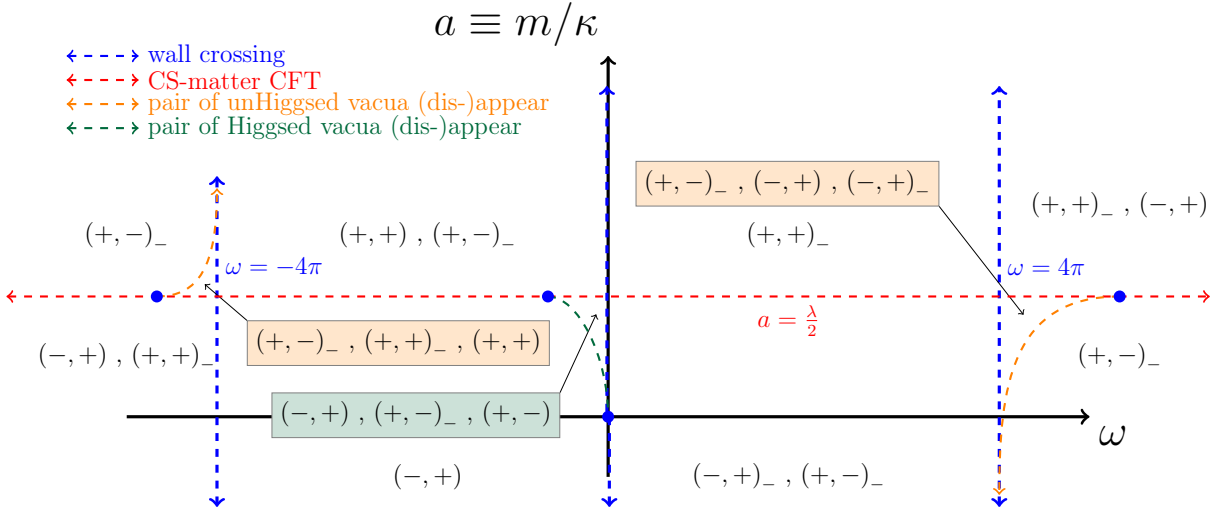

The phase diagrams - figures \ref{fig: lambda equals 0 phase diagram} and \ref{fig: IR phase diagram minwalla} - should capture the UV ($|M|\gg |\kappa|$) and respectively the IR ($|M|\ll|\kappa|$) regimes of the theory \eqref{eq:UV_action} to all orders in $\lambda$. This is possible because gauge dynamics decouple in the UV, while in the IR the theory is described by the simpler Chern-Simons theory, where the analysis of \cite{Dey:2019ihe,Inbasekar:2015tsa} holds. The full phase diagram of the theory should, in some way, ``interpolate'' between these two. In subsection \ref{subsec:Conjectured Phase diagram for General lambda}, we will present our conjecture for the simplest phase diagram that is consistent with both limits, but it would be helpful to arrive at it directly at least in some limit.

Indeed, at small $\lambda$ it's possible to find the diagram by using the effective superpotential at leading order in $\lambda$ which was calculated in section \ref{sec: computing the superpotential}. As discussed in subsection \ref{subsec: Validity of the Calculation}, this superpotential can be trusted, at small $\lambda$ and ``infinite'' $N$ for all values of the VEV, giving a complete phase diagram.

As discussed in subsection \ref{subsec: Higgesd branch of the superpotential} this effective superpotential in the Higgsed branch $W^{\rm H}\left(\phisq\right)$ is simply the tree level superpotential with the 1-loop correction that was computed in \cite{Choi:2018ohn}, and recomputed in the appendix \ref{sec:Detailed Higgsed Phase Computations}, given by \eqref{eq: Higgs branch superpotential small lambda}.

Similarly, the effective superpotential in the unHiggsed branch $W(M)$ was computed in subsection \ref{subsec:The UnHiggsed Branch of the Superpotential} and is given by \eqref{eq:final unHiggsed superpotential}.

The resulting phase diagram is given in figure \ref{fig: phase diagram infinite N leading lambda}. The mass $m$, corresponding to the vertical axis, is now measured in units of the YM mass $\kappa$, giving a dimensionless parameter $a\equiv m/\kappa$.

Due to the mass shift $m\to m-\frac{1}{2}\lambda\left|\kappa\right|$, the line of Chern-Simons-matter fixed points, represented by a dotted red line in figure \ref{fig: phase diagram infinite N leading lambda}, is shifted vertically with respect to the $m=0$ axis. Concretely, when on the red line ($m-\frac{1}{2}\lambda\left|\kappa\right|=0$), $M=0$ and $\phisq=0$ are solutions to the F-term equations. In its vicinity, $M,\phisq\ll\kappa$, so that there is a separation of scales and the vacua are effectively in the IR Chern-Simons theory.

The red line can be viewed as the $m=0$ axis of figure \ref{fig: IR phase diagram minwalla}. Phases in its vicinity should be consistent with those of figure \ref{fig: IR phase diagram minwalla}. However, the wall crossings of figure \ref{fig: IR phase diagram minwalla} no longer exist. This is as expected, since before a vacuum could appear from / escape to infinity (in field space), it would depart the IR regime, so that figure \ref{fig: IR phase diagram minwalla} would no longer be applicable. Instead, as expected, the wall crossings of the full theory are those of the UV theory, so their position is exactly that of figure \ref{fig: lambda equals 0 phase diagram}.

In keeping with the the notion, discussed in subsection \ref{subsec: witten index and vacua}, that jumps in the Witten index are ``UV effects'', that is, that they occur at infinity in field space, the Witten index is continuous along the red line. As in the $m=0$ lines of figures \ref{fig: classical phase diagram}, \ref{fig: lambda equals 0 phase diagram} and \ref{fig: IR phase diagram minwalla}, the equation which ensures this continuity in the Witten index is \eqref{eq:Witten index continuity identitiy} and variants thereof obtained with \eqref{eq:Witten index statistics flip}. The phase transition along the red line is ``at the origin of field space'', in the sense that $\sigma=M=\phisq=0$, and involves the coalescence of Higgsed and unHiggsed vacua.

Certain points along the red line, which we dub ``special points'', correspond to the position of the would be wall crossings of figure \ref{fig: IR phase diagram minwalla}. In the strict IR limit $\kappa \to \infty$, the theory has a moduli space of vacua at these points, as discussed in subsection \ref{subsec: lambda equals 0 phase diagram}. In the full theory the moduli space is lifted and these become endpoints of lines in parameter space, where pairs of vacua of opposite statistics form or annihilate at some finite position in field space. These lines, indicated by green (for Higgsed) and orange (for unHiggsed) dashed lines in figure \ref{fig: phase diagram infinite N leading lambda}, are akin to second order phase transition lines, in that they delineate two phases, and are characterized by a non-trivial CFT, but since they do not involve a change in the phase over a particular vacuum that exists on either of their sides, \textbf{we will not refer to them as phase transitions}. Instead, we'll refer to them simply as \textbf{orange and green} lines in all that follows. The CFT is simply that of a massless singlet matter sector - the massless particle being $\sigma$ for the orange lines, and the longitudinal mode $h$ for the green lines.

The reason for the appearance of these lines could be understood intuitively: consider starting in the vicinity of one of the ``special points'', and then tuning $\omega$ up or down, so as to approach the would-be wall crossing. If the IR theory predicts that a vacuum should \textbf{disappear} to infinity, then in the full theory it should ascend in field space to a point $\geq|\kappa|$, where the IR theory is no longer valid and then either:
\begin{itemize}
    \item stay there, or
    \item meet with another identical vacuum with opposite statistics, and ``annihilate'' with it - both ceasing to exist without changing the Witten index.
\end{itemize}
Similarly, if the IR theory predicts that a vacuum should \textbf{appear} from infinity, then in the full theory said vacuum should either:
\begin{itemize}
    \item already exist at values $\geq|\kappa|$ and simply descend into the IR regime, or
    \item form together with an identical vacuum of opposite statistics and then descend, while its partner remains outside the IR regime.
\end{itemize}
Which of these two options is realized depends on whether the vacuum is expected to exist in the UV regime.

In the vicinity of the special points, parts of the phase diagram contain vacua that should not exist in the IR theory, but nevertheless have a VEV $\ll|\kappa|$. For instance, on top of the orange / green lines described above, the formation / annihilation point is located at a finite point in field space, which approaches 0 as one moves along the line towards the special point at its end. This would appear to be a contradiction, but note that when $\kappa$ is taken to $\infty$, these ``unexpected'' vacua all scale to $\infty$ as well, albeit a ``smaller`` infinity. Hence the IR phase diagram of figure \ref{fig: IR phase diagram minwalla} is reproduced as expected. The existence of these ``unexpected'' vacua at VEVs $\ll|\kappa|$ for \textbf{finite} $\kappa$ can be viewed as a consequence of the moduli space of vacua that exists at the special points in the strict $\kappa \to \infty$ limit. That is, since both the relevant ($\sim \kappa M,\sim \kappa \phisq$) and marginal ($\sim M^2,\sim \left(\phisq\right)^2$) terms in the superpotential vanish at the special points, the dominant term becomes the ``dangerously irrelevant'' term $\sim \frac{M^3}{\kappa},\sim \frac{\left(\phisq\right)^3}{\kappa}$, making it so that the Yang-Mills interaction influences the phase structure well below the Yang-Mills scale.

The position of the special points is not exactly the one indicated in figure \ref{fig: IR phase diagram minwalla}. For instance, the position of the leftmost wall crossing in figure \ref{fig: IR phase diagram minwalla}, corresponding to the vacuum $\left(+,+ \right)$, is $\omega =  -4\pi \left(1 - \frac{1}{2} \lambda \right)$, which lies to the right of the actual wall crossing, while in figure \ref{fig: phase diagram infinite N leading lambda} it lies to the right of it. This is because the position of the special points is drawn as a function of the \textbf{UV parameter} $\omega$. As discussed in subsection \ref{subsec: Dimensional vs Yang-Mills Regularization}, the positions of wall crossings found by \cite{Dey:2019ihe} and indicated in figure \ref{fig: IR phase diagram minwalla} are computed in the dimensional reduction scheme, while the positions we find are computed in Yang-Mills regularization. In subsection \ref{subsec: Dimensional vs Yang-Mills Regularization} we discuss the redefinition of parameters that relates the two regularizations. Although from the perspective of the IR theory this distinction is ``artificial'', in the full theory it has a meaningful effect on the phase diagram.

\begin{figure}
\begin{centering}
\includegraphics[width=0.5\textwidth]{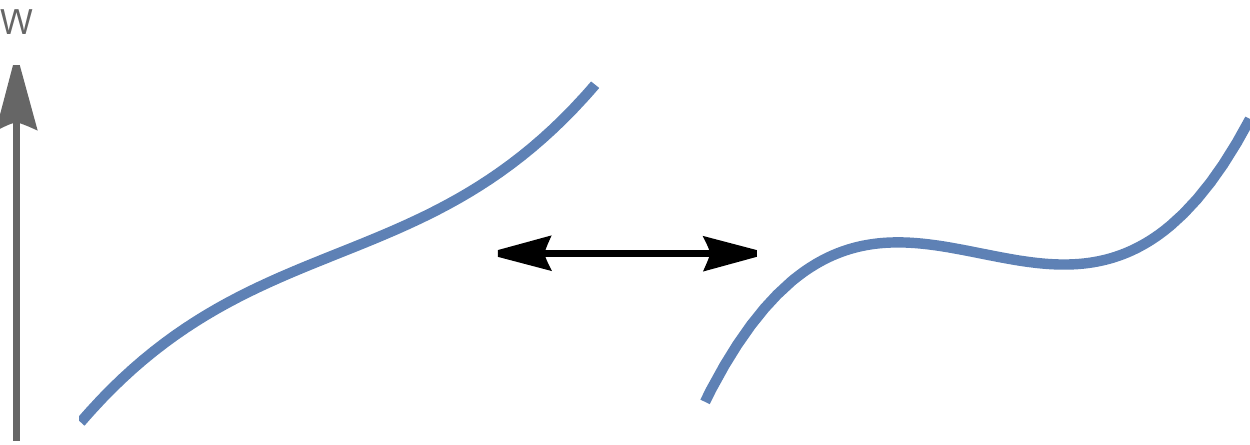}
\par\end{centering}
\caption{\label{fig: illustration of orange line} Change in the form of the superpotential as one crosses an orange or green line in figure \ref{fig: phase diagram infinite N leading lambda}. A pair of vacua of opposite statistics (singlet mass) (dis-)appear at a finite point in field-space.}

\end{figure}

The equation governing the continuity of the Witten index across the orange / green lines is \eqref{eq:Witten index statistics flip}. This formation / annihilation of vacua reflects the formation / annihilation of saddle points of the superpotential, as in figure \ref{fig: illustration of orange line}. The figure also illustrates why the two vacua have opposite statistics. The curvature of the graph at the saddle point is simply the mass of the singlet $\sigma$ in the unHiggsed phase, or of the singlet longitudinal mode $h$ of \eqref{eq: decomposition of phi to h and K} in the Higgsed. Since the two saddles on the right side in figure \ref{fig: illustration of orange line} have opposite curvatures, their effect on the statistics of the vacuum is the reverse of each other. Furthermore, precisely on top of the line, there is a single saddle that is also an inflection point, indicating that the CFT is that of a topological Chern-Simons theory with a decoupled singlet massless field. We can in principle find the parametrization of the green (or orange) lines $m(\omega)$ by solving $W'=0$ and $W''=0$ for $m$ and $\phisq$ (or $M$).

What of the UV endpoint of the orange and green lines of figure \ref{fig: phase diagram infinite N leading lambda}? Moving along the \textbf{green} line away from the IR special point, the VEV of the gapless vacuum increases, so that at $\phisq\to\infty$ it approaches the origin $\omega=m=0$. This is no surprise, as that is simply the ``UV special point'' where a moduli space develops in the UV theory. Much like in the IR, in the UV the leading and subleading terms in the Higgs branch superpotential \eqref{eq: Higgs branch superpotential small lambda} are the tree level terms, while the loop correction is sub-sub-leading. As discussed in subsection \ref{subsec: Validity of the Calculation} - higher order corrections are also expected to be sub-sub-leading. This arrangement also ensures that the position of the ``wall crossing within the wall crossing'' is unchanged, as expected.

In contrast, the orange lines of figure \ref{fig: phase diagram infinite N leading lambda} do not end at the expected ``special points''. This is because the 2-loop contribution to the unHiggsed branch \eqref{eq:final unHiggsed superpotential} is \textbf{subleading} in the UV, scaling as $O\left(M\log M\right) > O(M)$, so that it does shift the position of the ``wall crossing within the wall crossing''. Since the 2-loop correction doesn't vanish anywhere on the phase diagram, there is no longer such a ``wall crossing within the wall crossing'', and the orange lines simply asymptote along the wall crossings. Although this is un-intuitive given the decoupling of the gauge sector in the UV, the deviation from the UV diagram of figure \ref{fig: lambda equals 0 phase diagram} is minimal. The orange lines diverge logarithmically and hug the wall crossings very closely. When crossing the blue lines of figure \ref{fig: phase diagram infinite N leading lambda}, the unHiggsed vacuum that ``should have'' disappeared to infinity, will instead reach exponentially large VEVs and then annihilate with a statistics-reversed partner that genuinely appears from infinity. The jump in the Witten index is the same, due to \eqref{eq:Witten index statistics flip}, and occurs at precisely the same points $\omega=\pm 4 \pi$.

As discussed in subsection \ref{subsec: Validity of the Calculation}, the higher order corrections in $\lambda$ are not expected to introduce even faster growing terms to the superpotential in either branch at infinite $N$. The coefficient of the $\propto M\log M$ term may be a general function of $\lambda$, the sign of which will determine whether the orange line diverges upwards or downwards, but at leading order in $\lambda$ it is given as in \eqref{eq:final unHiggsed superpotential}.

Bounded by the orange (or green), red and blue lines in figure \ref{fig: phase diagram infinite N leading lambda} are new ``special phases'' with 3 vacua, that are a novelty of the full theory, as opposed to the IR and UV limits of figures \ref{fig: IR phase diagram minwalla} and \ref{fig: lambda equals 0 phase diagram}. 2 of the vacua are always statistics-flipped pairs. There are 3 such phases, one for each type of vacuum: $\left(+,+\right),\left(-,+\right),\left(+,-\right)$.

\subsection{\label{subsec:Conjectured Phase diagram for General lambda}Conjectured Phase Diagram for General \texorpdfstring{$\lambda$}{lambda}}

\begin{figure}
\begin{centering}
\includegraphics[width=0.8\textwidth]{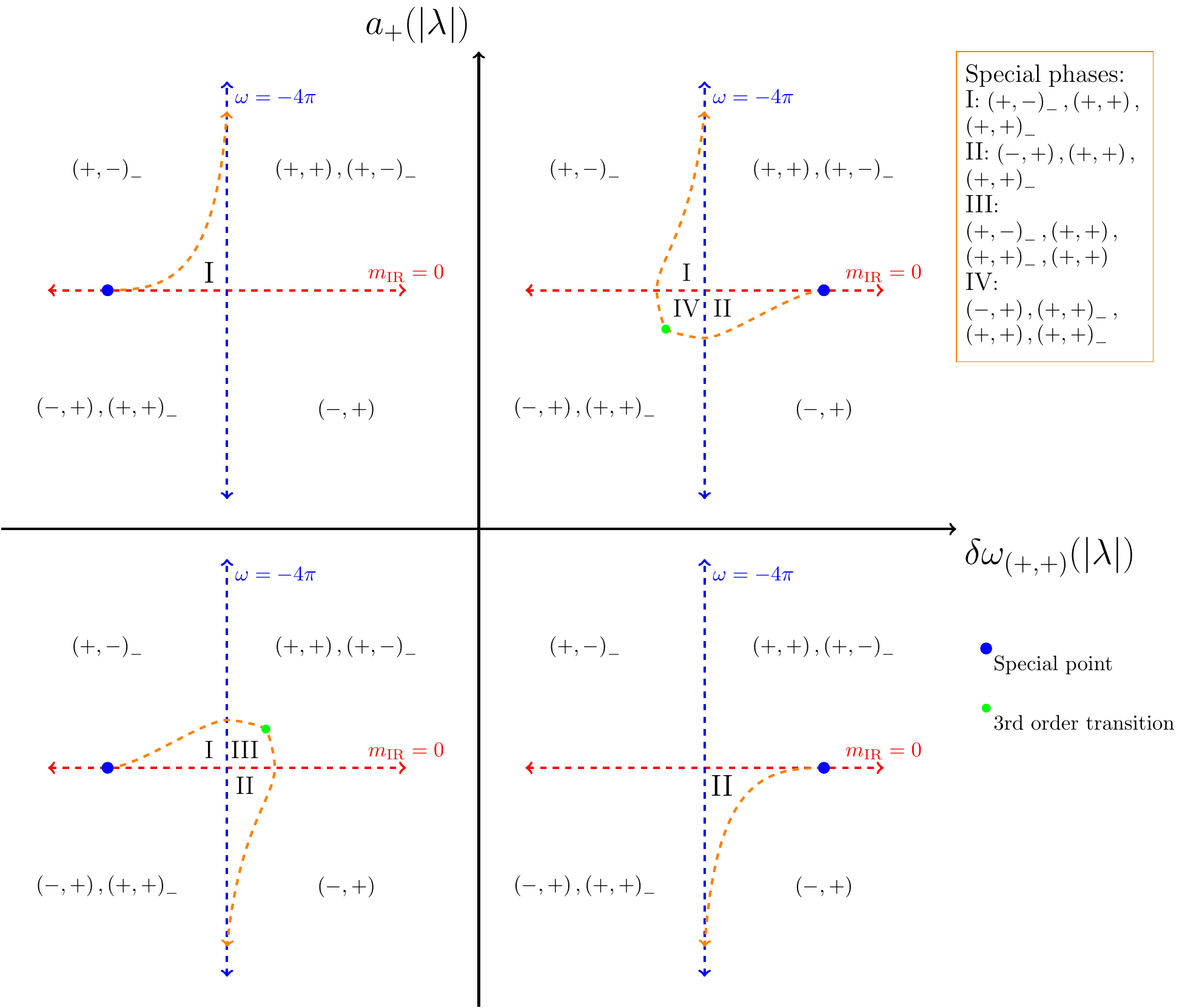}
\par\end{centering}
\caption{\label{fig: plus plus phase diagram general lambda}  Possible changes to the $\left(+,+\right)$ ``special phase'' at large values of $\lambda$.}

\end{figure}

\begin{figure}
\begin{centering}
\includegraphics[width=.8\textwidth]{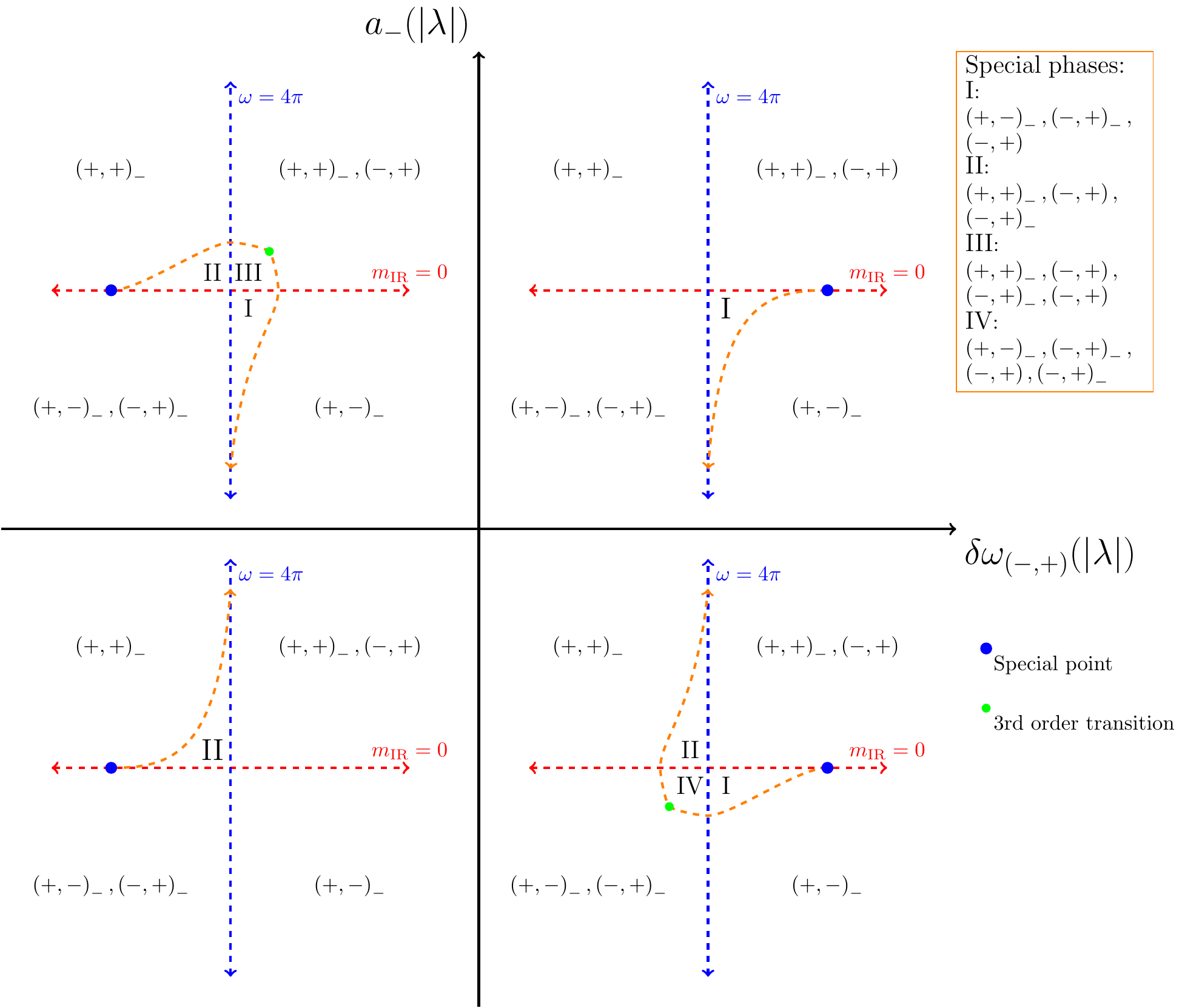}
\par\end{centering}
\caption{\label{fig: minus plus phase diagram general lambda}  Possible changes to the $\left(-,+\right)$ ``special phase'' at large values of $\lambda$.}

\end{figure}

\begin{figure}
\begin{centering}
\includegraphics[width=.7\textwidth]{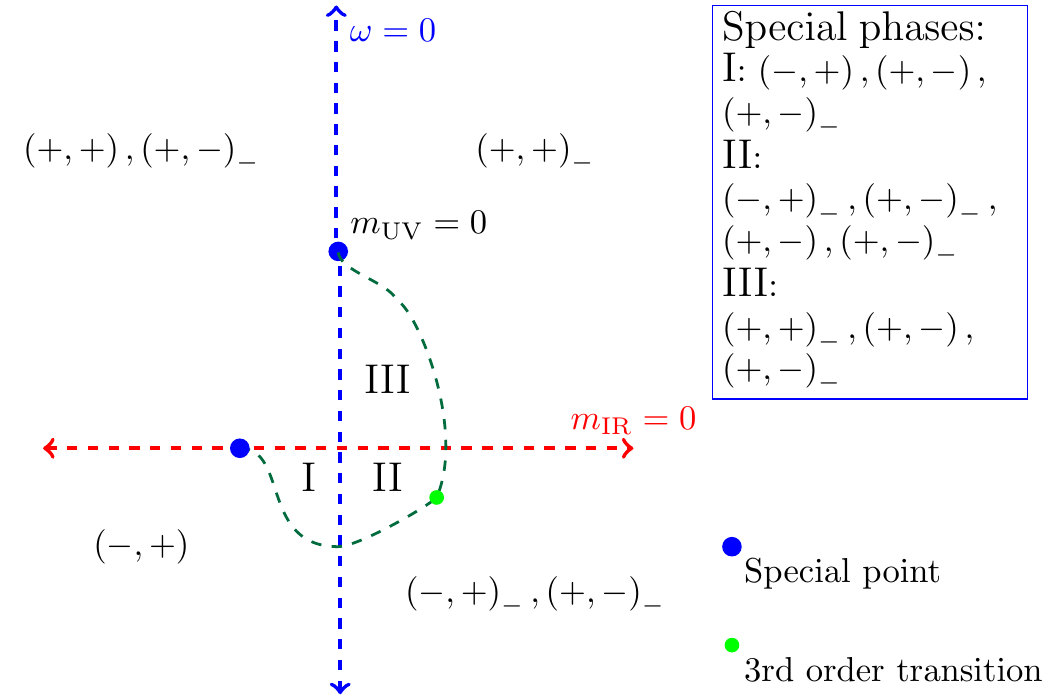}
\par\end{centering}
\caption{\label{fig: plus minus phase diagram general lambda}  The conjectured $\left(+,-\right)$ ``special phases'' in one scenario - $q<0,\delta\omega_{\left(+,-\right)}<0,a_3>0$.}
\end{figure}

When $\lambda$ is no longer small the theory at intermediate ranges $\phisq,|M|\approx |\kappa|$ becomes intractable. Large $N$ techniques are not sufficient to tame the full Yang-Mills theory - it being a theory of genuine adjoint propagating fields - but only the Chern-Simons + fundamental matter of the IR and the vector model in the UV. Nevertheless, we will make a conjecture for the form of the phase diagram. In general, as we argue below, the phase diagram should remain the same as in the small $\lambda$ case of figure \ref{fig: phase diagram infinite N leading lambda}, except near the ``special phases''. Thus in this section we focus attention to the vicinity of these phases and graph them separately, as in figures \ref{fig: plus plus phase diagram general lambda}, \ref{fig: minus plus phase diagram general lambda} and \ref{fig: plus minus phase diagram general lambda}.

Several things should remain unchanged as we move away from small $\lambda$. The rationale behind many of the features of the small $\lambda$ diagram - figure \ref{fig: phase diagram infinite N leading lambda}, discussed in subsection \ref{subsec: small lambda diagram}, remains unchanged at general $\lambda$. Hence:
\begin{enumerate}
    \item There should be a single, \textbf{horizontal}, ``red'' line of Chern-Simons-matter fixed points, as in figure \ref{fig: phase diagram infinite N leading lambda}. This is simplest to see from the unevaluated all-orders form of the Higgsed superpotential \eqref{eq: formal all orders higgsed superpotential}. The red line is defined by the equation:
    \begin{align}
        0=\left(W^{\rm H}\phisqbr\right)'\Big|_{\phisq=0},\label{eq: Higgsed red line equation}
    \end{align}
    which, using \eqref{eq: formal all orders higgsed superpotential}, can be written:
    \begin{align}
        m=\lambda |\kappa| q(|\lambda|), 
    \end{align}
    for some function $q$. This always has a single, $\omega$-independent solution for $m$. At leading order we found $q(0)=1/2$.
    The same equation should emerge also from either branch ($\left(+,+\right)$ or $\left(-,+\right)$) of the equivalent unHiggsed expression:
    \begin{align}
        0=\left(W^{\rm uH}(M)\right)'\Big|_{M=0^\pm},\label{eq: unHiggsed red line equation}
    \end{align}
    where $W^{\rm uH}$ is given by \eqref{eq: formal all orders unHiggsed superpotentail}. As discussed in subsubsection \ref{subsubsec:Feynman-Rules}, the ``counterterm'' $q$ in \eqref{eq: formal all orders unHiggsed superpotentail} is to be chosen to ensure that $M$ coincides with the pole mass in the IR regime. Equivalently it is chosen to cancel the term $\propto |M|$ in the IR. Stated a third way, it ensures that:
    \begin{align}
        \left(W^{\rm uH}(M)\right)'\Big|_{M=0^+}=\left(W^{\rm uH}(M)\right)'\Big|_{M=0^-}.
    \end{align}
    Indeed, we found at leading order $q(0)=1/2$. The ``counterterm'' $J(|\lambda|)$ in \eqref{eq: formal all orders unHiggsed superpotentail} was chosen primarily to cancel a divergence, but also to cancel the term $\propto M$ in the IR (apart from the tree level one). This choice also serves to incorporate shifts to the pole mass into $M$ (essentially by redefining $\sigma$ - a shift to $\sigma$ is equivalent to the introduction of a linear term $\delta J \sigma$ modulo redefinitions of other parameters in the Lagrangian \eqref{eq:UV_action}). A third way to justify this choice is that it reconciles the equations for the red line \eqref{eq: Higgsed red line equation} and \eqref{eq: unHiggsed red line equation}. We can stipulate therefore that both counterterms are defined to all orders in $\lambda$ by the above equivalent criteria.
    \item The positions of wall crossings are unaffected by gauge dynamics, and remain the same as in figures \ref{fig: phase diagram infinite N leading lambda} and \ref{fig: lambda equals 0 phase diagram}. 
    \item The position of the ``wall crossing within the wall crossing'' at $\omega=m=0$ won't change. This is because, as discussed in subsection \ref{subsec: Validity of the Calculation}, the contribution of the gauge sector to the Higgsed superpotential should remain sub-sub-leading in the UV.
    \item The phases in the vicinity of the red line should be consistent with the IR phase diagram - figure \ref{fig: IR phase diagram minwalla}. In particular, along the red line, there will be 3 ``special points'', corresponding to the phase transitions of figure \ref{fig: IR phase diagram minwalla}, as in figure \ref{fig: phase diagram infinite N leading lambda}. From these special points, in place of the wall crossings that exist in the strict IR limit $\kappa\to\infty$ (those of figure \ref{fig: IR phase diagram minwalla}), new lines will emerge, like the orange and green lines of figure \ref{fig: phase diagram infinite N leading lambda}, where pairs of vacua form / annihilate. These lines will \textbf{most likely}\footnote{\label{footnote: direction of emergence of orange lines}This is what would happen in the simplest possible phase diagram, and is what happened at small $\lambda$ in figure \ref{fig: phase diagram infinite N leading lambda}, but we cannot rule out that the lines meander further in the intermediate energy range $\sim|\kappa|$.} emerge in a direction dictated by consistency with the IR and UV theories.
    \item The ``green'' lines, corresponding to the \textbf{Higgsed} phase, should end ``in the UV'' at the special point $\omega=m=0$, as in figure \ref{fig: phase diagram infinite N leading lambda}.
    \item Since the sub-leading term in the UV limit of the unHiggsed superpotential has the form $\propto M\log M$, and vanishes nowhere in the phase diagram, the ``orange'' lines, corresponding to the \textbf{unHiggsed} phase, should asymptote along the wall crossings as in figure \ref{fig: phase diagram infinite N leading lambda}.
    \item Far from the red line (in the UV), and away from the wall crossings of the unHiggsed phase at $\omega=\pm4\pi$, the phases should be the same as in the UV diagram - figure \ref{fig: lambda equals 0 phase diagram}, as was the case at small $\lambda$ in figure \ref{fig: phase diagram infinite N leading lambda}.
\end{enumerate}

We conjecture then that the general form of the phase diagram at general $\lambda$ is the same as at small $\lambda$, presented in figure \ref{fig: phase diagram infinite N leading lambda}. At most, changes will occur concerning the ``special'' phases, and we will discuss these possible changes below. It cannot be ruled out that additional phases could appear in other regions of the phase diagram in the intermediate energy range $|M|,\phisq\approx |\kappa|$, but they have no reason to appear. Hypothetically, a phase could exist in which $\left(-,-\right)$ vacua form in pairs, confined entirely to the intermediate energy range. Again - we focus on the simplest possible phase diagram, and so we ignore these.

The following things can change at general $\lambda$:
\begin{enumerate}
    \item The position of the IR ``special points'' relative to the wall crossings could change. The \textbf{sign} of this relative position would have to change the behavior of the orange and green lines, and of the ``special'' phases. In particular, the \textbf{most probable} (see the footnote \ref{footnote: direction of emergence of orange lines})  \textbf{direction} at which the lines emerge from the special points is dictated by this relative position - ``up'' for a negative relative position and ``down'' for a positive one - as in figure \ref{fig: phase diagram infinite N leading lambda}. The position of the IR special points $\omega_{\left(+,+\right)}$, $\omega_{\left(-,+\right)}$ and $\omega_{\left(+,-\right)}$ is defined by:
    \begin{align}
        0&=W^{\rm uH}(M)'\Big|_{\omega=\omega_{\left(+,+\right)},M=0^+}=W^{\rm uH}(M)''\Big|_{\omega=\omega_{\left(+,+\right)},M=0^+} \\
        0&=W^{\rm uH}(M)'\Big|_{\omega=\omega_{\left(-,+\right)},M=0^-}=W^{\rm uH}(M)''\Big|_{\omega=\omega_{\left(-,+\right)},M=0^-} \\
        0&=W^{\rm H}\phisqbr'\Big|_{\omega=\omega_{\left(+,-\right)},\phisq=0}=W^{\rm H}\phisqbr''\Big|_{\omega=\omega_{\left(+,-\right)},\phisq=0}.
    \end{align}
    To quantify the shift relative to the wall crossings we define:
    \begin{align}
        \delta\omega_{\left(+,+\right)}&\equiv\omega_{\left(+,+\right)}-(-4\pi)\\
        \delta\omega_{\left(-,+\right)}&\equiv\omega_{\left(-,+\right)}-4\pi\\
        \delta\omega_{\left(+,-\right)}&\equiv\omega_{\left(+,-\right)}-0.
    \end{align}
    The position of the special points was computed by \cite{Dey:2019ihe} to all orders in $\lambda$, and is given by the formulae in figure \ref{fig: IR phase diagram minwalla}. The problem is that, as discussed in subsection \ref{subsec: small lambda diagram}, those are not the positions of the special points in terms of our UV parameter $\omega$ but in terms of its IR counterpart, which is renormalized non-trivially by the YM interaction, as discussed in subsection \ref{subsec: Dimensional vs Yang-Mills Regularization} at leading order in $\lambda$. Hence we cannot know the sign of the shifts without explicit higher-order computations.  
    \item The position of the red line $q(|\lambda|)$. The sign of $q$ is consequential for Higgsed ``special phases'', and the corresponding green line, since the latter begins on the red line and terminates at $m=0$.
    \item The sign of the coefficient of the sub-leading $M\log M$ term. In fact, in general, based on the discussion in subsection \ref{subsec: Validity of the Calculation}, and using parity and dimensional considerations, we expect the sub-leading term in the unHiggsed branch to have the form (in the UV):
    \begin{align}
        \frac{1}{8\pi}a_1(|\lambda|)|\kappa||\lambda|M \log |M|+\frac{1}{8\pi}a_2(|\lambda|)|\kappa|\lambda|M| \log |M|,
    \end{align}
    where at leading order we found $a_1(0)=1,a_2(0)=0$. In particular, in the $\left( \pm,+\right)$ branch we should care about the \textbf{sign} of the coefficient:
    \begin{align}
        a_\pm(|\lambda|) \equiv a_1 \pm a_2.
    \end{align}
    This sign will determine in \textbf{which direction} along - and on what \textbf{side} of -  the wall crossing do the orange lines, corresponding to the unHiggsed special phases, \textbf{asymptote} along the wall crossings as in figure \ref{fig: phase diagram infinite N leading lambda}. 
    \item The sign of the coefficient of the sub-sub-leading term in the Higgsed phase:
    \begin{align}
        \frac{1}{8\pi}a_3(|\lambda|){\rm sign}(\lambda)|\kappa|^{3/2}\sqrt{\phisq}.
    \end{align}
    At leading order we found $a_3(|\lambda|)\approx-|\lambda|^{1/2}$. The sign of $a_3$ will determine the direction of approach of the green line to its UV endpoint $\omega=m=0$.
\end{enumerate}

For concreteness, let's focus on the $\left(+,+\right)$ special phase. That is, let's focus on a region of parameter space $\omega\approx -4\pi$ that includes the leftmost wall crossing and IR special point. The parameters whose signs matter here are $a_+$ and $\delta\omega_{\left(+,+\right)}$. In each quadrant in the $\left(a_+, \delta\omega_{\left(+,+\right)} \right)$-plane there is a different ``minimal'' phase diagram, as depicted in figure \ref{fig: plus plus phase diagram general lambda}. The second quadrant is the one we explicitly found at small $\lambda$, and so matches figure \ref{fig: phase diagram infinite N leading lambda}.

The diagrams in the first and third quadrants in figure \ref{fig: plus plus phase diagram general lambda} have many new phases, including some with three different $\left(+,+\right)$ vacua. These can be understood intuitively as follows. Consider, for concreteness, the scenario depicted in the third quadrant of figure \ref{fig: plus plus phase diagram general lambda}, and consider starting at a point just above the red line and moving to the \textbf{left} towards the wall crossing (blue line). Since we remain to the right of the special point, a $\left(+,+\right)_+$ vacuum is expected to continue to exist at low energies. However, as we near the wall crossing, and since $a_+<0$, the UV theory ``demands a sacrifice'' in the form of a $\left(+,+\right)_+$ vacuum escaping to infinity. The contradiction can be resolved by a \textbf{new} $\left(+,+\right)_+$ vacuum appearing at intermediate energies together with a statistics-flipped $\left(+,+\right)_-$ vacuum. As one continues to move towards the wall crossing this new $\left(+,+\right)_+$ vacuum escapes to infinity while the old one remains at low energies.

Finally, note that by considering the trajectories of different vacua as one moves around the diagram in the first and third quadrant, one is led to the conclusion that all three $\left(+,+\right)$ vacua must ``collide'' at a particular point along the orange line, at some intermediate energy scale. This implies that at said point, drawn in green in figure \ref{fig: plus plus phase diagram general lambda}, the 3rd derivative of the superpotential vanishes at the vacuum (the orange line is already defined by the vanishing of the second derivative at the vacuum). This indicates a 3rd order phase transition takes place. At large but finite $N$, the CFTs along the orange line will be interacting CFTs, so it's possible that this 3rd order phase transition point will correspond to a free CFT - which is unstable in 3d.

The analysis of phases associated to $\left(-,+\right)$ unHiggsed vacua, near the $\omega=4\pi$ wall crossing, is almost identical, and is depicted in figure \ref{fig: minus plus phase diagram general lambda}. The main difference is that the small lambda regime of figure \ref{fig: phase diagram infinite N leading lambda} now corresponds to the \textbf{first} quadrant of figure \ref{fig: minus plus phase diagram general lambda}.

The general picture in the Higgsed phase is quite similar, but now depends on the three unknowns $a_3,\delta\omega_{\left(+,-\right)},q$. This gives rise to $2^3$ possible scenarios, each with their own phase diagram. For brevity, we present only the case $q<0,\delta\omega_{\left(+,-\right)}<0,a_3>0$ in figure \ref{fig: plus minus phase diagram general lambda}.

\subsection{\label{subsec: adding a sigma cubed coupling}Adding a \texorpdfstring{$\sigma^3$}{sigma cubed} Coupling}

\begin{figure}
\begin{centering}
\includegraphics[width=.65\textwidth]{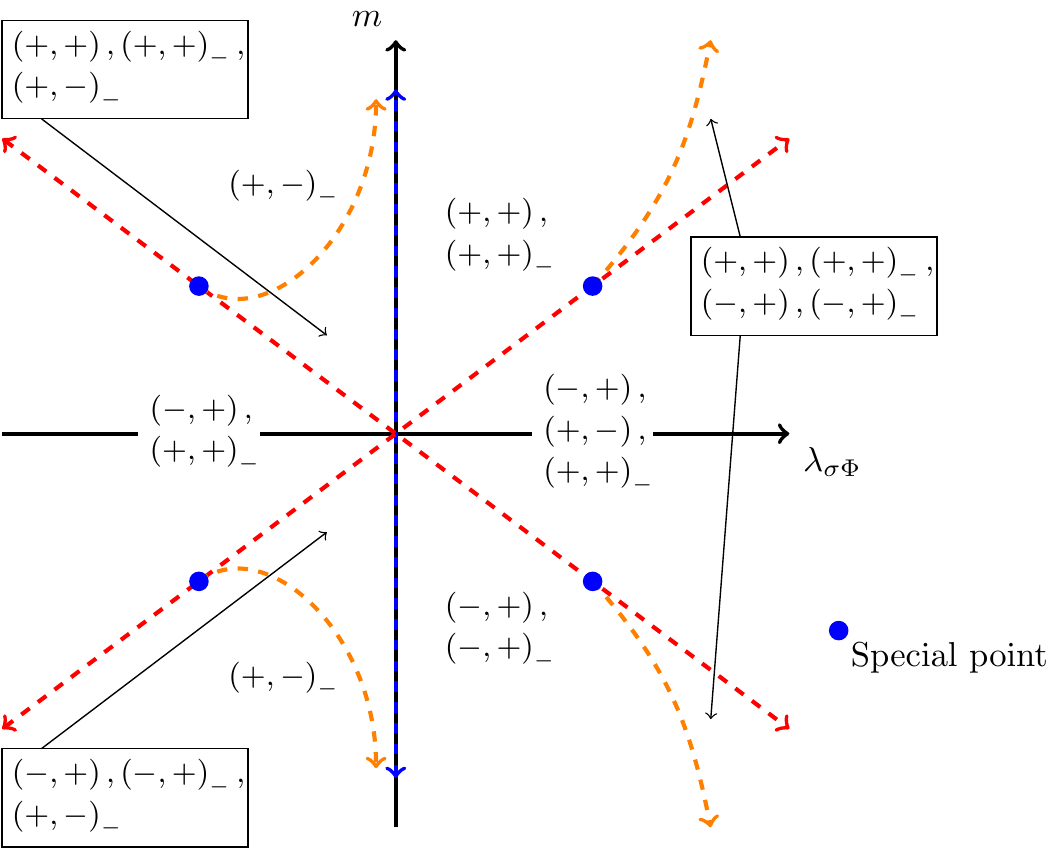}
\par\end{centering}
\caption{\label{fig: sigma cubed phase diagram}Phase diagram of the theory \ref{eq:UV_action} with decoupled gauge sector $\lambda\to0$ in the case ${\rm sign}\left(j_\sigma \lambda_\sigma \right)=-1$. More general cases with small $\lambda$ are governed by the superpotential \eqref{eq: complete superpotential small lambda sigma cubed}.}

\end{figure}

In this section we generalize our results to the most general case of a super-renormalizable theory with the field content of \eqref{eq:UV_action}, which is also the most general asymptotically free (see the footnote \ref{footnote:super-renormalizable}). This is achieved by replacing the quadratic self-interaction superpotential of $\sigma$ with a general \textbf{cubic} polynomial:
\begin{align}
    \frac{1}{6}\lambda_{\sigma}\sigma^{3}+\frac{1}{2}m_\sigma \sigma^2+j_{\sigma}\sigma.
\end{align}

By a field redefinition we can always cancel the quadratic term, while the linear term can only be cancelled for ${\rm sign}\left( j_\sigma \lambda_\sigma \right)<0$. So W.L.O.G. we can write:
\begin{align}
W_{{\rm tree}}=\frac{1}{6}\lambda_{\sigma}\sigma^{3}+j_{\sigma}\sigma+\lambda_{\sigma\Phi}\sigma\bar{\Phi}\Phi+m\bar{\Phi}\Phi,
\end{align}
where we now canonically normalize $\sigma$ so that its mass dimension is $1/2$ and it is defined by the EOM:
\begin{align}
\left\langle \sigma^{2}\right\rangle =-2\frac{j_{\sigma}+\lambda_{\sigma\Phi}\bar{\Phi}\Phi}{\lambda_{\sigma}}.
\end{align}
This is in contrast to the normalization convention used so far in this paper, given in \eqref{eq:UV_action_defs}. It is given by rescaling $\sigma\to \lambdasigph \sigma$. Furthermore we relate $\frac{1}{6}\lambdasig =\lambda_3 \lambda^3_{\sigma\Phi} $.

At leading order in $N$, as discussed in subsubsection \ref{subsubsec:N Counting}, $\sigma$ is essentially classical so the computation of the superpotential goes through exactly the same as in section \ref{sec: computing the superpotential}. By the same arguments, this superpotential can be trusted in every part of the phase space, and is given formally (at leading order in $N$) by:
\begin{align}
W^{{\rm uH}} & =\frac{1}{6}\lambda_{\sigma}\sigma^{3}+j_{\sigma}\sigma+|\kappa| \delta J(|\lambda|)M+\frac{1}{N{\rm Vol}^{3\mid2}}\log\Big[\intop D\bar{\Phi}^{i}D\Phi_{i}D\Gamma\label{eq: complete superpotential small lambda sigma cubed}\\
 & \exp\left(N\intop d^{5}z\left(\bar{\Phi}\nabla^{2}\Phi+M\phisq+q(|\lambda|)\lambda\left|\kappa\right|\phisq\right)+N\left(S_{{\rm YM}}+S_{{\rm CS}}\right)\right)\Big]\nonumber\\
W^{{\rm H}} & =\frac{1}{6}\lambda_{\sigma}\sigma^{3}+j_{\sigma}\sigma+\lambda_{\sigma\Phi}\sigma\bar{\Phi}\Phi+m\bar{\Phi}\Phi\nonumber\\&+\frac{1}{N{\rm Vol}^{3\mid2}}\log\left(\intop D\Gamma\exp\left(NS_{\rm YM+CS}- N\intop d^{5}z \phisq\left(\Gamma^{2}\right)_{1}^{1}\right)\right)\nonumber\\
M & \equiv m-q(|\lambda|)\lambda\left|\kappa\right|+\lambda_{\sigma\Phi}\sigma,\nonumber
\end{align}
and at leading order in $\lambda$ by:

\begin{align}
W^{{\rm uH}} & =\frac{1}{6}\lambda_{\sigma}\sigma^{3}+j_{\sigma}\sigma-\frac{1}{8\pi}M\left|M\right|\\&+\frac{1}{16\pi}\left(2M+\kappa\right)\lambda\kappa\log\left(1+\frac{2\left|M\right|}{\left|\kappa\right|}\right)-\frac{1}{8\pi}\lambda\left|\kappa\right|\left|M\right|\nonumber\\
W^{{\rm H}} & =\frac{1}{6}\lambda_{\sigma}\sigma^{3}+j_{\sigma}\sigma+\lambda_{\sigma\Phi}\sigma\bar{\Phi}\Phi+m\bar{\Phi}\Phi-\frac{\kappa\left|\kappa\right|}{8\pi}\sqrt{1+8\pi\lambda\frac{\bar{\Phi}\Phi}{\kappa}}\nonumber\\
M & \equiv m-\frac{1}{2}\lambda\left|\kappa\right|+\lambda_{\sigma\Phi}\sigma.\nonumber
\end{align}

The phase diagram of this theory differs significantly from that of the degenerate case $\lambda_\sigma=0$, but not in ways that affect the phases in the vicinity of the Chern-Simons-matter fixed points (the ``red line''). The main difference is that there are now \textbf{zero or two} such red lines, located at:
\begin{align}
    m=q(|\lambda|)\lambda\left|\kappa\right|\pm \sqrt{-\frac{2j_{\sigma}}{\lambda_{\sigma}}}\lambda_{\sigma\Phi},\label{eq: red line position sigma cubed}
\end{align}
when the expression is real. These are essentially the saddles of the cubic polynomial. In much of the parameter space, the Witten indices from vacua in the vicinity of these fixed points cancel one another, and for sufficiently large $|j_\sigma|$ all such vacua cancel and SUSY is \textbf{spontaneously broken}. Furthermore, the Witten index is controlled by the couplings $\lambda_\sigma, \lambda_{\sigma \Phi}$, so for a given $\lambda_\sigma$ the two wall crossings associated to the unHiggsed vacua in figure \ref{fig: phase diagram infinite N leading lambda} will cease to exist. Instead, they will become ``orange'' lines that extend into the classical regime and have no ``UV endpoint''. The wall crossing associated to Higgsed vacua will continue to exist.

For general parameter values the phase diagram can be quite complicated. We display it, in figure \ref{fig: sigma cubed phase diagram}, for the special case where the gauge sector decouples ($\lambda=0^+$) and ${\rm sign}(\lambda_\sigma)=-{\rm sign}(j_\sigma)=1$, as a function of the parameters $m,\lambda_{\sigma\Phi}$. Note that in the $\lambda_{\sigma\Phi}>0$ half-plane the Witten index vanishes. By varying $j_\sigma$ one can reach a phase where many parts of this half-plane have no SUSY vacua, as discussed above.

The red lines in figure \ref{fig: sigma cubed phase diagram}, unlike in previous diagrams such as figure \ref{fig: phase diagram infinite N leading lambda}, are not horizontal. This is because the ``UV'' mass of $\Phi$ around these vacua, as indicated by \eqref{eq: red line position sigma cubed}, is given by:
\begin{align}
    m=\pm \sqrt{-\frac{2j_{\sigma}}{\lambda_{\sigma}}}\lambda_{\sigma\Phi}.
\end{align}
If one took the vertical axis to correspond to one of these choices, then one of the red lines in figure \ref{fig: sigma cubed phase diagram} would become horizontal. The crossing point of the two red lines is also nothing to worry about. The phases associated to each line are separated in field space since they correspond to $\sigma \approx \pm \sqrt{-\frac{2j_{\sigma}}{\lambda_{\sigma}}}$.

Reintroducing $\lambda\neq0$, provided $|\kappa|$ is smaller than the other scales would only effect the vicinity of the red lines in figure \ref{fig: sigma cubed phase diagram}, by vertically moving the red lines, changing the position of the special points along the red lines and introducing a small ``green line''. For general $\kappa$ more complicated phases could appear, but for sufficiently large $m$ there won't be any difference. In general, $\lambda\neq0$ would spoil the $m\to-m$ symmetry of figure \ref{fig: sigma cubed phase diagram}.

\subsection{\label{subsec: finite N}Finite \texorpdfstring{$N$}{N}}

\begin{figure}
\begin{centering}



\definecolor{cadmiumgreen}{rgb}{0.0, 0.42, 0.24}

\begin{tikzpicture}[scale=.8, every node/.style={transform shape}]


\draw[very thick,->] (-8,0) -- (8,0) node[anchor=north west] {\LARGE $\omega$};
\draw[ very thick,->] (0,-1) -- (0,4.5+1.5) node[black, anchor=south east] {\LARGE  $a \equiv m/ \kappa $};

\draw[blue,dashed, very thick,<->] (0.02,-1.5) -- (0.02,0) -- (-0.02,0) -- (-0.02,4+1.5);
\draw[blue,dashed, very thick,<->] (-6.5,-1.5)  -- (-6.5,4+1.8-2);
\draw[] (-6.5,2.5) node[blue,anchor=west]{$\omega = -4 \pi $};
\draw[blue,dashed, very thick,<->] (6.5,-1.5)  -- (6.5,4+1.5);
\draw[] (6.5,2.5) node[blue,anchor=west]{$\omega = 4 \pi $};

\draw (3,3) node[black] {$ \left(+,+ \right)_- $};
\draw (-3,3) node[black] {$ \left(+,+ \right) \, , \, \left(+,- \right)_{-} $};
\draw (-8,3) node[black] {$  \left(+,- \right)_{-} $};
\draw (3,-1) node[black] {$ \left(-,+ \right)_- \, , \, \left(+,- \right)_-  $};
\draw (-3,-1) node[black] {$ \left(-,+ \right) $};
\draw (-8,1) node[black] {$\left(-,+\right)\,,\,\left(+,+\right)_{-}$};
\draw (8.5,3.5) node[black] {$\left(+,+\right)_-\,,\,\left(-,+\right)$};
\draw (8.5,1) node[black] {$\left(+,-\right)_-$};
\draw (3,2) node[red,anchor=north] {$ a \sim \frac{\lambda}{2} $};

\draw[very thin, ->] (-.8,.4) node[rectangle,anchor=north east,fill=cadmiumgreen!20, draw=gray,very thin, text=black] {$ \left(-,+ \right) \, , \, \left(+,- \right)_{-}\, , \, \left(+,- \right) $} -- (-.2,1.7);

\draw[very thin, ->] (-6,.8+.8) node[rectangle,anchor=north west,fill=orange!20, draw=gray,very thin, text=black] {$ \left(+,- \right)_{-} \, , \, \left(+,+ \right)_{-}\, , \, \left(+,+ \right) $} -- (-6.7,2.2);

\draw[very thin, ->] (5.5,4.5-1) node[rectangle,anchor=south east,fill=orange!20, draw=gray,very thin, text=black] {$ \left(+,- \right)_- \, , \, \left(-,+ \right)\, , \, \left(-,+ \right)_- $} -- (6.9,1.7);

\draw[thick,<->,blue,dashed] (-9,6) -- (-7.8,6) node[anchor=west] {wall crossing};
\draw[thick,<->,red,dashed] (-9,6-.4) -- (-7.8,6-.4) node[anchor=west] {CS-matter CFT};
\draw[thick,<->,orange,dashed] (-9,6-.8) -- (-7.8,6-.8) node[anchor=west] {pair of unHiggsed vacua (dis-)appear};
\draw[thick,<->,cadmiumgreen,dashed] (-9,6-1.2) -- (-7.8,6-1.2) node[anchor=west] {pair of Higgsed vacua (dis-)appear};

\draw[red,thick] (-9,4.8-.4) circle (0.08cm)node[black, anchor=west ] {unstable fixed point};
\draw[red,thick,fill=red] (-9,4.5-.5) circle (0.08cm)node[black, anchor=west ] {stable fixed point};	

\draw[red,thick,dashed, <->] (-10,2) -- (10,2);
\draw[cadmiumgreen,thick,dashed] (-2,2) node[] {} .. controls (-.5,2) and (0,1) .. (0,0)node[blue,circle,fill,inner sep=2pt] {};
\draw[orange,thick,dashed,->] (9.5,2) node[] {} .. controls (6.7,2) and (6.5,.5) .. (6.5,-1.3);
\draw[orange,thick,dashed,->] (-8,2) node[] {} .. controls (-6.8,2) and (-6.5,2.5) .. (-6.5,3.7);

\draw[blue,thick,fill=blue] (-1,2) circle (0.09cm);
\draw[blue,thick,fill=blue] (8.5,2) circle (0.09cm);
\draw[blue,thick,fill=blue] (-7,2) circle (0.09cm);

\draw[red,thick,fill=red] (2,2) node[anchor=south]{$\mathcal{N}=2$} circle (0.09cm);
\draw[red,thick,fill=red] (-4,2) circle (0.09cm);

\draw[red,thick] (-2,2) circle (0.09cm);
\draw[red,thick] (9.5,2) circle (0.09cm);
\draw[red,thick] (-8,2) circle (0.09cm);

\end{tikzpicture}

\par\end{centering}
\caption{The proposed phase diagram at large but finite $N$ for small $\lambda$. The red line is not exactly horizontal, and its slope must be calculated. }
\label{fig:finite_N} 
\end{figure}
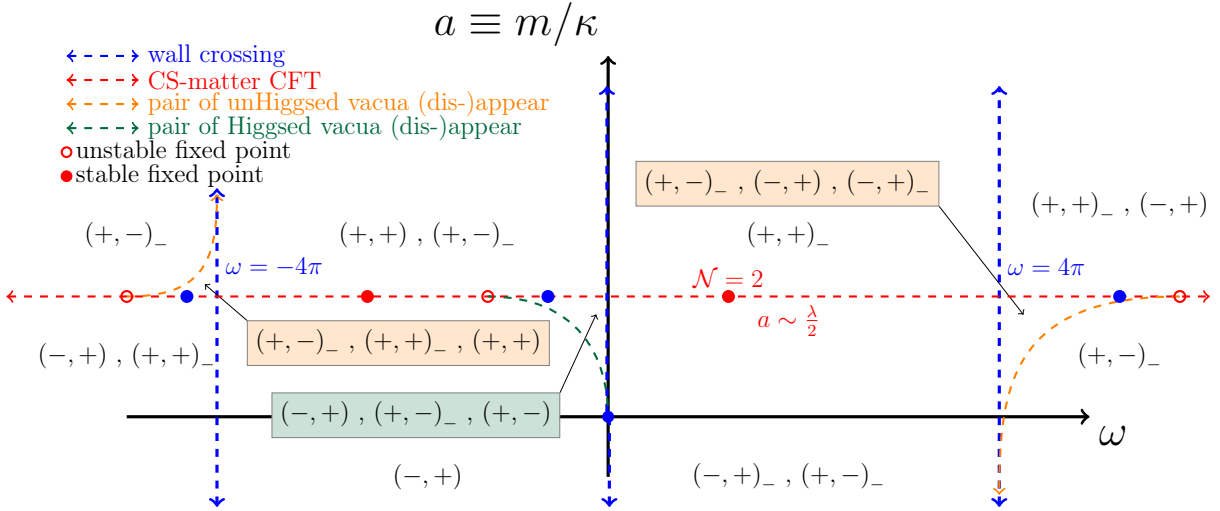

Finally, we discuss the expected phase diagram at large but finite $N$. There are several changes we must take into account. 
\begin{enumerate}
    \item First, the quartic coupling $w$ develops a beta function at order $1/N$, and so the conformal manifold is lifted and we are left with a finite number of fixed points. The authors of \cite{Aharony:2019mbc} showed that for small and large $\lambda$ there exist 6 fixed points which can be arrived at by tuning $m$. Three of these are stable and three are unstable to deformations by the quartic coupling $\omega$. Starting near an unstable fixed point, we expect the theory to flow towards a stable fixed point, and so the phase around an unstable fixed point is determined by the stable fixed point we flow to.
    \item Next, the field $\sigma$ now becomes dynamical, and we must include it in loop diagrams for the effective superpotential. In particular, this means that we have an additional parameter in the system, which is schematically the coefficient of the kinetic term $\lambda_{\sigma\Phi}$ in equation \eqref{eq:UV_action_defs} (we effectively set this parameter to be infinite up to now). This means that we cannot trust our effective potentials when the effective mass of $\sigma$ is small. 
\end{enumerate}
Assuming that $\lambda_{\sigma\Phi}$ is still very large, the main changes in the phase diagram thus amount to a shift of the phase transition lines due to the appearance of the unstable fixed points. The other various lines in the diagram will also be shifted due to subleading corrections in $1/N$. Most notably, the orange and purple lines no longer end at the ``special points'' of figure \ref{fig: phase diagram infinite N leading lambda}, but rather at unstable fixed points along the red line. For example, at two loops we expect a shift to the IR mass which is proportional to $\omega^2$, and so the ``red line'' will no longer be horizontal. We schematically plot the new form of the phase diagram in figure \ref{fig:finite_N}. The precise values of the fixed points are known only for small or large $\lambda$, but their schematic positions for all $\lambda$ were conjectured in \cite{Aharony:2019mbc}. We emphasize that at finite $N$ the IR theory along the red line is given by the result at the corresponding stable fixed point. In particular, there is no longer a conformal manifold. Note also that one of these stable fixed points is the $\mathcal{N}=2$ fixed point corresponding to $m_{\rm IR} =0, \omega_{\rm IR} = 2\pi \lambda$, the rightmost stable fixed point in figure \ref{fig:finite_N}.

Note that unlike the infinite-$N$ case, we can no longer trust the results to arbitrarily large $\omega$, since it is no longer enough to add the quartic coupling at tree level. We can thus trust this phase diagram only for small enough $\omega$.
Also note that at large $\lambda$, the position of the leftmost unstable fixed point changes, and it appears on the right side of the wall crossing (this can be seen using the duality transformation \eqref{eq:bosonization_transformation}). This will change the phase diagram considerably, depending on which extension of the phase diagram is the correct one for finite $\lambda$.

\section{\label{sec:summary}Summary}

We studied an asymptotically free UV completion of $\mathcal{N}=1$ supersymmetric $SU(N)$ Yang-Mills-Chern-Simons theory coupled to a fundamental scalar superfield, given by the Lagrangian \eqref{eq:UV_action}, at large $N$. At small `t Hooft coupling $\lambda$, we computed the effective superpotential and used it to find the phase diagram of the theory (see figure \ref{fig: phase diagram infinite N leading lambda}). We argued that this is also most likely the phase diagram of the theory at finite $\lambda$, with possibly a few modifications, described in subsection \ref{subsec:Conjectured Phase diagram for General lambda}.

This UV completion then gives us adjustable parameters, $m$ and $\omega$, that we can tune to reach the various fixed points of \cite{Aharony:2019mbc}. The phase diagram \ref{fig: phase diagram infinite N leading lambda} generalizes that of the theory at $\omega=0$ in \cite{Choi:2018ohn}, for the case $N_{\rm f}=1$, and embeds the phase diagram of the Chern-Simons theory, studied in \cite{Dey:2019ihe}, as the IR regime of a larger theory.

Contrary to the CS theory of \cite{Dey:2019ihe}, the YM-CS theory has new ``special phases'', where multiple vacua of the same type $\left(\delta N,\delta k\right)$ exist, distinguished by the VEV of $\phisq$ and the pole masses of excitations. The special phases that we found are essentially the minimal special phases needed to ``interpolate'' between the UV and IR regimes. We also further refined the question of wall-crossings in this theory where the Witten index jumps, and found generalizations of those discussed in e.g. \cite{Bashmakov:2018wts,Choi:2018ohn,Dey:2019ihe}.

One can further generalize our results by exploring the theory with additional flavors $N_{\rm f}>1$, other gauge groups, or other matter representations. Such theories may have more varied quartic interactions, and so higher-dimensional parameter spaces. There may be other interesting limiting cases, such as large $N_{\rm f}$ but finite, or even small $N$. The vacua of the UV completion \eqref{eq:UV_action} can also be studied in greater detail on the torus by computing the effective holonomy potential, using the techniques of \cite{Ghim:2019rol}.

\section*{Acknowledgments}

We would like to thank Ofer Aharony for much discussion and guidance. This work was supported in part by an Israel Science Foundation center for excellence grant (grant number 2289/18), by the Minerva foundation with funding from the Federal German Ministry for Education and Research, and by grant no. 2018068 from the United States-Israel Binational Science Foundation (BSF).

\newpage
\appendix
\addtocontents{toc}{\protect\setcounter{tocdepth}{3}}

\section{Conventions}\label{app:conventions}

\subsection{Superspace Conventions}\label{app:superspace_conventions}

In this appendix we summarize our conventions for $3d$ $\mathcal N =1$ superspace. We mainly follow \cite{Gates:1983nr,Inbasekar:2015tsa}.
A (Majorana) spinor is $\psi^\alpha$, $\alpha=1,2$, and all other representations will be denoted using spinor indices (e.g. a vector is a symmetric $V^{\alpha\beta}$). Indices are raised and lowered using  
\begin{equation}
    C_\ab=-C^\ab=  (\sigma_2)_\ab \; \Longrightarrow\; C_\ab C^{\gamma\delta}=\delta_\al{}^\delta\delta_\be{}^\delta-\delta_\be{}^\delta\delta_\al{}^\delta\quad.
\end{equation}
We use the conventions that for a spinor 
\begin{equation}
\psi^\al=C^\ab\psi_\be~~,~~\psi_\be =\psi^\al C_\ab~~,~~\psi^2=\frac12\psi^\al\psi_\al=i\psi^1\psi^2\
\end{equation}

The basic SUSY representation is a real scalar multiplet $\Phi(x,\theta)$. Expanding in the Grassmann coordinates $\theta$, we find that it includes a real Boson, a two-component spinor $\psi$ and an auxiliary field $F$:
\begin{equation}
\Phi(x,\theta)=A(x)+\theta^\alpha \psi_\alpha(x)-\theta^2 F(x)\;.
\end{equation}
The action is 
\begin{equation}
S=\frac12\int d^5z\Phi D^2 \Phi + W(\Phi) \;,
\end{equation}
where $d^5z=d^3xd^2\theta$ and the superspace derivative is given by
\begin{equation}
 D_\alpha=\partial_\alpha+i\theta^\beta\partial_{\alpha\beta}\;.
\end{equation}
We summarize some important identities for the superspace derivative:
\begin{gather}\label{eq:identities}
        D_\alpha D_\beta =i\partial_{\alpha\beta} -C_{\alpha\beta} D^2~~,~~ D^\alpha D_\beta D_\alpha =0~~,~~ D^2 D_\alpha =- D_\alpha D^2 = i \partial_{\alpha\beta} D^\beta\;,\nonumber\\
       \partial^{\alpha\beta}\partial_{\gamma\beta}=\delta_\gamma{}^\alpha \square \quad,\quad(D^2)(D^2)=\square\quad,\quad \square \equiv \frac12 \partial^{\alpha\beta}\partial_{\alpha\beta}\;.
\end{gather}

For gauge theories, we define the covariant superspace derivative
\begin{equation}
    \nabla^{\alpha}=D^{\alpha}-i\Gamma^{\alpha}
\end{equation}
with $\Gamma$ the spinor gauge field. We also define the fields strength:
\begin{equation}
    W_{\alpha}=\frac12 D^{\beta}D_\alpha \Gamma_\beta-\frac{i}{2}[\Gamma^\beta,D_\beta \Gamma_\alpha]-\frac{1}{6}[\Gamma^\beta,{\Gamma_\beta,\Gamma_\alpha}]\;.
\end{equation}
Coupling matter to gauge fields simply requires replacing superspace derivatives with their covariant versions. We will also include  Yang-Mills and Chern-Simons terms for the gauge fields, which take the form
\begin{equation}
\begin{split}
\mathcal{L}_{{\rm YM}}&=\frac{1}{g^{2}}{\rm tr}W^{2}\;,\\
\mathcal{L}_{{\rm CS}}&=\frac{k}{4\pi}{\rm tr}\frac{1}{2}\Gamma^{\alpha}\left(W_{\alpha}-\frac{1}{6}\left\{ \Gamma^{\beta},\Gamma_{\alpha\beta}\right\} \right)\;.
\end{split}
\end{equation}

\subsection{Chern-Simons Conventions}\label{app:CS_conventions}

In discussing the Chern-Simons level, we must be careful about the regularization procedure. There are two common regularizations in the literature, and both will come into play in this paper: dimensional regularization and YM regularization. We will denote the corresponding levels by $\kappa$ and $k$ respectively, and we will focus on $SU(N)$ gauge theories. 

For non-supersymmetric CS theories, the relation between these two is that a Chern-Simons theory at level $\kappah$ in dimensional regularization is
the same as an identical theory with Yang-Mills regularization at level $k$, with
\begin{equation}
    \kappah = \text{sign}(k)(|k|+N)\;.
\end{equation}
The additional shift can be thought of as a result of integrating out the gluons. The 't Hooft coupling is $\lambda=\frac{N}{\kappah}$. 

For $\mathcal{N}=1$ supersymmetric theories, there is an additional subtelty due to the gaugino, which can shift the CS level when inegrated out. Since it is in the adjoint of the gauge group, it contributes a shift proportional to $\frac{N}{2}$. Thus an $\mathcal{N}=1$ CS theory with level $k^{\mathcal{N}=1}$ is equivalent to a non-supersymmetric CS theory with level $k$, where\footnote{We are assuming here that $k$ is large enough such that SUSY is not spontaneously broken.}
\begin{equation}
k=\text{sign}(k^{\mathcal{N}=1})(|k^{\mathcal{N}=1}|-\frac{N}{2}),\qquad\qquad \kappah = k^{\mathcal{N}=1}+\text{sign}(k^{\mathcal{N}=1})\frac{N}{2}\;.
\end{equation}

\section{Some Useful Integrals}

Here we compute some Feynman integrals in Euclidean signature and dimensional regularization. In general, in 3 dimensions 1-loop Feynman integrals have no logarithmic divergences, and so in particular are finite in dim-reg.

\subsection{1-loop Tadpole}

We will evaluate:
\begin{align}
\Lambda^{-\epsilon}\intop\frac{d^{3+\epsilon}k}{(2\pi)^{3+\epsilon}}\frac{1}{k^2+m^2} & =-\frac{|m|}{4\pi}.\label{eq:tadpole_1-loop}
\end{align}
Which is finite in dim reg.

The following integral is convergent in 3d:
\begin{equation}
\intop\frac{d^3k}{(2\pi)^3}\frac{1}{k^2\left(k^2+m^2\right)}=\frac{1}{4\pi|m|}.\label{eq:convergent_tadpole_1-loop}
\end{equation}
It can be evaluated simply in radial coordinates:
\begin{equation}
\intop\frac{d^3k}{(2\pi)^3}\frac{1}{k^2\left(k^2+m^2\right)}=\frac{4\pi}{(2\pi)^3}\intop_0^{\infty}dk\frac{1}{k^2+m^2}=\frac{4\pi}{(2\pi)^3}\frac{1}{|m|}\frac{\pi}{2}.
\end{equation}
Furthermore, in dim-reg, scale-less integrals vanish:
\begin{equation}
\Lambda^{-\epsilon}\intop\frac{d^{3+\epsilon}k}{(2\pi)^{3+\epsilon}}\frac{1}{k^2}=0,
\end{equation}
so we may write using \eqref{eq:convergent_tadpole_1-loop}:
\begin{align}
\Lambda^{-\epsilon}\intop\frac{d^{3+\epsilon}k}{(2\pi)^{3+\epsilon}}\frac{1}{k^2+m^2} & =\Lambda^{-\epsilon}\intop\frac{d^{3+\epsilon}k}{(2\pi)^{3+\epsilon}}\left(\frac{1}{k^2+m^2}-\frac{1}{k^2}\right)\\
 & =-m^2\Lambda^{-\epsilon}\intop\frac{d^{3+\epsilon}k}{(2\pi)^{3+\epsilon}}\frac{1}{k^2\left(k^2+m^2\right)}\\
 & =-m^2\frac{1}{4\pi|m|},
\end{align}
which gives \eqref{eq:tadpole_1-loop}.

\subsection{1-loop Determinant of Scalar Superfield}

Given a free, real scalar superfield $\phi$, standard Gaussian integration
gives:
\begin{equation}
\log\left(\intop D\phi\exp\left(\intop d^5z\frac{1}{2}\phi\left(D^2+m\right)\phi\right)\right)=-\frac{1}{2}{\rm Tr}\log\left(D^2+m\right)
\end{equation}
where the trace can be taken over momentum $k$ and Grassmann coordiantes
$\theta^\alpha$:
\begin{equation}
{\rm Tr}\left(f\left(D^2\right)\right)\equiv\intop\frac{d^3k}{(2\pi)^3}d^2\theta_1\delta^2\left(\theta_0-\theta_1\right)f\left(D^2\left(\theta_1,k\right)\right)\delta^2\left(\theta_1-\theta_0\right).\label{eq:supertrace_in_momentum_space}
\end{equation}
We will prove that:
\begin{equation}
{\rm Tr}\log\left(D^2+m\right)=\frac{1}{8\pi}m|m|.\label{eq:trace-log-scalar-formula}
\end{equation}
We work in the \textbf{dimensional reduction scheme}, where the integral is first reduced in 3d to a ``standard'' Feynman integral which contains no factors of $D^\alpha$, and then is evaluated in $3+\epsilon$ dimensions. Furthermore, the set of identities:
\begin{align}
0 & =\delta^2\left(\theta_0-\theta_1\right)\delta^2\left(\theta_1-\theta_0\right),\\
0 & =\delta^2\left(\theta_0-\theta_1\right)D^\alpha\delta^2\left(\theta_1-\theta_0\right),\\
\delta^2\left(\theta_1-\theta_0\right) & =\delta^2\left(\theta_0-\theta_1\right)D^2\delta^2\left(\theta_1-\theta_0\right),
\end{align}
when applied to \eqref{eq:supertrace_in_momentum_space}, gives:
\begin{equation}
{\rm Tr}\left(f\left(D^2\right)\right)=\Lambda^{-\epsilon}\intop\frac{d^{3+\epsilon}k}{(2\pi)^{3+\epsilon}}f\left(D^2\right)\big|_{D^2},\label{eq:supertrace_in_momentum_space_1}
\end{equation}
where $f\left(D^2\right)\big|_{D^2}$ means the part in $f\left(D^2\right)$
proportional to $D^2$ (which is itself only a function of the momentum
$k$). Finally, using the identity 
$\left(D^2\right)^2=-k^2 $ from equation \eqref{eq:identities},
we can write:
\begin{equation}
f\left(D^2\right)\big|_{D^2}=\frac{1}{\left|k\right|}\Im\left(f\left(i\left|k\right|\right)\right).\label{eq:D_squared_part_formula}
\end{equation}
We now apply this to prove \eqref{eq:trace-log-scalar-formula}. First,
using \eqref{eq:supertrace_in_momentum_space_1} and \eqref{eq:D_squared_part_formula}:
\begin{align}
{\rm Tr}\log\left(D^2+m\right) & =\Lambda^{-\epsilon}\intop\frac{d^{3+\epsilon}k}{(2\pi)^{3+\epsilon}}\frac{1}{\left|k\right|}\Im\left(\log\left(i\left|k\right|+m\right)\right)\\
 & =\Lambda^{-\epsilon}\intop\frac{d^{3+\epsilon}k}{(2\pi)^{3+\epsilon}}\frac{1}{\left|k\right|}\arctan\left(\frac{\left|k\right|}{m}\right).
\end{align}
The last integral can be reduced to \eqref{eq:tadpole_1-loop} by
integrating by parts along the radial direction:
\begin{align}
 & \Lambda^{-\epsilon}\intop\frac{d^{3+\epsilon}k}{(2\pi)^{3+\epsilon}}\frac{1}{\left|k\right|}\arctan\left(\frac{\left|k\right|}{m}\right)\\
= & \Lambda^{-\epsilon}\intop\frac{d\Omega^{2+\epsilon}}{(2\pi)^{3+\epsilon}}\intop_0^{\infty}dk\left|k\right|^{1+\epsilon}\arctan\left(\frac{\left|k\right|}{m}\right)\\
= & \frac{1}{2+\epsilon}\Lambda^{-\epsilon}\intop\frac{d\Omega^{2+\epsilon}}{(2\pi)^{3+\epsilon}}\underbrace{\cancel{\left(\left|k\right|^{2+\epsilon}\arctan\left(\frac{\left|k\right|}{m}\right)\right)\Big|_0^{\infty}}}_{=0-0}\\
 & -\frac{m}{2+\epsilon}\Lambda^{-\epsilon}\intop\frac{d\Omega^{2+\epsilon}}{(2\pi)^{3+\epsilon}}\intop_0^{\infty}dk\left|k\right|^{2+\epsilon}\frac{1}{k^2+m^2}\\
= & -\frac{m}{2+\epsilon}\Lambda^{-\epsilon}\intop\frac{d^{3+\epsilon}k}{(2\pi)^{3+\epsilon}}\frac{1}{k^2+m^2}\\
= & -\frac{m}{2}\left(-\frac{|m|}{4\pi}\right).
\end{align}
So we find:
\begin{equation}
{\rm Tr}\log\left(D^2+m\right)=\Lambda^{-\epsilon}\intop\frac{d^{3+\epsilon}k}{(2\pi)^{3+\epsilon}}\frac{1}{\left|k\right|}\arctan\left(\frac{\left|k\right|}{m}\right)=\frac{1}{8\pi}m|m|.
\end{equation}

\subsection{1-loop Propagator Correction}

We will compute the following convergent integral:
\begin{equation}
\intop\frac{d^3k}{(2\pi)^3}\frac{1}{k^2+m_1^2}\frac{1}{\left(k-p\right)^2+m_2^2}=\frac{1}{4\pi}\frac{1}{|p|}\arctan\left(\frac{|p|}{\left|m_1\right|+\left|m_2\right|}\right),\label{eq:1-loop_propagator_correction_integral}
\end{equation}
which also has the immediate consequence:
\begin{align}
& \intop\frac{d^{3}k}{\left(2\pi\right)^{3}}\frac{1}{k^{2}\left(k^{2}+\kappa^{2}\right)}\frac{1}{\left(k-p\right)^{2}+M^{2}} \\ = & \frac{1}{4\pi\kappa^{2}}\frac{1}{\left|p\right|}\left(\arctan\left(\frac{\left|p\right|}{\left|M\right|}\right)-\arctan\left(\frac{\left|p\right|}{\left|\kappa\right|+\left|M\right|}\right)\right).\label{eq:1-loop_propagator_correction_integral_variant_2}
\end{align}
Note that for $p\to0$ \eqref{eq:1-loop_propagator_correction_integral} reduces to:
\begin{equation}
\intop\frac{d^3k}{(2\pi)^3}\frac{1}{k^2+m_1^2}\frac{1}{k^2+m_2^2}=\frac{1}{4\pi}\frac{1}{\left|m_1\right|+\left|m_2\right|},
\end{equation}
which for $m_1=0,m_2=m$ reduces to \eqref{eq:convergent_tadpole_1-loop}.

The integral can be computed using Feynman parameters:
\begin{eqnarray}
 &  & \intop\frac{d^3k}{(2\pi)^3}\frac{1}{k^2+m_1^2}\frac{1}{\left(k-p\right)^2+m_2^2}\\
 & = & \intop_0^1dx\intop\frac{d^3k}{(2\pi)^3}\frac{1}{\left((1-x)k^2+(1-x)m_1^2+x\left(k-p\right)^2+xm_2^2\right)^2}\\
 & = & \intop_0^1dx\intop\frac{d^3k}{(2\pi)^3}\frac{1}{\left(k^2+p^2x(1-x)+xm_2^2+(1-x)m_1^2\right)^2}\\
 & = & \frac{1}{2\pi^2}\intop_0^1dx\intop dk\frac{k^2}{\left(k^2+p^2x(1-x)+xm_2^2+(1-x)m_1^2\right)^2}\\
 & = & \frac{1}{2\pi^2}\intop_0^1dx\frac{\pi}{4}\frac{1}{\sqrt{p^2x(1-x)+xm_2^2+(1-x)m_1^2}}\\
 & = & \frac{1}{8\pi}\intop_0^1dx\frac{1}{\sqrt{p^2x(1-x)+xm_2^2+(1-x)m_1^2}}\\
 & = & \frac{1}{8\pi}\frac{i\left(\log\left(2\sqrt{m_2^2}-\frac{i\left(m_1^2-m_2^2+p^2\right)}{|p|}\right)-\log\left(2\sqrt{m_1^2}+\frac{i\left(-m_1^2+m_2^2+p^2\right)}{|p|}\right)\right)}{|p|}\\
 & = & \frac{1}{4\pi}\frac{1}{|p|}\arctan\left(\frac{|p|\left(\sqrt{m_1^2}-\sqrt{m_2^2}+i|p|\right)}{i|p|\left(\sqrt{m_1^2}+\sqrt{m_2^2}\right)+m_1^2-m_2^2}\right).
\end{eqnarray}
This can be reduced algebraically:
\begin{eqnarray}
\dots & = & \frac{1}{4\pi}\frac{1}{|p|}\arctan\left(\frac{\begin{array}{c}
|p|\left(\sqrt{m_1^2}-\sqrt{m_2^2}+i|p|\right)\\
\times\left(-i|p|\left(\sqrt{m_1^2}+\sqrt{m_2^2}\right)+m_1^2-m_2^2\right)
\end{array}}{|p|^2\left(\sqrt{m_1^2}+\sqrt{m_2^2}\right)^2+\left(m_1^2-m_2^2\right)^2}\right)\\
 & = & \frac{1}{4\pi}\frac{1}{|p|}\arctan\left(\frac{\begin{array}{c}
|p|\left(\sqrt{m_1^2}-\sqrt{m_2^2}\right)\left(m_1^2-m_2^2\right)\\
+|p|\left(i|p|\right)\left(-i|p|\left(\sqrt{m_1^2}+\sqrt{m_2^2}\right)\right)\\
+|p|\left(i|p|\right)\left(m_1^2-m_2^2\right)\\
+|p|\left(\sqrt{m_1^2}-\sqrt{m_2^2}\right)\left(-i|p|\left(\sqrt{m_1^2}+\sqrt{m_2^2}\right)\right)
\end{array}}{|p|^2\left(\sqrt{m_1^2}+\sqrt{m_2^2}\right)^2+\left(m_1^2-m_2^2\right)^2}\right)\\
 & = & \frac{1}{4\pi}\frac{1}{|p|}\arctan\left(\frac{|p|\left(\sqrt{m_1^2}-\sqrt{m_2^2}\right)\left(m_1^2-m_2^2\right)+|p|^3\left(\sqrt{m_1^2}+\sqrt{m_2^2}\right)}{|p|^2\left(\sqrt{m_1^2}+\sqrt{m_2^2}\right)^2+\left(m_1^2-m_2^2\right)^2}\right)\\
 & = & \frac{1}{4\pi}\frac{1}{|p|}\arctan\left(\frac{|p|}{\sqrt{m_1^2}+\sqrt{m_2^2}}\frac{\left(\sqrt{m_1^2}-\sqrt{m_2^2}\right)^2+|p|^2}{|p|^2+\left(\sqrt{m_1^2}-\sqrt{m_2^2}\right)^2}\right)\\
 & = & \frac{1}{4\pi}\frac{1}{|p|}\arctan\left(\frac{|p|}{\sqrt{m_1^2}+\sqrt{m_2^2}}\right),
\end{eqnarray}
giving \eqref{eq:1-loop_propagator_correction_integral}.

\subsection{2-loop Logarithmically Divergent Vacuum Bubble}

We will compute:
\begin{align}
 & \Lambda^{-\epsilon}\intop\frac{d^{3+\epsilon}p}{(2\pi)^{3+\epsilon}}\frac{d^3k}{(2\pi)^3}\frac{1}{\left(k^2+m_1^2\right)\left((p-k)^2+m_2^2\right)\left(p^2+m_3^2\right)}\label{eq:2-loop_log_divergent_vacuum_bubble_general_case}\\
= & -\frac{1}{16\pi^2}\left(\frac{1}{\epsilon}+\log\left(\frac{\left|m_1\right|+\left|m_2\right|+\left|m_3\right|}{\Lambda}\right)-\frac{\log(4\pi)+\psi\left(\frac{3}{2}\right)}{2}\right),\nonumber 
\end{align}
where $\psi$ is the digamma function and we evaluate the convergent
$k$ integral in 3 dimensions and then the resulting $p$ integral
is evaluated in dim-reg. A special case is that with $m_1=\kappa,m_2=m_3=M$:
\begin{align}
 & \Lambda^{-\epsilon}\intop\frac{d^{3+\epsilon}p}{(2\pi)^{3+\epsilon}}\frac{d^3k}{(2\pi)^3}\frac{1}{\left(k^2+\kappa^2\right)\left((p-k)^2+M^2\right)\left(p^2+M^2\right)}\label{eq:2-loop_log_divergent_vacuum_bubble}\\
= & -\frac{1}{16\pi^2}\left(\frac{1}{\epsilon}+\log\left(\frac{\left|\kappa\right|}{\Lambda}\right)+\log\left(1+\frac{2|m|}{\left|\kappa\right|}\right)-\frac{\log(4\pi)+\psi\left(\frac{3}{2}\right)}{2}\right).\nonumber 
\end{align}
The $k$ integral can be evaluated using \eqref{eq:1-loop_propagator_correction_integral}:
\begin{align}
 & \Lambda^{-\epsilon}\intop\frac{d^{3+\epsilon}p}{(2\pi)^{3+\epsilon}}\frac{d^3k}{(2\pi)^3}\frac{1}{\left(k^2+m_1^2\right)\left((p-k)^2+m_2^2\right)\left(p^2+m_3^2\right)}\\
= & \frac{1}{4\pi}\Lambda^{-\epsilon}\intop\frac{d^{3+\epsilon}p}{(2\pi)^{3+\epsilon}}\frac{1}{p^2+m_3^2}\frac{1}{|p|}\arctan\left(\frac{|p|}{\left|m_1\right|+\left|m_2\right|}\right).
\end{align}
This can then be split into a divergent and a convergent piece:
\begin{align}
\dots= & \frac{1}{8}\Lambda^{-\epsilon}\intop\frac{d^{3+\epsilon}p}{(2\pi)^{3+\epsilon}}\frac{1}{p^2+m_3^2}\frac{1}{|p|}\\
+ & \frac{1}{4\pi}\intop\frac{d^3p}{(2\pi)^3}\frac{1}{p^2+m_3^2}\frac{1}{|p|}\left(\arctan\left(\frac{|p|}{\left|m_1\right|+\left|m_2\right|}\right)-\frac{\pi}{2}\right),
\end{align}
where the convergent piece has been analytically continued to 3 dimensions.
Now we can evaluate the convergent integral using integration by parts:
\begin{align}
\dots= & \frac{1}{(2\pi)^3}\intop_0^{\infty}d|p|\frac{|p|}{p^2+m_3^2}\left(\arctan\left(\frac{|p|}{\left|m_1\right|+\left|m_2\right|}\right)-\frac{\pi}{2}\right)\\
= & \frac{1}{2}\frac{1}{(2\pi)^3}\underbrace{\cancel{\left(\log\left(1+\frac{p^2}{m_3^2}\right)\left(\arctan\left(\frac{|p|}{\left|m_1\right|+\left|m_2\right|}\right)-\frac{\pi}{2}\right)\right)\Big|_0^{\infty}}}_{=0-0}\\
 & -\frac{1}{2}\frac{1}{(2\pi)^3}\left(\left|m_1\right|+\left|m_2\right|\right)\intop_0^{\infty}d|p|\log\left(1+\frac{p^2}{m_3^2}\right)\frac{1}{p^2+\left(\left|m_1\right|+\left|m_2\right|\right)^2}\\
= & 0-\frac{1}{4\pi}\frac{1}{(2\pi)^2}\left(\left|m_1\right|+\left|m_2\right|\right)\frac{\pi}{2}\frac{\log\left(\left(\frac{\left|m_1\right|+\left|m_2\right|+\left|m_3\right|}{\left|m_3\right|}\right)^2\right)}{\left|m_1\right|+\left|m_2\right|}\\
= & -\frac{1}{16\pi^2}\log\left(\frac{\left|m_1\right|+\left|m_2\right|+\left|m_3\right|}{\left|m_3\right|}\right).
\end{align}
While the divergent integral can be evaluated directly:
\begin{align}
\frac{1}{8}\Lambda^{-\epsilon}\intop\frac{d^{3+\epsilon}p}{(2\pi)^{3+\epsilon}}\frac{1}{p^2+m_3^2}\frac{1}{|p|} & =-\frac{1}{8}\frac{2^{-\epsilon-3}\pi^{-\frac{\epsilon}{2}-\frac{1}{2}}\Lambda^{-\epsilon}\csc\left(\frac{\pi\epsilon}{2}\right)\left|m_3\right|{}^{\epsilon}}{\Gamma\left(\frac{\epsilon+3}{2}\right)}\\
 & =-\frac{1}{16\pi^2\epsilon}-\frac{\log\left(\frac{\left|m_3\right|}{\Lambda}\right)}{16\pi^2}+\frac{\log(4\pi)+\psi\left(\frac{3}{2}\right)}{32\pi^2}+O\left(\epsilon\right).
\end{align}
Putting it all together, we arrive at \eqref{eq:2-loop_log_divergent_vacuum_bubble_general_case}.

\section{Supergraphs}\label{app:supergraphs}

We use $D$-algebra and the identitites:
\begin{align}
0 & =\delta^{2}\left(\theta_{0}-\theta_{1}\right)\delta^{2}\left(\theta_{1}-\theta_{0}\right)\\
0 & =\delta^{2}\left(\theta_{0}-\theta_{1}\right)D^{\alpha}\delta^{2}\left(\theta_{1}-\theta_{0}\right)\\
\delta^{2}\left(\theta_{1}-\theta_{0}\right) & =\delta^{2}\left(\theta_{0}-\theta_{1}\right)D^{2}\delta^{2}\left(\theta_{1}-\theta_{0}\right),
\end{align}
To reduce supergraphs to standard Feynman integrals. For a review
of 3d superspace and supergraphs see 

We start with propagator corrections to $\Phi$, which take the general
form $a\left(p^{2}\right)+b\left(p^{2}\right)D^{2}$ and are responsible
for renormalizing the mass and amplitude of $\Phi$. We then use these
to compute the vacuum diagrams used in subsubsection \ref{subsubsec:2-Loop-Analysis}.
We use the Feynman rules described in subsubsection \ref{subsubsec:Feynman-Rules},
with the propagators:
\begin{equation}
\left\langle \bar{\Phi}^{i}\Phi_{j}\right\rangle _{{\rm free}}=\incgraph[freescalarprop]=-\frac{1}{N}\frac{1}{D^{2}+M}\delta_{j}^{i}=\frac{1}{N}\frac{D^{2}-M}{p^{2}+M^{2}}\delta_{j}^{i},
\end{equation}
and (Landau gauge $\xi=0$):
\begin{equation}
\left\langle \Gamma^{\alpha a}\Gamma^{\beta b}\right\rangle =\frac{4\pi\lambda}{N}\kappa\frac{D^{\alpha}D^{\beta}\left(\kappa-D^{2}\right)}{p^{2}\left(\kappa^{2}+p^{2}\right)}\delta^{ab}.
\end{equation}
This choice satisfies ``transversality of the propagator'':
\begin{equation}
\Delta^{\alpha\beta}D_{\alpha}=D_{\beta}\Delta^{\alpha\beta}=0,
\end{equation}
which in particular means that $D^{\alpha}\Gamma_{\alpha}$ terms
decouple entirely. The vertices include a quartic vertex:
\begin{equation}
-\bar{\Phi}\Gamma^{2}\Phi,
\end{equation}
and a cubic vertex which (using propagator transversality) can be
written in a few different ways:
\begin{equation}
\frac{i}{2}D^{\alpha}\bar{\Phi}\Gamma_{\alpha}\Phi-\frac{i}{2}\bar{\Phi}\Gamma^{\alpha}D_{\alpha}\Phi\sim-i\bar{\Phi}\Gamma^{\alpha}D_{\alpha}\Phi\sim-iD_{\alpha}\bar{\Phi}\Gamma^{\alpha}\Phi.
\end{equation}

\subsection{\label{subsec: phi propagator corrections appendix}\texorpdfstring{$\Phi$}{Phi} Propagator Corrections}

We begin with the ``tadpole correction''.
\begin{align}
 & \text{\ensuremath{\begin{gathered}\incgraph[propcorrection1]\end{gathered}
}}\nonumber \\
= & \left(-1\right)\frac{4\pi\lambda}{N}\kappa\underbrace{\left(T^{a}T^{a}\right)_{i}^{j}}_{\approx\frac{N}{2}\delta_{j}^{i}}\intop\frac{d^{3}p}{\left(2\pi\right)^{3}}d^{2}\theta'\delta^{2}\left(\theta-\theta'\right)\frac{1}{2}\frac{C_{\beta\alpha}D^{\alpha}D^{\beta}\left(\kappa-D^{2}\right)}{p^{2}\left(\kappa^{2}+p^{2}\right)}\delta^{2}\left(\theta-\theta'\right)\nonumber \\
= & -2\pi\lambda\kappa^{2}\delta_{j}^{i}\intop\frac{d^{3}p}{\left(2\pi\right)^{3}}\frac{1}{p^{2}\left(\kappa^{2}+p^{2}\right)}\label{eq:phi propagator correction - gauge tadpole}\\
= & -2\pi\lambda\kappa^{2}\delta_{j}^{i}\left(\frac{1}{4\pi\left|\kappa\right|}\right)\nonumber \\
= & -\frac{1}{2}\lambda\left|\kappa\right|\delta_{j}^{i},\nonumber 
\end{align}
Next, we have:
\begin{equation}
\incgraph[propcorrection3].\label{eq:rainbow propagator correction diagram}
\end{equation}
We can use the transversality of the propagator to have both superderivatives
from the cubic vertices act on the scalar propagator, as indicated
in \eqref{eq:cubic vertex choices}. Then the scalar propagator with
superderivatives becomes:
\begin{align}
 & D_{\alpha}\left(\theta_{1},-\left(p-k\right)\right)D_{\beta}\left(\theta_{2},p-k\right)\frac{D^{2}-M}{\left(p-k\right)^{2}+M^{2}}\delta^{2}\left(\theta_{2}-\theta_{1}\right)\\
= & D_{\beta}\left(\theta_{2},p-k\right)\frac{D^{2}-M}{\left(p-k\right)^{2}+M^{2}}D_{\alpha}\left(\theta_{2},p-k\right)\delta^{2}\left(\theta_{2}-\theta_{1}\right)\\
= & -D_{\beta}D_{\alpha}\frac{D^{2}+M}{\left(p-k\right)^{2}+M^{2}}\delta^{2}\left(\theta_{2}-\theta_{1}\right),
\end{align}
where we brought it to a form where all superderivatives act to the
right. Then we can write, using $D^{\alpha}\left(p\right)D^{\beta}\left(p\right)=-C^{\alpha\beta}D^{2}+p^{\alpha\beta}$:
\begin{align}
 & \incgraph[propcorrection3]\\
= & \left(-i\right)^{2}N^{2}\frac{4\pi\lambda}{N}\kappa\underbrace{\left(T^{a}T^{a}\right)_{i}^{j}}_{\approx\frac{N}{2}\delta_{j}^{i}}\frac{1}{N}\intop\frac{d^{3}k}{\left(2\pi\right)^{3}}d^{2}\theta_{2}\\
\times & \frac{D^{\alpha}D^{\beta}\left(\kappa-D^{2}\right)}{k^{2}\left(\kappa^{2}+k^{2}\right)}\delta^{2}\left(\theta_{2}-\theta_{1}\right)\times-D_{\beta}D_{\alpha}\frac{D^{2}+M}{\left(p-k\right)^{2}+M^{2}}\delta^{2}\left(\theta_{2}-\theta_{1}\right)\\
= & N2\pi\lambda\kappa\delta_{j}^{i}\intop\frac{d^{3}k}{\left(2\pi\right)^{3}}d^{2}\theta_{2}\\
\times & \frac{-C^{\alpha\beta}\left(\kappa D^{2}+k^{2}\right)+k^{\alpha\beta}\left(\kappa-D^{2}\right)}{k^{2}\left(\kappa^{2}+k^{2}\right)}\delta^{2}\left(\theta_{2}-\theta_{1}\right)\\
\times & \frac{C^{\beta\alpha}\left(\left(p-k\right)^{2}-MD^{2}\right)+\left(p-k\right)^{\alpha\beta}\left(D^{2}+M\right)}{\left(p-k\right)^{2}+M^{2}}\delta^{2}\left(\theta_{2}-\theta_{1}\right)\\
= & N4\pi\lambda\kappa\delta_{j}^{i}\intop\frac{d^{3}k}{\left(2\pi\right)^{3}}\\
\times & \Big[\left(\frac{\left(\left(p-k\right)^{2}-D^{2}M\right)\left(\kappa D^{2}+k^{2}\right)+\left(\kappa-D^{2}\right)\left(D^{2}+M\right)k\cdot\left(p-k\right)}{k^{2}\left(\kappa^{2}+k^{2}\right)\left(\left(p-k\right)^{2}+M^{2}\right)}\right)\Big|_{D^{2}}\\
+ & \frac{\begin{array}{c}
\left(\left(p-k\right)^{2}-D^{2}M\right)\Big|_{D^{2}}\left(\kappa D^{2}+k^{2}\right)\Big|_{D^{2}}\\
+\left(\kappa-D^{2}\right)\Big|_{D^{2}}\left(D^{2}+M\right)\Big|_{D^{2}}k\cdot\left(p-k\right)
\end{array}}{k^{2}\left(\kappa^{2}+k^{2}\right)\left(\left(p-k\right)^{2}+M^{2}\right)}D^{2}\Big],
\end{align}
where in the last line we used integration by parts to transfer the
superderivatives from the gauge propagator to the scalar propagators
and carry out the $\theta_{2}$ integration. One term arises when
the derivatives hit the internal propagator (a ``mass correction''),
while the other comes from the derivatives hitting the external propagator
(``field amplitude renormalization''). The ``mixed term'' will
vanish when $\theta_{2}$ is integrated as it contains an odd number
of superderivatives. Simplifying further we arrive at:
\begin{align}
 & \incgraph[propcorrection3]\label{eq: rainbow propagator correction 1-1}\\
= & N4\pi\lambda\kappa\delta_{j}^{i}\intop\frac{d^{3}k}{\left(2\pi\right)^{3}}\Big[\left(\frac{-k^{2}M+\left(p-k\right)^{2}\kappa+\left(\kappa-M\right)k\cdot\left(p-k\right)}{k^{2}\left(\kappa^{2}+k^{2}\right)\left(\left(p-k\right)^{2}+M^{2}\right)}\right)\nonumber \\
+ & \frac{-M\kappa-k\cdot\left(p-k\right)}{k^{2}\left(\kappa^{2}+k^{2}\right)\left(\left(p-k\right)^{2}+M^{2}\right)}D^{2}\Big].\nonumber 
\end{align}

\subsection{Vaccuum Bubbles}

First, using \eqref{eq:phi propagator correction - gauge tadpole}:
\begin{align}
\frac{1}{N}\begin{gathered}\incgraph[vacuum1]\end{gathered}
 & =\begin{array}{c}
      {\displaystyle \frac{1}{N}\intop\frac{d^{3}p}{\left(2\pi\right)^{3}}\left(-\frac{1}{D\left(p\right)^{2}+M}\right)\Big|_{D^{2}} }  \\ \\
      {\displaystyle \times \left(-2\pi\lambda\kappa^{2}\delta_{i}^{i}\intop\frac{d^{3}k}{\left(2\pi\right)^{3}}\frac{1}{k^{2}\left(\kappa^{2}+k^{2}\right)}\right) } 
 \end{array} \nonumber\\
 & =-2\pi\lambda\kappa^{2}\intop\frac{d^{3}p}{\left(2\pi\right)^{3}}\frac{1}{p^{2}+M^{2}}\intop\frac{d^{3}k}{\left(2\pi\right)^{3}}\frac{1}{k^{2}\left(\kappa^{2}+k^{2}\right)}.\label{eq:2-loop vacuum bubble 1}
\end{align}
Next we will evaluate:
\begin{equation}
\frac{1}{N}\begin{gathered}\incgraph[vacuum2]\end{gathered}
.
\end{equation}
First note that:
\begin{equation}
\intop d^{2}\theta_{2}\delta^{2}\left(\theta_{2}-\theta_{1}\right)\left(a+bD^{2}\right)\frac{D^{2}-M}{p^{2}+M^{2}}\delta^{2}\left(\theta_{2}-\theta_{1}\right)=\frac{a-bM}{p^{2}+M^{2}}.
\end{equation}
With this and with \eqref{eq: rainbow propagator correction 1-1},
and including a symmetry factor of $1/2$:
\begin{align}
 & \frac{1}{N}\begin{gathered}\incgraph[vacuum2]\end{gathered}
\\
= & 2\pi\lambda\kappa\delta_{j}^{i}\intop\frac{d^{3}p}{\left(2\pi\right)^{3}}\frac{d^{3}k}{\left(2\pi\right)^{3}}\frac{\begin{array}{c}
-k^{2}M+\left(p-k\right)^{2}\kappa+\left(\kappa-M\right)k\cdot\left(p-k\right)\\
-M\left(-M\kappa-k\cdot\left(p-k\right)\right)
\end{array}}{k^{2}\left(\kappa^{2}+k^{2}\right)\left(\left(p-k\right)^{2}+M^{2}\right)\left(p^{2}+M^{2}\right)}
\end{align}
Now we can decompose, the numerator into a sum of inverse propagators,
using the identity:
\begin{equation}
k\cdot\left(p-k\right)=\frac{1}{2}\left(-\left(p-k\right)^{2}-M^{2}-k^{2}+p^{2}+M^{2}\right),
\end{equation}
which gives:
\begin{align}
\dots= & -\left(2M+\kappa\right)\pi\lambda\kappa\intop\frac{d^{3}p}{\left(2\pi\right)^{3}}\frac{d^{3}k}{\left(2\pi\right)^{3}}\frac{1}{\left(\kappa^{2}+k^{2}\right)\left(\left(p-k\right)^{2}+M^{2}\right)\left(p^{2}+M^{2}\right)}\\
+ & \pi\lambda\kappa^{2}\intop\frac{d^{3}p}{\left(2\pi\right)^{3}}\frac{d^{3}k}{\left(2\pi\right)^{3}}\left(\frac{1}{k^{2}\left(\kappa^{2}+k^{2}\right)\left(p^{2}+M^{2}\right)}+\frac{1}{k^{2}\left(\kappa^{2}+k^{2}\right)\left(\left(p-k\right)^{2}+M^{2}\right)}\right)\\
= & -\left(2M+\kappa\right)\pi\lambda\kappa\intop\frac{d^{3}p}{\left(2\pi\right)^{3}}\frac{d^{3}k}{\left(2\pi\right)^{3}}\frac{1}{\left(\kappa^{2}+k^{2}\right)\left(\left(p-k\right)^{2}+M^{2}\right)\left(p^{2}+M^{2}\right)}\\
+ & 2\pi\lambda\kappa^{2}\intop\frac{d^{3}p}{\left(2\pi\right)^{3}}\frac{1}{p^{2}+M^{2}}\intop\frac{d^{3}k}{\left(2\pi\right)^{3}}\frac{1}{k^{2}\left(\kappa^{2}+k^{2}\right)},
\end{align}
where in the third integral we used a shift $p\to p+k$. So we find:
\begin{align}
 & \frac{1}{N}\begin{gathered}\incgraph[vacuum2]\end{gathered}
\nonumber \\
= & -\left(2M+\kappa\right)\pi\lambda\kappa\intop\frac{d^{3}p}{\left(2\pi\right)^{3}}\frac{d^{3}k}{\left(2\pi\right)^{3}}\frac{1}{\left(\kappa^{2}+k^{2}\right)\left(\left(p-k\right)^{2}+M^{2}\right)\left(p^{2}+M^{2}\right)}\nonumber \\
+ & 2\pi\lambda\kappa^{2}\intop\frac{d^{3}p}{\left(2\pi\right)^{3}}\frac{1}{p^{2}+M^{2}}\intop\frac{d^{3}k}{\left(2\pi\right)^{3}}\frac{1}{k^{2}\left(\kappa^{2}+k^{2}\right)}.\label{eq:2 loop vacuum bubble 2}
\end{align}
Putting \eqref{eq:2-loop vacuum bubble 1} and \eqref{eq:2 loop vacuum bubble 2}
together we get:
\begin{align}
 & \frac{1}{N}\begin{gathered}\incgraph[vacuum1]\end{gathered}
+\frac{1}{N}\begin{gathered}\incgraph[vacuum2]\end{gathered}
\label{eq:2 loop sum of vacuum bubbles}\\
= & -\left(2M+\kappa\right)\pi\lambda\kappa\intop\frac{d^{3}p}{\left(2\pi\right)^{3}}\frac{d^{3}k}{\left(2\pi\right)^{3}}\frac{1}{\left(\kappa^{2}+k^{2}\right)\left(\left(p-k\right)^{2}+M^{2}\right)\left(p^{2}+M^{2}\right)}\nonumber 
\end{align}

\section{\label{sec:Field-Amplitude-Renormalization}Field Amplitude Renormalization
of \texorpdfstring{$\Phi$}{Phi}}

Equations \eqref{eq: M_IR definition} and \eqref{eq:m_IR definition}
imply that the singlet field $\sigma$ is renormalized differently
in the two regularization schemes:
\begin{equation}
\sigma_{{\rm YM-reg}}=\left(1-2\left|\lambda\right|\right)^{-1/2}\sigma_{{\rm dim-reg}}.
\end{equation}
We can check whether this field amplitude renormalization is reproduced,
to leading order in $\lambda$, in a standard computation. $\sigma$
is defined via its equation of motion:
\begin{equation}
\left\langle \sigma\right\rangle =\omega\left\langle \phisq\right\rangle ,
\end{equation}
and $\omega$ and $m$ via \eqref{eq:UV_action_defs}:
\begin{equation}
\mathcal{L}=-\frac{\sigma^{2}}{2\omega}+\left(\sigma+m\right)\phisq+\dots
\end{equation}
Suppose the amplitude of $\Phi$ is renormalized by $Z^{1/2}$:
\begin{equation}
\phisq\to Z\phisq,
\end{equation}
then we must simultaneously transform \footnote{This is essentially the ``anomalous dimesion'' $\gamma$ term of
the Callan-Symanzik equation, responsible for the renormalization
of a coupling due to field amplitude renormalizations.}:
\begin{align}
m & \to Z^{-1}m\\
\sigma & \to Z^{-1}\sigma\\
\omega & \to Z^{-2}\omega,
\end{align}
or, one could say:
\begin{align}
m_{{\rm IR}} & =Zm\\
\sigma_{{\rm IR}} & =Z\sigma\\
\omega_{{\rm IR}} & =Z\omega.
\end{align}
So based on \eqref{eq: M_IR definition}, \eqref{eq:m_IR definition},
\eqref{eq:omega_IR definition} and \eqref{eq:Phi_IR definition}
we expect:
\begin{equation}
Z\approx\sqrt{1-2\left|\lambda\right|}\approx1-\left|\lambda\right|.
\end{equation}
To check this we need to compute $Z$. Since $\Phi$ is defined by
canonical normalization of the kinetic term, we can find $Z$ through
the propagator:
\begin{equation}
\left\langle \phisq\right\rangle =Z\left\langle \bar{\Phi}_{{\rm IR}}\Phi_{{\rm IR}}\right\rangle =-\frac{Z}{D^{2}+M+\delta M},
\end{equation}
where $M$ here should be thought of as the pole mass around the chosen
vacuum, rather then some dynamical variable. Suppose we write using
some small $a,b$:
\begin{align}
\left\langle \phisq\right\rangle  & =-\frac{1}{D^{2}\left(1+a\right)+M\left(1+b\right)}\approx-\frac{1}{D^{2}+M}+\frac{1}{D^{2}+M}\frac{aD^{2}+bM}{D^{2}+M}+\dots\\
Z & \approx1-a\\
\delta M & \approx M\left(b-a\right),
\end{align}
so we expect:
\begin{equation}
a_{{\rm YM-reg}}-a_{{\rm dim-reg}}=\left|\lambda\right|.\label{eq:amplitude renormalization discrepancy}
\end{equation}

$a$ gets a contribution from \eqref{eq: rainbow propagator correction 1-1}:
\footnote{The other propagator correction, \eqref{eq:phi propagator correction - gauge tadpole},
does not contribute to $a$ as there is no term proportional to $D^{2}$
(it is simply a mass correction). }
\begin{align}
 & \incgraph[propcorrection3]\\
= & N4\pi\lambda\kappa\delta_{j}^{i}\intop\frac{d^{3}k}{\left(2\pi\right)^{3}}\Big[\left(\frac{-k^{2}M+\left(p-k\right)^{2}\kappa+\left(\kappa-M\right)k\cdot\left(p-k\right)}{k^{2}\left(\kappa^{2}+k^{2}\right)\left(\left(p-k\right)^{2}+M^{2}\right)}\right)\\
 & +\frac{-M\kappa-k\cdot\left(p-k\right)}{k^{2}\left(\kappa^{2}+k^{2}\right)\left(\left(p-k\right)^{2}+M^{2}\right)}D^{2}\Big].
\end{align}
We focus on the $a$ term. We want to take the $p\to0$ limit, so
that we're measuring the generation of the kinetic term, rather than
some higher-derivative term:
\begin{eqnarray*}
 &  & 4\pi\lambda\kappa\intop\frac{d^{3}k}{\left(2\pi\right)^{3}}\frac{-\kappa M+k^{2}}{k^{2}\left(\kappa^{2}+k^{2}\right)\left(k^{2}+M^{2}\right)}D^{2}\\
 & = & 4\pi\lambda\intop\frac{d^{3}k}{\left(2\pi\right)^{3}}\frac{-M\left(k^{2}+\kappa^{2}\right)+k^{2}\left(\kappa+M\right)}{k^{2}\left(\kappa^{2}+k^{2}\right)\left(k^{2}+M^{2}\right)}D^{2}\\
 & = & \lambda\left(-\frac{M}{\left|M\right|}+\frac{M+\kappa}{\left|M\right|+\left|\kappa\right|}\right)D^{2}\\
\kappa\to\infty\Rightarrow & \to & \lambda\left(-{\rm sign}\left(M\right)+{\rm sign}\left(\lambda\right)\right)D^{2}.
\end{eqnarray*}
Had we instead taken $\kappa\to\infty$ \textbf{prior} to integration:
\begin{eqnarray*}
 &  & 4\pi\lambda\intop\frac{d^{3}k}{\left(2\pi\right)^{3}}\frac{-M\left(k^{2}+\kappa^{2}\right)+k^{2}\left(\kappa+M\right)}{k^{2}\left(\kappa^{2}+k^{2}\right)\left(k^{2}+M^{2}\right)}D^{2}\\
 & \to & -4\pi\lambda M\intop\frac{d^{3}k}{\left(2\pi\right)^{3}}\frac{1}{k^{2}\left(k^{2}+M^{2}\right)}D^{2}\\
 & = & -\lambda{\rm sign}\left(M\right)D^{2}.
\end{eqnarray*}
It appears that:
\begin{align}
a_{{\rm YM-reg}} & =-\lambda{\rm sign}\left(M\right)+\left|\lambda\right|\\
a_{{\rm dim-reg}} & =-\lambda{\rm sign}\left(M\right)\\
a_{{\rm YM-reg}}-a_{{\rm dim-reg}} & =\left|\lambda\right|,
\end{align}
which matches \eqref{eq:amplitude renormalization discrepancy} as
expected.

\section{Gauge Propagator}\label{app:Gauge_propagator}

We review some aspects of the gauge propagator in $\xi$ gauge and
in Landau gauge, which are relevant to the computation in the unHiggsed
phase; and in unitary gauge, which is relevant in the Higgsed phase.

\subsection{\texorpdfstring{$\xi$}{xi} and Landau Gauge}

The quadratic part of the gauge field action with gauge fixing terms
in $\xi$-gauge is given by:
\begin{align}
\mathcal{L}_{2\,{\rm CS}+{\rm YM}} & =-\frac{k}{16\pi}{\rm tr}\left(\Gamma^{\alpha}\left(D_{\beta}D_{\alpha}+\frac{1}{\xi}D_{\alpha}D_{\beta}\right)\Gamma^{\beta}\right)\\
 & -\frac{1}{4g^{2}}{\rm tr}\left(\Gamma^{\alpha}\left(D_{\beta}D_{\alpha}-\frac{1}{\xi}D_{\alpha}D_{\beta}\right)D^{2}\Gamma^{\beta}\right),
\end{align}
which gives a propagator:
\begin{eqnarray}
\Delta^{\alpha\beta ab} & \equiv & \left\langle \Gamma^{\alpha a}\left(-p\right)\Gamma^{\beta b}\left(p\right)\right\rangle \nonumber \\
 & = & \incgraph[freegaugeprop]\nonumber \\
 & = & \delta^{ab}\frac{4\pi\lambda}{N}\kappa\frac{D^{\alpha}D^{\beta}\left(\kappa-D^{2}\right)+\xi\left(\kappa+D^{2}\right)D^{\beta}D^{\alpha}}{p^{2}\left(\kappa^{2}+p^{2}\right)},\label{eq:xi gauge propagator}
\end{eqnarray}
where $\kappa\equiv\frac{kg^{2}}{4\pi}$ is the Yang Mills mass. Note
that here the superderivatives are acting ``to the right''. The
form which acts to the left can be obtained simply by the identity:
\begin{equation}
D^{\alpha}\left(\theta_{2},p\right)\delta^{2}\left(\theta_{2}-\theta_{1}\right)=-D^{\alpha}\left(\theta_{1},-p\right)\delta^{2}\left(\theta_{2}-\theta_{1}\right).
\end{equation}
One finds:
\begin{align}
D^{\alpha}\left(\theta_{2},p\right)D^{\beta}\left(\theta_{2},p\right) & \leftrightarrow-D^{\beta}\left(\theta_{1},-p\right)D^{\alpha}\left(\theta_{1},-p\right)\\
D^{2}\left(\theta_{2},p\right) & \leftrightarrow D^{2}\left(\theta_{1},-p\right),
\end{align}
etc. The overall minus sign is a manifestation of Fermi statistics.

We will work in Landau gauge $\xi=0$ where the propagator becomes:
\begin{equation}
\left\langle \Gamma^{\alpha a}\Gamma^{\beta b}\right\rangle _{{\rm free}}=\incgraph[freegaugeprop][.25]=\frac{4\pi\lambda}{N}\kappa\frac{D^{\alpha}D^{\beta}\left(\kappa-D^{2}\right)}{p^{2}\left(\kappa^{2}+p^{2}\right)}.\label{eq:Landau_gauge_propagator-1}
\end{equation}
This choice satisfies ``transversality of the propagator'':
\begin{equation}
\Delta^{\alpha\beta}D_{\alpha}=D_{\beta}\Delta^{\alpha\beta}=0,
\end{equation}
which in particular means that $D^{\alpha}\Gamma_{\alpha}$ terms
decouple entirely. This is because:
\begin{align}
\left\langle D_{\alpha}\Gamma^{\alpha}\Gamma^{\beta}\right\rangle _{{\rm free}} & \sim D_{\alpha}\left(\theta_{1},-p\right)D^{\alpha}\left(\theta_{2},p\right)D^{\beta}\left(\theta_{2},p\right)\delta^{2}\left(\theta_{2}-\theta_{1}\right)\\
 & =D^{\alpha}\left(\theta_{2},p\right)D^{\beta}\left(\theta_{2},p\right)D_{\alpha}\left(\theta_{1},-p\right)\delta^{2}\left(\theta_{2}-\theta_{1}\right)\\
 & =-D\left(\theta_{2},p\right)D^{\beta}\left(\theta_{2},p\right)D_{\alpha}\left(\theta_{2},p\right)\delta^{2}\left(\theta_{2}-\theta_{1}\right)
\end{align}
which vanishes due to the identity:
\begin{equation}
D^{\alpha}D^{\beta}D_{\alpha}=0.
\end{equation}

At higher loops one would have to include Faddeev-Popov ghost superfields. For a discussion of this, see \cite{Gates:1983nr,Choi:2018ohn}.

\subsection{\label{subsec:Unitary Gauge}Unitary Gauge}

It's not possible to impose both $\xi$ gauge and unitary gauge simultaneously,
but suppose we included the $\xi$-gauge fixing terms in the Lagrangian,
then the quadratic part of the gauge field path integral is:
\begin{equation}
\intop D\Gamma\exp\left(\intop d^{5}z\mathcal{L}_{{\rm quad}}\right)
\end{equation}
\begin{align}
\mathcal{L}_{{\rm quad}}= & -\frac{1}{2}\frac{N}{16\pi\lambda\kappa}\left(\Gamma^{\alpha a}\left(D_{\beta}D_{\alpha}+\frac{1}{\xi}D_{\alpha}D_{\beta}\right)\kappa\Gamma^{\beta a}\right)\label{eq:Higgsed quadratic Lagrangian}\\
+ & \intop-\frac{1}{2}\frac{N}{16\pi\lambda\kappa}\left(\Gamma^{\alpha a}\left(D_{\beta}D_{\alpha}-\frac{1}{\xi}D_{\alpha}D_{\beta}\right)D^{2}\Gamma^{\beta a}\right)\nonumber \\
+ & N\intop-\frac{1}{4}C_{\beta\alpha}\Gamma^{a\alpha}\phi^{ab}\Gamma^{b\beta},\nonumber 
\end{align}
where, in analogy with \cite{Choi:2018ohn}, we defined:
\begin{equation}
\phi^{ab}=\bar{\Phi}T^{(a}T^{b)}\Phi.
\end{equation}
This gives a propagator:
\begin{equation}
\left\langle \Gamma^{\alpha}\Gamma^{\beta}\right\rangle =\frac{4\pi\lambda}{N}\kappa\frac{D^{\alpha}D^{\beta}\left(\kappa-D^{2}\right)+\xi D^{\beta}D^{\alpha}\left(\kappa+D^{2}\right)-8\pi\lambda\kappa\xi\phi^{ab}C^{\alpha\beta}}{\left(\left(D^{2}+\frac{1}{2}\kappa\right)^{2}-\frac{1}{4}\left(\kappa^{2}+16\pi\lambda\kappa\phi^{ab}\right)\right)\left(\left(D^{2}-\frac{1}{2}\kappa\right)^{2}-\frac{1}{4}\left(\kappa^{2}+16\pi\lambda\kappa\xi\phi^{ab}\right)\right)}.\label{eq:xi + unitary propagator}
\end{equation}
For $\phi^{ab}=0$ we reproduce the unHiggsed propagator \eqref{eq:xi gauge propagator}:
\begin{eqnarray*}
\left\langle \Gamma^{\alpha}\Gamma^{\beta}\right\rangle  & = & \frac{4\pi\lambda}{N}\kappa\frac{D^{\alpha}D^{\beta}\left(\kappa-D^{2}\right)+\xi D^{\beta}D^{\alpha}\left(\kappa+D^{2}\right)}{\left(\left(D^{2}+\frac{1}{2}\kappa\right)^{2}-\frac{1}{4}\kappa^{2}\right)\left(\left(D^{2}-\frac{1}{2}\kappa\right)^{2}-\frac{1}{4}\kappa^{2}\right)}\\
 & = & \frac{4\pi\lambda}{N}\kappa\frac{D^{\alpha}D^{\beta}\left(\kappa-D^{2}\right)+\xi D^{\beta}D^{\alpha}\left(\kappa+D^{2}\right)}{D^{2}\left(D^{2}+\kappa\right)D^{2}\left(D^{2}-\kappa\right)}\\
 & = & \frac{4\pi\lambda}{N}\kappa\frac{D^{\alpha}D^{\beta}\left(\kappa-D^{2}\right)+\xi D^{\beta}D^{\alpha}\left(\kappa+D^{2}\right)}{p^{2}\left(\kappa^{2}+p^{2}\right)}.
\end{eqnarray*}
In \eqref{eq:xi + unitary propagator} we see the emergence of the
two expected massive polarization states with masses:
\begin{equation}
\frac{\kappa}{2}\left(1\pm\sqrt{1+16\pi\lambda\frac{\phi^{\cdot\cdot}}{\kappa}}\right),\label{eq:Higgsed gauge masses-1}
\end{equation}
as well as two \textbf{unphysical} states with gauge-dependent masses:
\begin{equation}
\frac{\kappa}{2}\left(-1\pm\sqrt{1+16\pi\lambda\xi\frac{\phi^{\cdot\cdot}}{\kappa}}\right).
\end{equation}
For $\phi^{ab}\neq0$ it's possible to take the limit $\xi\to\infty$
which corresponds to not imposing $\xi$ gauge in the first place:
\begin{equation}
\left\langle \Gamma^{\alpha}\Gamma^{\beta}\right\rangle =-\frac{1}{N}\frac{D^{\beta}D^{\alpha}\left(\kappa+D^{2}\right)-8\pi\lambda\kappa\phi^{ab}C^{\alpha\beta}}{\left(\left(D^{2}+\frac{1}{2}\kappa\right)^{2}-\frac{1}{4}\left(\kappa^{2}+16\pi\lambda\kappa\phi^{ab}\right)\right)\phi^{ab}}.\label{eq:broken gauge propagator}
\end{equation}
This reflects the fact that unitary gauge is imposed, so the propagator
remains invertible, and the two unphysical states are gone. In reality
we should impose $\xi$ gauge along the \textbf{unbroken} directions,
and unitary gauge along the \textbf{broken} directions, so that \eqref{eq:broken gauge propagator}
governs the propagation of $W$ and $Z$ superfields. The splitting
into $W$ and $Z$ superfields can be made more precise, but we will
only need \eqref{eq:broken gauge propagator} in the following. Furthermore,
one should include ghost superfields, but those
play no part in leading order computations.

\section{\label{sec:Detailed Higgsed Phase Computations}Detailed Higgsed Phase Computations}

We rederive \eqref{eq: W 1 loop Higgsed adar chang-ha} which was obtained by \cite{Choi:2018ohn}. As discussed in subsubsection \ref{subsubsec: N counting Higgsed}, we use unitary gauge and obtain the same result as \cite{Choi:2018ohn} in a slightly different method. The computations in \cite{Choi:2018ohn} had a few numerical and sign errors that luckily cancel.

We go over the computation, and compare our results to those of \cite{Choi:2018ohn} at various steps as a sanity check. Portions of the calculation are relegated to the subsections \ref{subsec:The Trace Over Spinor Indices} and \ref{subsec:The Trace Over Adjoint Indices}.

From setting $\xi\to\infty$ in \eqref{eq:Higgsed quadratic Lagrangian},
we see that the 1-loop contribution to the superpotential is given
by the Pfaffian:
\begin{equation}
W_{1-{\rm loop}}=\frac{1}{N}{\rm Pf}\left(\delta^{ab}D_{\beta}D_{\alpha}\left(\kappa+D^{2}\right)+8\pi\lambda\kappa C_{\beta\alpha}\phi^{ab}\right),\label{eq: w 1 loop pfaffian form}
\end{equation}
where, in analogy with \cite{Choi:2018ohn}, we defined:
\begin{equation}
\phi^{ab}=\bar{\Phi}T^{(a}T^{b)}\Phi.
\end{equation}
\eqref{eq: w 1 loop pfaffian form} is simply the Gaussian approximation to the path integral in \eqref{eq: formal all orders higgsed superpotential}. Rather than evaluating this directly, it would be simpler to take
a derivative with respect to the magnitude of $\phi^{ab}$. This is
somewhat like computing the F-term equation directly. We do this in subsection \ref{subsec:The Trace Over Spinor Indices}, and arrive at:
\begin{align}
W_{\text{1-loop}} & =  -\frac{1}{2N}{\rm Tr}\left(D^{2}+\frac{\kappa}{2}\left(1+\sqrt{1+16\pi\lambda\frac{\phi^{\cdot\cdot}}{\kappa}}\right)\right)\label{eq: Higgsed 1-loop as 2 polarizations}\\
   & -\frac{1}{2N}{\rm Tr}\left(D^{2}+\frac{\kappa}{2}\left(1-\sqrt{1+16\pi\lambda\frac{\phi^{\cdot\cdot}}{\kappa}}\right)\right),\nonumber
\end{align}
where the trace ${\rm Tr}$ is over superfield and adjoint indices, but not spinor indices. In essence, we have carried out only the trace over spinor indices. \eqref{eq: Higgsed 1-loop as 2 polarizations} has the form of a pair of contributions, each from a different
polarization state, with the anticipated physical masses \eqref{eq:Higgsed gauge masses}.
Although it might seem we have obtained an $O\left(1/N\right)$ order
contribution, after taking the trace over Adjoint indices and keeping
the leading order that depends on $\phisq$, we should get
an $O\left(1\right)$ answer.

Note that using \eqref{eq:trace-log-scalar-formula} to carry out
the trace over superfield indices (the 1-loop integral) we get:
\begin{align}
W_{\text{1-loop}}  = & -\frac{1}{2N}\kappa\left|\kappa\right|{\rm Tr}_{{\rm adj}}\sqrt{1+16\pi\lambda\frac{\phi^{\cdot\cdot}}{\kappa}},\label{eq:evaluated 1 loop superpotential Higgsed adjoint trace}
\end{align}
in accordance with \cite{Choi:2018ohn}.

We evaluate the trace over adjoint indices in subsection \ref{subsec:The Trace Over Adjoint Indices} and obtain:
\begin{eqnarray}
W_{\text{1-loop}} & = & -{\rm Tr}\left(D^{2}+\frac{\kappa}{2}\left(1+\sqrt{1+8\pi\lambda\frac{\phisq}{\kappa}}\right)\right)\label{eq:W 1 loop Higgsed unevaluated appendix}\\
 &  & -{\rm Tr}\left(D^{2}+\frac{\kappa}{2}\left(1-\sqrt{1+8\pi\lambda\frac{\phisq}{\kappa}}\right)\right),\nonumber 
\end{eqnarray}
where now the trace is only over superfield indices ($\theta,p$).
The evaluated form is then (as in \eqref{eq:evaluated 1 loop superpotential Higgsed adjoint trace}):
\begin{equation}
W_{\text{1-loop}}=-\kappa\left|\kappa\right|\sqrt{1+8\pi\lambda\frac{\phisq}{\kappa}}.
\end{equation}

\subsection{\label{subsec:The Trace Over Spinor Indices}The Trace Over Spinor Indices}

We wish to evaluate:
\begin{equation}
W_{1-{\rm loop}}=\frac{1}{N}{\rm Pf}\left(\delta^{ab}D_{\beta}D_{\alpha}\left(\kappa+D^{2}\right)+8\pi\lambda\kappa C_{\beta\alpha}\phi^{ab}\right).
\end{equation}
Rather than evaluating this directly, it would be simpler to take
a derivative with respect to the magnitude of $\phi^{ab}$. Imagine setting
$\phi^{ab}\to v\phi^{ab}$ in \eqref{eq:Higgsed quadratic Lagrangian},
and then computing:
\begin{align}
\frac{\partial W_{1-{\rm loop}}}{\partial v}= & \frac{\partial}{\partial v}\frac{1}{{\rm Vol}^{3\mid2}}\log\left(\intop D\Gamma\exp\left(\intop d^{5}z\mathcal{L}_{{\rm quad}}\right)\right)\\
= & -N\frac{1}{4}C_{\beta\alpha}\phi^{ab}\frac{1}{{\rm Vol}^{3\mid2}}\intop d^{5}z\left\langle \Gamma^{\alpha a}\Gamma^{\beta a}\right\rangle _{\text{free}}\\
= & -N\frac{1}{4}C_{\beta\alpha}\phi^{ab}\left\langle \Gamma^{\alpha a}\Gamma^{\beta a}\right\rangle _{\text{free}},
\end{align}
Where we've omitted all non-quadratic parts of the Lagrangian, ignored
the overall scale of the fields and defined:
\begin{equation}
\left\langle \mathcal{O}\right\rangle _{\text{free}}=\frac{\intop D\Gamma\mathcal{O}\exp\left(\intop d^{5}z\mathcal{L}_{{\rm quad}}\right)}{\intop D\Gamma\exp\left(\intop d^{5}z\mathcal{L}_{{\rm quad}}\right)}.
\end{equation}
To evaluate this we can use the propagator \eqref{eq:broken gauge propagator}:
\begin{align}
 & -N\frac{1}{4}C_{\beta\alpha}\phi^{ab}\left\langle \Gamma^{\alpha a}\Gamma^{\beta a}\right\rangle _{\text{free}}\\
= & -N\frac{1}{4}C_{\beta\alpha}\phi^{ab}\left(\intop\frac{d^{3}p}{\left(2\pi\right)^{3}}\left(-\frac{1}{N}\frac{D^{\beta}D^{\alpha}\left(\kappa+D^{2}\right)-8\pi\lambda\kappa v\phi^{\cdot\cdot}C^{\alpha\beta}}{\left(\left(D^{2}+\frac{1}{2}\kappa\right)^{2}-\frac{1}{4}\left(\kappa^{2}+16\pi\lambda\kappa v\phi^{\cdot\cdot}\right)\right)v\phi^{\cdot\cdot}}\right)\Big|_{D^{2}}\right)_{ba}\\
= & -\frac{1}{2}\phi^{ab}\left(\intop\frac{d^{3}p}{\left(2\pi\right)^{3}}\left(\frac{D^{2}\left(\kappa+D^{2}\right)-8\pi\lambda\kappa v\phi^{\cdot\cdot}}{\left(D^{2}\left(\kappa+D^{2}\right)-4\pi\lambda\kappa v\phi^{\cdot\cdot}\right)v\phi^{\cdot\cdot}}\right)\Big|_{D^{2}}\right)_{ba}\\
= & 2\pi\lambda\kappa\phi^{ab}\left(\intop\frac{d^{3}p}{\left(2\pi\right)^{3}}\left(0+\frac{1}{\left(D^{2}\left(\kappa+D^{2}\right)-4\pi\lambda\kappa v\phi^{\cdot\cdot}\right)}\Big|_{D^{2}}\right)\right)_{ba}\\
= & 2\pi\lambda\kappa\phi^{ab}\left(\frac{1}{-4\pi\lambda\kappa\phi^{\cdot\cdot}}\frac{1}{\partial v}\intop\frac{d^{3}p}{\left(2\pi\right)^{3}}\left(\log\left(D^{2}\left(\kappa+D^{2}\right)-4\pi\lambda\kappa v\phi^{\cdot\cdot}\right)\right)\Big|_{D^{2}}\right)_{ba}\\
= & -\frac{1}{2}\left(\frac{1}{\partial v}\intop\frac{d^{3}p}{\left(2\pi\right)^{3}}\left(\log\left(D^{2}\left(\kappa+D^{2}\right)-4\pi\lambda\kappa v\phi^{\cdot\cdot}\right)\right)\Big|_{D^{2}}\right)_{aa}\\
= & -\frac{1}{2}{\rm Tr}_{{\rm adj}}\left(\frac{1}{\partial v}\intop\frac{d^{3}p}{\left(2\pi\right)^{3}}\left(\log\left(D^{2}\left(\kappa+D^{2}\right)-4\pi\lambda\kappa v\phi^{\cdot\cdot}\right)\right)\Big|_{D^{2}}\right).
\end{align}
By taking an anti-derivative and setting $v=1$:
\begin{align}
NW_{\text{1-loop}}  = & -\frac{1}{2}{\rm Tr}\left(\log\left(D^{2}\left(\kappa+D^{2}\right)-4\pi\lambda\kappa\phi^{\cdot\cdot}\right)\right),
\end{align}
Where here the ${\rm Tr}$ represents a trace over all the indices
(adjoint, momentum, $\theta$) \textbf{besides} the spinor indices. For
greater clarity using:
\begin{equation}
D^{2}\left(\kappa+D^{2}\right)-4\pi\lambda\kappa\phi^{\cdot\cdot}=\left(D^{2}+\frac{1}{2}\kappa\right)^{2}-\frac{1}{4}\left(\kappa^{2}+16\pi\lambda\kappa v\phi^{\cdot\cdot}\right),
\end{equation}
we get:
\begin{eqnarray}
W_{\text{1-loop}} & = & -\frac{1}{2N}{\rm Tr}\left(D^{2}+\frac{\kappa}{2}\left(1+\sqrt{1+16\pi\lambda\frac{\phi^{\cdot\cdot}}{\kappa}}\right)\right) \label{eq:higgsed 1-loop traced over spinor}\\
 &  & -\frac{1}{2N}{\rm Tr}\left(D^{2}+\frac{\kappa}{2}\left(1-\sqrt{1+16\pi\lambda\frac{\phi^{\cdot\cdot}}{\kappa}}\right)\right),\nonumber
\end{eqnarray}
which has the form of a pair of contributions, each from a different
polarization state, with the anticipated physical masses \cite{Dunne:1998qy}.

\subsection{\label{subsec:The Trace Over Adjoint Indices}The Trace Over Adjoint Indices}

To evaluate the trace over adjoint indices in \eqref{eq:higgsed 1-loop traced over spinor}, we should expand the argument
of the trace in matrix powers of $\phi^{\cdot\cdot}$, and use the
approximate Fierz identity:
\begin{equation}
\left(T^{a}\right)_{j}^{i}\left(T^{b}\right)_{k}^{l}=\frac{1}{2}\delta_{k}^{i}\delta_{j}^{l}-\frac{1}{2N}\delta_{j}^{i}\delta_{k}^{l}\approx\frac{1}{2}\delta_{k}^{i}\delta_{j}^{l}.\label{eq:approximate Fierz}
\end{equation}
Also recall that $\phi$ can be written:
\begin{equation}
\phi^{ab}=\bar{\Phi}T^{(a}T^{b)}\Phi={\rm Tr}_{\text{fund}}\left(T^{(a}T^{b)}\bar{\Phi}\otimes\Phi\right),
\end{equation}
where $\bar{\Phi}\otimes\Phi$ is the matrix defined by $\left(\bar{\Phi}\otimes\Phi\right)_{j}^{i}=\bar{\Phi}^{i}\Phi_{j}$.
At $n$-th order this gives:
\begin{align}
 & {\rm Tr}_{\text{adj}}\left({\rm Tr}_{\text{fund}}\left(T^{(a}T^{b)}\bar{\Phi}\otimes\Phi\right)\right)^{n}\\
\approx & \frac{1}{2^{n}}\sum_{k=0}^{n}{\binom{n}{k}}{\rm Tr}_{{\rm fund}}\left(\bar{\Phi}\otimes\Phi\right)^{k}{\rm Tr}_{{\rm fund}}\left(\bar{\Phi}\otimes\Phi\right)^{n-k}\\
= & \frac{1}{2^{n}}{\rm Tr}_{{\rm bi-fund}}\left(\left(\bar{\Phi}\otimes\Phi\right)\otimes I_{{\rm fund}}+I_{{\rm fund}}\otimes\left(\bar{\Phi}\otimes\Phi\right)\right)^{n},
\end{align}
where in the second line, we expanded all the symmetrizations $T^{(a}T^{b)}\to T^{a}T^{b}+T^{b}T^{a}$,
giving $2^{n}$ terms, and applied the approximate Fierz identity
\eqref{eq:approximate Fierz} $n$ times. The result organizes into
a sum over the ``number of index switches'' $k$. In the third line
we simply wrote this as a trace over bi-fundamental indices. If $\Phi$
were an $N_{{\rm f}}\times N$ matrix whose main diagonal consisted
of the entries $\phi_{i},\,i=1,\dots,N_{{\rm f}}$ (and write $\phi_{i}=0$
for $N_{{\rm f}}<i\leq N$) and $\bar{\Phi}\otimes\Phi$ was defined
by contracting the flavour indices, this would give us:
\begin{equation}
{\rm Tr}_{{\rm adj}}\left(f\left(\phi^{\cdot\cdot}\right)\right)=\sum_{i,j=1}^{N}f\left(\frac{1}{2}\left(\phi_{i}^{2}+\phi_{j}^{2}\right)\right),
\end{equation}
in accordance with equation (4.2) of \cite{Choi:2018ohn}. This can be interpreted as a sum over the modes of the gauge field that diagonalize $\phi^{ab}$, replacing it with its \textbf{eigenvalues}. In our case,
of course, $N_{{\rm f}}=1$, so $\phi_{i}=\sqrt{\phisq}\delta_{i1}$
and we obtain:
\begin{equation}
f\left(\phisq\right)+2\left(N-1\right)f\left(\frac{1}{2}\phisq\right)+\left(N-1\right)^{2}f\left(0\right),\label{eq:3 contributions to the trace over ajoint indices}
\end{equation}
which has the expected form of a contribution from 1 $Z$ superfield, $N-1$ complex $W$ superfields and $\left(N-1 \right)^2$ unbroken gauge superfields. The leading $\Phi$-dependent contribution comes from
$\approx N$ complex W-Bosons. So we can write:
\begin{eqnarray}
W_{\text{1-loop}} & = & -{\rm Tr}\left(D^{2}+\frac{\kappa}{2}\left(1+\sqrt{1+8\pi\lambda\frac{\phisq}{\kappa}}\right)\right)\label{eq:W 1 loop Higgsed unevaluated appendix -1}\\
 &  & -{\rm Tr}\left(D^{2}+\frac{\kappa}{2}\left(1-\sqrt{1+8\pi\lambda\frac{\phisq}{\kappa}}\right)\right),\nonumber 
\end{eqnarray}
where now the trace is only over superfield indices ($\theta,p$).

\newpage
\raggedright
\printbibliography

\end{document}